\documentclass[prd,aps,amsmath,amssymb,
preprint,
reprint,
showpacs]{revtex4-1}

\usepackage{atlasphysics}
\usepackage{subfigure}
\usepackage{mathrsfs}
\usepackage{xspace}
\usepackage[hyperindex,breaklinks]{hyperref} 
\usepackage{graphicx}
\usepackage{float}

\newcommand{\dmttbar}{\ensuremath{{ d}m_{\ttbar}}}
\newcommand{\yttbar}{\ensuremath{y_{\ttbar}}}
\newcommand{\absyttbar}{\ensuremath{\left|y_{\ttbar}\right|}\xspace}

\newcommand{\dabsyttbar}{\ensuremath{{ d}\absyttbar}\xspace}
\newcommand{\mttbar}{\ensuremath{m_{\ttbar}}}
\newcommand{\mt}{\ensuremath{m_{\mathrm{t}}}}

\newcommand{\dptttbar}{\ensuremath{{ d}\pt^{\ttbar}}}
\newcommand{\dsigma}{\ensuremath{{ d}\sigma}}
\newcommand{\ptt}{\ensuremath{\pt^t}}
\newcommand{\dptt}{\ensuremath{{ d}\pt^t}}
\newcommand{\dX}{\ensuremath{{ d}X}}
\newcommand{\ptttbar}{\ensuremath{\pt^{\ttbar}}}
\newcommand{\Etmiss}{\ensuremath{\ET^{\rm miss}}}

\newcommand{\ejets}{{$e+$jets}\xspace}
\newcommand{\mujets}{{$\mu+$jets}\xspace}
\newcommand{\ljets}{{$\ell+$jets}\xspace}

\newcommand{\sigtt}{\ensuremath{\sigma_{\ttbar}}\xspace}

\newcommand{\mtw}{\ensuremath{m_{\mathrm{T}}^W}}

\newcommand{\lumitot}{\mbox{4.6\,fb$^{-1}$}}

\newcommand{\Alpgen}{{\sc Alpgen}\xspace}
\newcommand{\Powheg}{{\sc Powheg}\xspace}

\newcommand{\MCFM}{{\sc MCFM}\xspace}
\newcommand{\Jimmy}{{\sc Jimmy}\xspace}
\newcommand{\Pythia}{{\sc Pythia}\xspace}
\newcommand{\Herwig}{{\sc Herwig}\xspace}

\newcommand{\McAtNlo}{{\sc MC{@}NLO}\xspace}

\newcommand{\AcerMC}{{\sc AcerMC}\xspace}

\usepackage{preprintcover}  
\PreprintCoverPaperTitle{Measurements of normalized differential cross-sections for $t\bar{t}$ production in $pp$ collisions at $\sqrt{s}=7\,\TeV$ using the ATLAS detector}  
\PreprintIdNumber{CERN-PH-EP-2014-099}  
\PreprintCoverAbstract{Measurements of normalized differential cross-sections for top-quark pair production are presented as a~function of
the top-quark transverse momentum, and of the mass, transverse momentum, and rapidity of the \ttbar{} system, in proton--proton collisions at a~center-of-mass energy of $\sqrt{s}=7\,{\rm TeV}$. The dataset corresponds to an integrated luminosity of \mbox{4.6\,fb$^{-1}$}, recorded in 2011 with the ATLAS detector at the CERN Large Hadron Collider. Events are selected in the lepton+jets channel, requiring exactly one lepton and at least four jets with at least one of the jets tagged as originating from a~$b$-quark. The measured spectra are corrected for detector efficiency and resolution effects and are compared to several Monte Carlo simulations and theory calculations. The results are in fair agreement with the predictions in a~wide kinematic range. Nevertheless, data distributions are softer than predicted for higher values of the mass of the \ttbar{} system and of the top-quark transverse momentum. The measurements can also discriminate among different sets of parton distribution functions.}  
\PreprintJournalName{Physical Review D}  

\begin{document}

\title{Measurements of normalized differential cross-sections for $t\bar{t}$ production in $pp$ collisions at $\sqrt{s}=7\,\TeV$ using the ATLAS detector}

\author{The ATLAS Collaboration}

\begin{abstract}
Measurements of normalized differential cross-sections for top-quark pair production are presented as a~function of
the top-quark transverse momentum, and of the mass, transverse momentum, and rapidity of the \ttbar{} system, in proton--proton collisions at a~center-of-mass energy of $\sqrt{s}=7\,{\rm TeV}$. The dataset corresponds to an integrated luminosity of \lumitot{}, recorded in 2011 with the ATLAS detector at the CERN Large Hadron Collider. Events are selected in the lepton+jets channel, requiring exactly one lepton and at least four jets with at least one of the jets tagged as originating from a~$b$-quark. The measured spectra are corrected for detector efficiency and resolution effects and are compared to several Monte Carlo simulations and theory calculations. The results are in fair agreement with the predictions in a~wide kinematic range. Nevertheless, data distributions are softer than predicted for higher values of the mass of the \ttbar{} system and of the top-quark transverse momentum. The measurements can also discriminate among different sets of parton distribution functions.

\end{abstract}

\maketitle

\section{Introduction}\label{sec:Introduction}

Top-quark measurements have entered a~high-precision era at the Large Hadron Collider (LHC) where the cross-sections for single top-quark and top-quark pair (\ttbar{}) production at a~center-of-mass energy $\sqrt{s}=7\,{\rm TeV}$ are factors of 40 and 20 higher than at the Tevatron. The large number of $t\bar{t}$ events makes it possible to measure precisely the $\ttbar{}$ production cross-sections differentially, providing precision tests of current predictions based on perturbative Quantum Chromodynamics (QCD). The top quark plays an important role in many theories beyond the Standard Model (SM)~\cite{MochUwer2008} and differential measurements have been proposed to be sensitive to new-physics effects~\cite{Frederix:2009}. 

The inclusive cross-section for \ttbar{} production (\sigtt) in proton--proton ($pp$) collisions at a~center-of-mass energy $\sqrt{s}=7\,{\rm TeV}$ has been measured by both the ATLAS and CMS experiments with increasing precision in a~variety of channels~\cite{atlasXsec1,atlasXsec2,atlasXsec3,cmsXsec1,cmsXsec2,cmsXsec3,cmsXsec4}. The CMS Collaboration has published~\cite{cmsDiff} differential cross-sections using the full dataset collected in 2011 at $\sqrt{s}=7$ \TeV{} and corresponding to an integrated luminosity of 5.0 fb$^{-1}$. The ATLAS Collaboration has published~\cite{atlasDiff} the differential cross-sections as a~function of the mass ($\mttbar$), the transverse momentum ($\ptttbar$), and the rapidity ($\yttbar$) of the $\ttbar$ system with a~subset of the data collected in 2011 at $\sqrt{s}=7$ \TeV{} corresponding to an integrated luminosity of 2.05 fb$^{-1}$. The measurements shown here improve the statistical precision of the previous ATLAS results by including the full 2011 dataset (\lumitot). Furthermore, improved reconstruction algorithms and calibrations are used, thereby significantly reducing the systematic uncertainties affecting the measurements. The rapidity distribution is symmetrized and presented as $\absyttbar$ and in addition to the variables previously shown, this paper also presents a~measurement of the cross-section as a~function of the top-quark transverse momentum ($\ptt$).

In the SM, the top quark decays almost exclusively into a~\Wboson{} boson and a~$b$-quark. The signature of a~\ttbar{} decay is therefore determined by the \Wboson{} boson decay modes. This analysis makes use of the lepton$+$jets decay mode, where one \Wboson{} boson decays into an electron or muon and a~neutrino and the other \Wboson{} boson decays into a~pair of quarks, with the two decay modes referred to as the $e$+jets and $\mu$+jets channel, respectively. Events in which the \Wboson{} boson decays to an electron or muon through a~$\tau$~decay are also included.

Kinematic reconstruction of the \ttbar{} system is performed using a~likelihood fit. The results are unfolded to the parton level after QCD radiation, and the normalized differential cross-section measurements are compared to the predictions of Monte Carlo (MC) 
generators and next-to-leading-order (NLO) QCD calculations. The \ptt{}, \mttbar{} and \ptttbar{} spectra are also compared to NLO QCD calculations including next-to-next-to-leading-logarithmic (NNLL) effects, namely Ref.~\cite{NNLO_calc} for \ptt, Ref.~\cite{nnloMtt} for \mttbar{} and Ref.~\cite{PhysRevLett.110.082001,PhysRevD.88.074004} for \ptttbar{}.

The paper is organized as follows. Section~\ref{sec:Detector} briefly describes the ATLAS detector, while Secs~\ref{sec:DataSamples} and~\ref{sec:Simulation} describe the data and simulation samples used in the measurements. The reconstruction of physics objects, the event selection and the kinematic reconstruction of the events are explained in Sec.~\ref{sec:EventReco}. Section~\ref{sec:BackgroundDetermination} discusses the background processes affecting these measurements. Event yields for both the signal and background samples, as well as distributions of measured quantities before unfolding, are shown in Sec.~\ref{sec:YieldsAndPlots}. The measurements of the cross-sections, including the unfolding and combination procedures, are described in Sec.~\ref{sec:XSDetermination}. Statistical and systematic uncertainties are discussed in Sec.~\ref{sec:Uncertainties}. The results are presented in Sec.~\ref{sec:Results} and the comparison with theoretical predictions is discussed in Sec.~\ref{sec:Interpretation}.

\section{The ATLAS Detector}\label{sec:Detector}

The ATLAS detector~\cite{atlasDetector3} is cylindrically symmetric and has a~barrel and two endcaps
~\footnote{ATLAS uses a~right-handed coordinate system with its origin at the nominal interaction point (IP) in the center of the detector and the $z$-axis along the beam pipe. The $x$-axis points from the IP to the center of the LHC ring, and the $y$-axis points upward. Cylindrical coordinates $(r,\phi)$ are used in the transverse plane, $\phi$ being the azimuthal angle around the beam pipe. The pseudorapidity is defined in terms of the polar angle $\theta$ as $\eta=-\ln\tan(\theta/2)$. The distance in $\eta$--$\phi$ coordinates is $\Delta R=\sqrt{(\Delta\eta)^2+(\Delta\phi)^2}$, also used to define cone radii.}.
The inner detector (ID) is nearest to the interaction point and contains three subsystems providing high-precision track reconstruction: a~silicon pixel detector (innermost), a~silicon microstrip detector, and a~transition radiation tracker (outermost), which also helps to discriminate electrons from hadrons. The ID covers a~range of $|\eta|<2.5$. It is surrounded by a~superconducting solenoid, which produces a~2\,T axial field within the ID. Liquid argon (LAr) sampling electromagnetic (EM) calorimeters cover $|\eta|<4.9$, while the hadronic calorimeter uses scintillator tiles within $|\eta|<1.7$ and LAr within $1.7<|\eta|<4.9$. The outermost detector is the muon spectrometer, which employs three sets of air-core toroidal magnets with eight coils each and is composed of three layers of chambers for triggering ($|\eta| <$ 2.4) and precision track measurements ($|\eta| <$ 2.7). 

The trigger is divided into three levels referred to as Level 1 (L1), Level 2 (L2), and Event Filter (EF). The L1 trigger uses custom-made hardware and low-granularity detector data. The L2 and EF triggers are implemented as software algorithms. The L2 trigger has access to the full detector granularity, but only retrieves data for regions of the detector identified by L1 as containing interesting objects, while the EF system utilizes the full detector readout to reconstruct an event.

\section{Data Sample} \label{sec:DataSamples}

The dataset used in this analysis was recorded during $pp$ collisions at $\sqrt{s}=7\,{\rm TeV}$ in 2011. It only includes data recorded with stable beam conditions and with all relevant subdetector systems operational. The number of $pp$ collisions per bunch crossing significantly increased during the data taking, reaching mean values up to 20 in the last part of the 2011 LHC run. 

Single-muon and single-electron triggers were used to select the data. The single-muon trigger required at least one muon with transverse momentum (\pt{}) of at least $18\,$GeV and the single-electron trigger required at least one electron with a~\pt{} threshold of either $20$ or $22\,$GeV. The \pt{} threshold increased during data taking to cope with increased luminosity. With these requirements the total integrated luminosity of the dataset is \lumitot{} with an uncertainty of 1.8\%~\cite{lumi2011}.

\section{Simulation} \label{sec:Simulation}

Simulated \ttbar{} events with up to five additional light partons were generated using \Alpgen{}~\cite{ALPGEN} (v2.13) with the leading-order (LO) CTEQ6L1~\cite{cteq6l1} parton distribution functions (PDF). \Herwig{}~\cite{HERWIG} (v6.520) was used for parton showering and hadronization and \Jimmy{}~\cite{JIMMY} (v4.31) was used for the modeling of multiple parton interactions. The ATLAS AUET2 tune~\cite{tunesAUET2} was used for the simulation
of the underlying event. 
The \Alpgen{} generator uses tree-level matrix elements with a~fixed number of partons in the final state, with the MLM matching scheme~\cite{Mangano:2001xp} to avoid double counting between partons created in the hard process or in the subsequent parton shower. 

Two other generators, which make use of NLO QCD matrix elements with the NLO CT10 PDF~\cite{CT10}, are used for comparisons with the final measured results, namely \McAtNlo{}~\cite{MCATNLO} (v4.01) and \Powheg{}~\cite{POWHEGBOX} (\Powheg{}-hvq, patch4). Both are interfaced to \Herwig{} and \Jimmy{} with the ATLAS AUET2 tune. The \McAtNlo{} generator is also used for the evaluation of systematic uncertainties along with additional generators and simulation samples discussed in Sec.~\ref{sec:signalmodeling}. 
As an additional comparison the \Powheg{} generator is also interfaced to \Pythia{}6~\cite{Sjostrand:2006za},
with the Perugia 2011C tune~\cite{perugia}.

All of the simulation samples were generated assuming a~top-quark mass, $m_t$, equal to $172.5\,$GeV. The \ttbar{} samples are normalized to 
a~cross-section of $\sigtt = 167^{+17}_{-18}$~pb, obtained from approximate NNLO QCD calculations~\cite{HATOR} for $pp$ 
collisions at $\sqrt{s} = 7 \tev$,  again using $m_t=172.5\,$GeV. During the completion of this analysis, a~calculation of the inclusive cross-section to full NNLO precision with additional NNLL corrections was published~\cite{Czakon:2013goa} and gives a~cross-section of $\sigma_{t\bar{t}} = 177.3^{+11.5}_{-12.0}$~pb at $\sqrt{s} = 7 \tev$ for the same top-quark mass. 
This change would only affect the results presented here by increasing the normalization of the dilepton \ttbar{} background. The corresponding effect on the final results would be at the sub-percent level and is covered by the assigned systematic uncertainties.

Single top-quark events produced via electroweak interactions were simulated using the \AcerMC{} generator~\cite{ACERMC} (v3.8) interfaced to \Pythia{}6 with the MRSTMCal PDF~\cite{MRST2007LO} for the $t$-channel process and \McAtNlo{} for the $s$-channel and $\Wboson t$-channel processes. The production of \Wboson{}/\Zboson{} bosons in association with jets (\Wboson+jets or \Zboson+jets) was simulated using \Alpgen{}+\Herwig{}. \Wboson{}+jets events containing heavy-flavor quarks ($Wbb$+jets, $Wcc$+jets, and $Wc$+jets) were generated separately using leading-order matrix elements with massive $b$- and $c$-quarks. An overlap-removal procedure was used to avoid double counting of heavy-flavor quarks between the matrix element and the parton shower evolution. Diboson events (\Wboson{}\Wboson{}, \Wboson{}\Zboson{}, \Zboson{}\Zboson{}) were generated using \Herwig{} with the MRSTMCal PDF. 

All the simulation samples account for multiple $pp$ interactions per bunch crossing (pile-up), including both the in-time (additional collisions within the same bunch crossing) and out-of-time (collisions from neighboring bunch crossings) contributions, using \Pythia{}6 and the ATLAS AMBT2B CTEQ6L1 tune~\cite{atlastune2} to simulate minimum bias events. The events were reweighted so that the distribution of the average number of interactions per bunch crossing matches that observed in the data. The samples were processed through the GEANT4~\cite{GEANT4} simulation of the ATLAS detector~\cite{ATLASsim} and the standard ATLAS reconstruction software. Simulated events were corrected so that the trigger efficiency and physics object identification efficiencies, energy scales and energy resolutions match those determined in data control samples, with the exception of the electrons and jets, the energies of which were scaled in data to match the simulation.

\section{Event Reconstruction} \label{sec:EventReco}

The lepton+jets \ttbar{} decay mode is characterized by a~high-\pt{} lepton, two jets originating from $b$-quarks, two jets from the hadronic \Wboson{} boson decay, and missing transverse momentum due to the neutrino.

\subsection{Object Reconstruction and Identification}\label{sec:ObjectDef}

Primary vertices in the event are formed from reconstructed tracks such that they are spatially compatible with the luminous interaction region. The hard-scatter primary vertex is chosen to be the vertex with the highest $\sum \pt^2$ where the sum extends over all associated tracks with $\pt > 0.4\,$GeV.

The same electron definition as was used in the \ttbar{} cross-section measurement with 2010 data~\cite{atlas_ttxsec_2010} is adopted in this analysis, but optimized for the higher pile-up conditions of the 2011 data~\cite{Aad:2014fxa}. Strict quality requirements are applied to the shape of the energy deposition in the EM calorimeters and to the electron track variables~\cite{atlasElecPerf}. 
The resulting electron candidates are required to have transverse energy $\ET>25 \,$GeV and $|\eta_{\rm cluster}|< 2.47$, where $|\eta_{\rm cluster}|$ is the pseudorapidity of the EM cluster associated with the electron. In order to ensure high-quality reconstruction, candidates in the transition region between the barrel and endcap calorimeters, $1.37 < |\eta_{\rm cluster}| < 1.52$, and candidates matching the criteria for converted photons are rejected.

Muon candidates are reconstructed by combining track segments in different layers of the muon chambers~\cite{muonReso,Aad:2014zya}.
Such segments are assembled starting from the outermost layer, with a~procedure that takes material effects into account, and are then matched with tracks found in the ID. The candidates are then re-fitted using all hits from both the muon spectrometer and the ID, and are required to have $\pt >25\,$GeV and $|\eta|<2.5$.

Electron and muon candidates are required to be isolated in order to reduce the backgrounds from hadrons mimicking lepton signatures and leptons from heavy-flavor decays. 

For electrons, the isolation requirements are similar to the ones tuned for 2010 data~\cite{PUB2011006:elecperf} but optimized for the 2011 running conditions. The total transverse energy deposited in the calorimeter, in a~cone of size $\Delta R = 0.2$ around the electron candidate, is considered. 
The energy associated with the electron is subtracted, and corrections are made to account for the energy deposited by pile-up interactions. An analogous isolation requirement is applied using the sum of track \pt{} (excluding the electron track) in a~cone of $\Delta R = 0.3$ around the electron direction.
Isolation requirements on both the transverse energy and momentum are tuned as a~function of $\eta_{\rm cluster}$ and $\ET$ in order to ensure a~uniform $90\%$ efficiency for electrons from $Z\to ee$ decays satisfying the electron definition described above. 

For muon candidates, after subtracting the contributions from the muon itself, the total energy deposited in the calorimeter in a~cone of size  $\Delta R = 0.2$ around the muon direction is required to be below $4\,$GeV and the sum of track transverse momenta for tracks with $\pt>1\,$GeV in a~cone of $\Delta R = 0.3$ around the muon direction is required to be below $2.5\,$GeV. The above set of cuts has an efficiency of $88\%$ for simulated \ttbar{} signal events in the \mujets{} channel with a~negligible dependence on the pile-up conditions.

Jets are reconstructed from topological clusters~\cite{Lampl:2008zz} of energy depositions using the anti-$k_{t}$ algorithm~\cite{akt1} with a~radius parameter of $R=0.4$. The jet energy is first corrected for pile-up effects and then to the hadronic scale corresponding to the particle-level jets using energy and $\eta$-dependent correction factors derived from simulation~\cite{jer_2}. The energies of jets in data are further corrected, using in situ measurements, to match simulation~\cite{jes:2013}. Only jets with $\pt >25\,$GeV and $|\eta|<2.5$ are considered in the analysis. To suppress jets from in-time pile-up, the jet vertex fraction, defined as the sum of the \pt{} of tracks associated with the jet and originating from the primary vertex divided by the sum of the \pt{} from all tracks associated with the jet, is required to be greater than~0.75. 

The missing transverse momentum vector, {\bf \Etmiss{}}, is derived from the vector sum of calorimeter cell energies within $|\eta| < 4.9$ and corrected on the basis of the dedicated calibrations of the associated physics objects~\cite{atlasEtmisPerf}, including muons.
Calorimeter cells containing energy depositions above noise and not associated with high-\pt{} physics objects (referred to as the unassociated-cell term) are also included.

The identification of \ttbar{} events is improved by tagging jets originating from $b$-quarks using a~combination of three $b$-tagging algorithms~\cite{btag}. 
The results of the three taggers are combined using a~neural network resulting in a~single discriminating variable. The combined tagger operating point chosen for this analysis corresponds to a~tagging efficiency of 70\% for $b$-jets in simulated \ttbar{} events, while $c$-jets are suppressed by a~factor of five and light-flavor- and gluon-initiated jets are suppressed by a~factor of about 100.

\subsection{Event Selection}\label{sec:EventSelection}

Events are first required to pass either a~single-electron or single-muon trigger and the hard-scatter primary vertex is required to be constructed from at least five tracks with $p_{\rm T} > 0.4\,$GeV.

Leptons and jets are required to be well separated from each other to minimize ambiguities, background and systematic uncertainties. 
First, jets within  $\Delta R = 0.2$ of an electron satisfying the requirements described in Sec.~\ref{sec:ObjectDef}, 
but with the $p_{\rm T}$ threshold lowered to $15\,$GeV, are removed. If there is another jet found within $\Delta R = 0.4$, the electron is discarded. Finally muons within $\Delta R = 0.4$ of the axis of a~jet are removed. 

Events are required to contain exactly one isolated lepton and this lepton is required to have fired the trigger. Four or more jets where at least one jet is $b$-tagged are also required. In addition, events must satisfy $\Etmiss > 30\,$GeV and $\mtw > 35\,$GeV, where $\Etmiss$ is the magnitude of the missing transverse momentum vector {\bf \Etmiss{}} and the \Wboson{} boson transverse mass, $\mtw$, is defined as
\begin{equation}
\mtw = \sqrt{2p_{\rm T}^{\ell}p_{\rm T}^{\nu}(1-\cos(\phi^{\ell}-\phi^{\nu}))}\,,
\end{equation}
where $p_{\rm T}^{\ell}$ and $\phi^{\ell}$ are, respectively, the transverse momentum and the azimuthal angle of the lepton, $p_{\rm T}^{\nu}$ is identified at the reconstruction level with \Etmiss{} and $\phi^{\nu}$ is the azimuthal angle of {\bf \Etmiss{}}.

\subsection{Kinematic Reconstruction of the \ttbar{} System} \label{sec:TopSystemReconstruction}

A~kinematic likelihood fit~\cite{KLFit:2013} is used to fully reconstruct the $\ttbar$ kinematics. The algorithm relates the measured kinematics of the reconstructed objects (lepton, jets and {\bf \Etmiss{}}) to the leading-order representation of the $\ttbar$ system decay. The event likelihood ($\mathscr{L}$) is constructed as the product of Breit--Wigner (BW) distributions and transfer functions (TF)

\begin{equation}
\begin{split}
\mathscr{L} \equiv  {\rm TF}(\tilde{E}^{\ell},E^{\ell}) &\cdot \left( \prod_{i=1}^4 {\rm TF}(\tilde{E}_{{\rm jet}~i}, E_{{\rm quark}~i}) \right) \\
&\cdot\, {\rm TF}(E_x^{\rm miss}|p_x^{\nu}) \cdot \, {\rm TF}(E_y^{\rm miss}|p_y^{\nu}) \\
&\cdot \, {\rm BW}(m_{jj}|m_W) \cdot \, {\rm BW}(m_{\ell \nu}|m_W) \\
&\cdot \, {\rm BW}(m_{jjj}|m_{t}) \cdot {\rm BW}(m_{\ell \nu j}|m_{t})\,,
\end{split}
\end{equation}
where the Breit--Wigner distributions associate the {\bf \Etmiss{}}, lepton, and jets with \Wboson{} bosons and top quarks, making use of their known widths and masses. The top-quark mass used is $172.5\,$GeV. The transfer functions, derived from the \McAtNlo{}+\Herwig{} simulation of the \ttbar{} signal, represent the experimental resolutions in terms of the probability that the observed energy at reconstruction level ($\tilde{E}$) is produced by a~parton-level object with a~certain energy $E$. Transverse energy is used to parameterize the muon momentum resolution while lepton energy is used in the electron channel.

The missing transverse momentum is used as a~starting value for the neutrino \pt{}, with its longitudinal component ($p_z^{\nu}$) as a~free parameter in the kinematic likelihood fit. Its starting value is computed from the \Wboson{} mass constraint. If there are no real solutions for $p_z^{\nu}$ then zero is used as a~starting value. Otherwise, if there are two real solutions, the one giving the larger likelihood is used. The five highest-$\pt$ jets (or four if there are only four jets in the event) are used as input to the likelihood fit and the best four-jet combination is selected.

The likelihood is maximized as a~function of the energies of the $b$-quarks, the quarks from the hadronic \Wboson{}~boson decay, the charged lepton, and the components of the neutrino three-momentum. The maximization is performed by testing all possible permutations, assigning jets to partons. The likelihood is combined with the probability for a~jet to be $b$-tagged, given the parton from the top-quark decay it is associated with, to construct an event probability. The $b$-tagging efficiencies and rejection factors are used to promote permutations for which a~$b$-tagged jet is assigned to a~$b$-quark and penalize those where a~$b$-tagged jet is assigned to a~light quark. The permutation of jets with the highest event probability is retained. 

\begin{figure*}[!htbp]
\centering
\subfigure[]{ \includegraphics[width=0.45\textwidth]{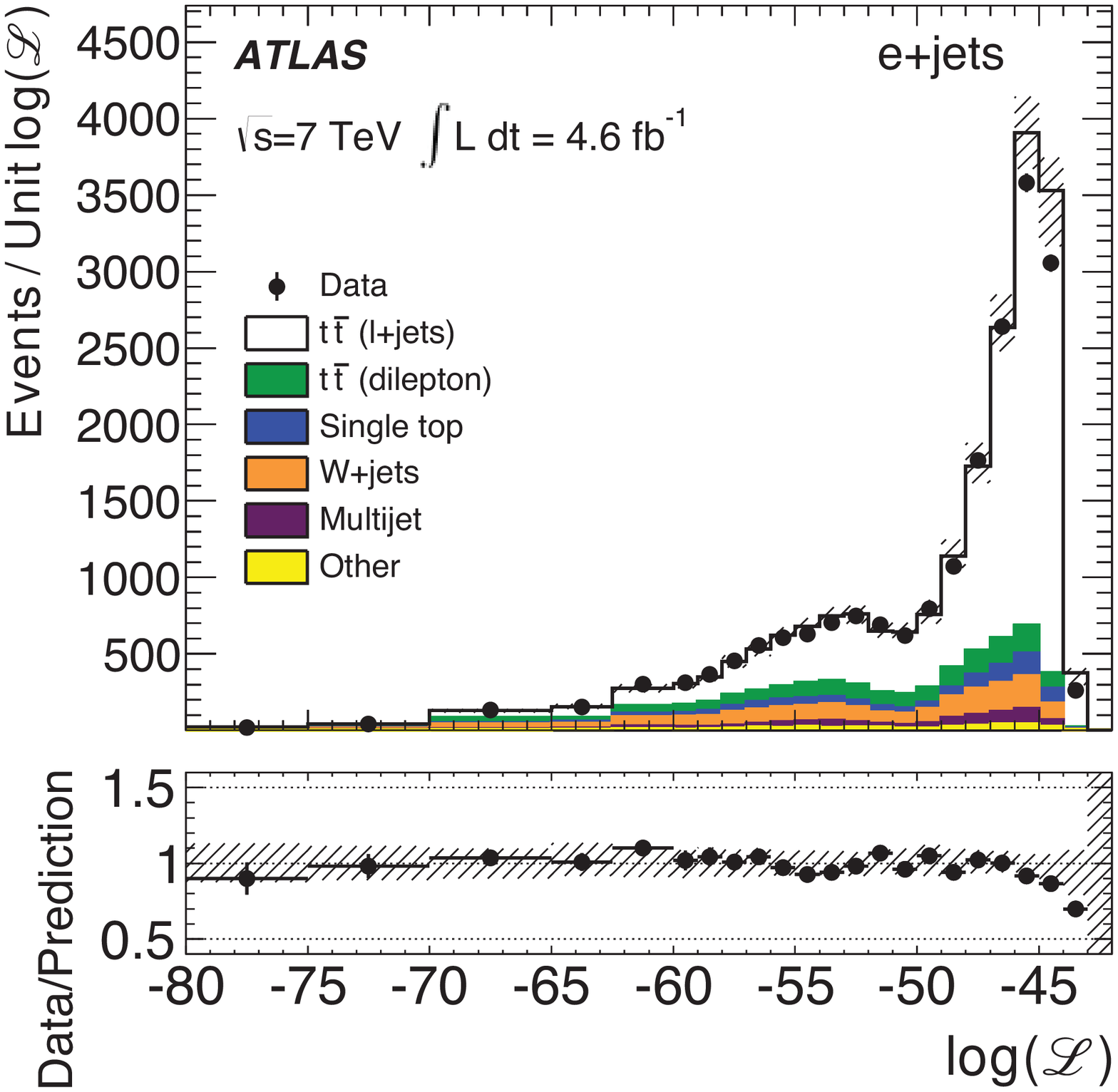}\label{lhood_el}}
\subfigure[]{ \includegraphics[width=0.45\textwidth]{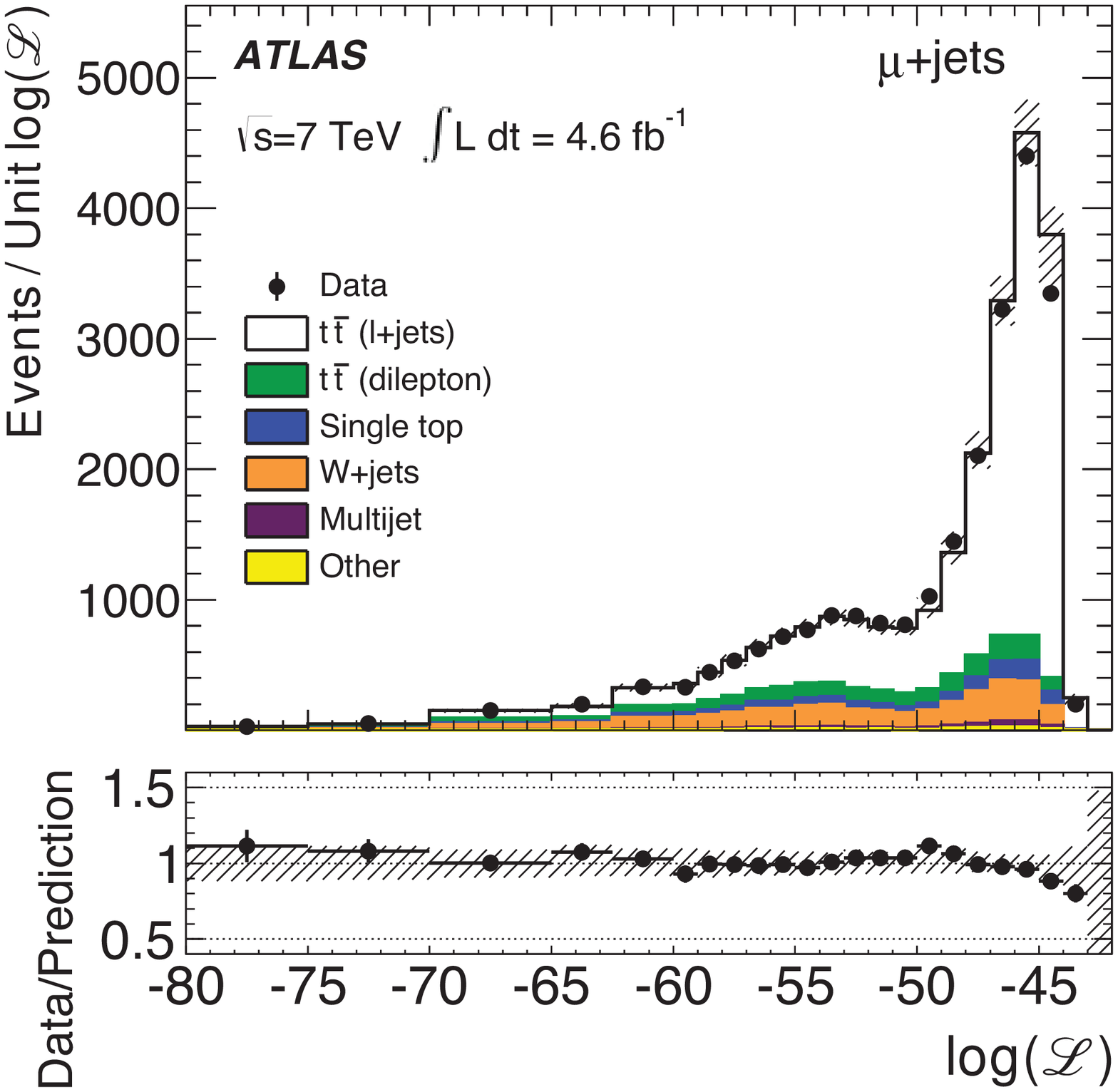}\label{lhood_mu}}
\caption{(Color online)  Distribution of the logarithm of the likelihood ($\log(\mathscr{L})$) obtained from the kinematic fit in the \subref{lhood_el}~\ejets{} and \subref{lhood_mu}~\mujets{} channels. Data distributions are compared to predictions, using \Alpgen{}+\Herwig{} as the $t\bar{t}$ signal model. The hashed area indicates the combined statistical and systematic uncertainties in the prediction, excluding systematic uncertainties related to the modeling of the $\ttbar$ system. Signal and background processes are shown in different colors, with ``Other" including the small backgrounds from diboson and $Z$+jets production. The lower parts of the figures show the ratios of data to the predictions.}
\label{fig:lhood_tagged}
\end{figure*}

The event likelihood must satisfy $\log{(\mathscr{L})} > -50$. This requirement provides a~good separation between properly and poorly-reconstructed events. 
Distributions of $\log{(\mathscr{L})}$ for data and simulation events  are shown in Fig.~\ref{fig:lhood_tagged} separately for the \ejets{} and \mujets{} channels. 
The data-to-MC ratio of the efficiency of the likelihood requirement is found to be 0.98 and the simulation is corrected for this difference. 
The full event selection, including this final requirement on the likelihood, is summarized in Table~\ref{tab:selectionList}.

Once the best likelihood is found, the four-momenta of both top quarks in the event are formed from their decay products as determined by the kinematic likelihood fit. One top quark is reconstructed from the fitted charged lepton, neutrino and one of the $b$-partons. This is referred to as the leptonically decaying top quark. The other, referred to as the hadronically decaying top quark, is reconstructed from the other three partons. 
The hadronically decaying top quark is selected to represent the top-quark \pt{} because the final result for this variable has smaller systematic uncertainties than the leptonically decaying top quark. The two spectra were compared and their results are compatible. 
The \ttbar{} system is the combination of the leptonically and hadronically decaying top quarks.

\begin{table}[!ht]
\begin{center}
\begin{tabular}{ l | l}
Event selection \\
\hline
\hline
Trigger & {\ }Single lepton \\
Primary vertex & {\ }$\ge5$ tracks with \pt{} $ >0.4\,$GeV \\
Exactly one & {\ }Muons: \pt{} $> 25\,$GeV, $|\eta| <  2.5$ \\
isolated lepton & {\ }Electrons: \pt{} $>$ $25\,$GeV \\
 & {\ }$|\eta|< 2.47$, excluding $1.37 < |\eta| < 1.52$ \\
$\ge4$ jets & {\ }\pt{} $> 25\,$GeV, $|\eta| < 2.5$ \\
$b$-tagging & {\ }$\geq 1$ $b$-tagged jet at $\epsilon_b=70\%$ \\
\Etmiss{} & {\ }\Etmiss{} $>$ $30\,$GeV \\
\mtw & {\ }\mtw{} $>$ $35\,$\GeV \\
Kinematic fit & {\ } $\log (\mathscr{L}) > -50$ \\
\hline
\end{tabular}
\end{center}
\caption{Summary of all requirements included in the event selection.}
\label{tab:selectionList}
\end{table}

\section{Background Determination} \label{sec:BackgroundDetermination}

After the event selection is applied, the largest background process is \Wboson{}+jets. Other backgrounds are due to multijet production, single top-quark electroweak production, diboson production, $Z$+jets production and the other decay channels associated with \ttbar{} production: the dilepton channel, which gives a~significant contribution, and the all-hadronic channel, which is found to be negligible. The \Wboson{}+jets and multijet backgrounds are determined using a~combination of simulation and data-driven techniques. The other backgrounds are determined from simulation and normalized to higher-order theoretical predictions.

\subsection{Simulated Background Contributions} \label{sec:SimulatedBackgrounds}

The single top-quark, dilepton \ttbar, \Zboson+jets, and diboson contributions are estimated from simulations and normalized to theoretical calculations of the inclusive cross-sections as follows. The single top-quark cross-section is normalized to the NLO+NNLL prediction: the $t$-channel to $64.6{}^{+2.6}_{-1.7}\,{\rm pb}$~\cite{Kidonakis:2011wy}, the $s$-channel to $4.6{}\pm 0.2\,{\rm pb}$~\cite{Kidonakis:2010tc}, and the $Wt$-channel to $15.7{} \pm 1.2 \,{\rm pb}$~\cite{Kidonakis:2010ux}. The dilepton \ttbar{} background is normalized to the same inclusive cross-section given in Sec.~\ref{sec:Simulation} for the signal $t\bar{t}\rightarrow \ell+$jets sample. The \Zboson+jets background is normalized to the NNLO QCD calculation for inclusive \Zboson{}~production~\cite{Anastasiou:2003ds} 
and the diboson background is normalized to the NLO QCD cross-section prediction~\cite{Campbell:2011bn}.

\subsection{\Wboson+jets Background}\label{sec:WjetsBackground}

At the LHC the rate of $W^+$+jets events is larger than that of $W^-$+jets as
the up-quark density in the proton is larger than the down-quark one.
Exploiting the fact that the ratio of $W^+$+jets to $W^-$+jets cross-sections
is predicted more precisely than the total $W$+jets cross-section \cite{Kom:2010mv},
the charge asymmetry in $W$+jets production can be used to estimate the total
$W$+jets background from the data. Considering that processes other than
$W$+jets give, to a~good approximation, equal numbers of positively and
negatively charged leptons, the total number of $W$+$n$-jets events before
requiring a~$b$-tagged jet (pretag sample) can be estimated as
\begin{equation}
\begin{split}
N_{n_{\rm jets}}^{\rm W, pretag} &= N^{W^+}_{n_{\rm jets}} + N^{W^-}_{n_{\rm jets}} \\ 
                          &= \left( \frac{r^{\rm MC}_{n_{\rm jets}}+1}{r^{\rm MC}_{n_{\rm jets}}-1} \right) (D_{n_{\rm jets}}^+ - D_{n_{\rm jets}}^-)\,,
\end{split}
\label{eq:wnorm}
\end{equation}
where $n_{\rm jets}$ is the number of jets, $D_{n_{\rm jets}}^+$ ($D_{n_{\rm jets}}^-$) the total numbers of events with positively (negatively) charged leptons in data
meeting the selection criteria described in Sec.~\ref{sec:EventSelection} with the appropriate $n_{\rm jets}$ requirement and 
without the $b$-tagging requirement, and
$r^{\rm MC}_{n_{\rm jets}}$ is the ratio of $\sigma(pp \rightarrow W^+ + n$-jets$)$ to
$\sigma(pp \rightarrow W^- + n$-jets$)$ estimated from simulation.
Small additional sources of charge asymmetry in data, mainly due to the single top-quark
contribution, are estimated from the simulation and subtracted from data.
The largest uncertainties in the ratio come from the PDFs and the
heavy-flavor fractions in $W$+jets events.

The jet flavor composition of the pretag sample is the other important element
needed to estimate the number of events after the requirement of at least one
$b$-tagged jet. It is evaluated using a~combination of data- and simulation-driven approaches starting
from the estimation of the flavor fractions from data for the two-jet sample.
\begin{equation}
\begin{split}
N^{W,{\rm tag}}_2 &= N^{W,{\rm pretag}}_2 (F_{bb,2}P_{bb,2}+F_{cc,2}P_{cc,2}\\
               &+F_{c,2}P_{c,2}+F_{\rm light,2}P_{\rm light,2})\,,
\end{split}
\label{eq:flavfrac}
\end{equation}
where $N^{W,{\rm tag}}_2$ is the number of $W$+jets events after
the $b$-tagging requirement in the two-jet sample, evaluated
from data after subtracting all non-$W$ events (including the multijet background, estimated using the
data-driven method described in Sec.~\ref{sec:FakeLeptonBackground}, the \ttbar{} signal and the other backgrounds, estimated from simulation);
$N^{W,{\rm pretag}}_2$ is the number of events before the $b$-tagging
requirement estimated from data using Eq.(\ref{eq:wnorm})
for the background-dominated two-jet sample.
The quantities $F_{x,2}$ (with $x=bb/cc/c/{\rm light}$, where {\rm light} refers to $u/d/s$-quark- and gluon-initiated jets) represent the flavor fractions in the
two-jet sample and the $P_{x,2}$ the respective $b$-tagging probabilities taken from the simulation.
The flavor fractions add up to unity for each jet multiplicity 
\begin{equation}
F_{bb,2}+k_{cc \rightarrow bb} \cdot F_{bb,2}+F_{c,2}+F_{\rm light,2}=1
\label{eq:scnorm}
\end{equation}
with $F_{cc,2}$ constrained by $F_{bb,2}$ using the ratio $k_{cc \rightarrow bb}$ between the two
fractions taken from simulation.
The $Wc$+jets events have a~different charge asymmetry with respect to
$Wbb$/$Wcc$/$W+{\rm light}$-jets events. This is because, at leading order, the former is dominated by gluon-$s$ and gluon-$\bar{s}$ scattering, which involve symmetric $s$- and $\bar{s}$-quark PDF, while the latter are dominated by $u$-$\bar{d}$ and $d$-$\bar{u}$ scattering, which are asymmetric because they involve the $u$- and $d$-valence-quark PDF.
The flavor fractions can therefore be determined by applying Eq.(\ref{eq:flavfrac})
and Eq.(\ref{eq:scnorm}) separately for events with positive and negative leptons. These flavor fractions are
used to re-determine the overall normalization and the procedure is iterated until no significant changes
are observed. They are then used to correct the flavor fractions in the simulation.

Finally the number of events after the $b$-tagging and requiring $\ge 4$-jets 
is estimated using the number of pretag events, $N^{W,{\rm pretag}}_{\ge 4}$, measured from the
charge asymmetry method of Eq.(\ref{eq:wnorm}), as
\begin{equation}
N^{W,{\rm tag}}_{\ge 4}=N^{W,{\rm pretag}}_{\ge 4} \cdot f^{\rm tag}_2 \cdot f_{2\rightarrow \ge 4}^{\rm tag}\,,
\end{equation}
where $f^{\rm tag}_2$ is the fraction of events in the two-jet sample that are $b$-tagged
and $f_{2\rightarrow \ge 4}^{\rm tag}$ the ratio between the $b$-tagged event fractions in the $\ge 4$-jet
and two-jet samples evaluated using simulated $W$+jets events with corrected flavor fractions. 
The correction factors for a~selection 
requiring $\geq 4$ jets are obtained from the ones of the two-jet
sample by applying an overall normalization factor in order to preserve the requirement
that the flavor fractions add up to unity. 

This method has the advantage
that $f^{\rm tag}_2$ is evaluated from the data in a~sample dominated by the $W$+jets background
and that it relies on the ratio between the tagging fractions in the two-jet and $\ge 4$-jet
samples, strongly reducing the systematic uncertainties due to the $b$-tagging efficiencies and
the heavy-flavor components of the $W$+jets background.

\subsection{Multijet Background} \label{sec:FakeLeptonBackground}

The multijet background is characterized by jets that are misidentified as isolated prompt leptons, or non-prompt leptons that are misidentified as isolated leptons. These are referred to as ``fake leptons".

The rate of identifying such a~fake lepton as a~real one 
is calculated from data by defining two control samples. The first sample uses the lepton definition described in Sec.~\ref{sec:ObjectDef}, which is referred to as the tight selection. To define the second sample, a~loose selection is used, for which the identification criteria are relaxed and the isolation requirements are removed. Using these samples, the number of fake leptons passing the tight selection is given by
\begin{equation}
\label{eq:multijet_estimation}
N^{\rm tight}_{\rm fake} = \frac{\epsilon_{\rm fake}}{\epsilon_{\rm real}-\epsilon_{\rm fake}} (N^{\rm loose}\epsilon_{\rm real}-N^{\rm tight})\,,
\end{equation}
where $N^{\rm tight}$ and $N^{\rm loose}$ are the numbers of events with a~tight or loose lepton, respectively, and $\epsilon_{\rm real}$ and $\epsilon_{\rm fake}$ are the fractions of real and fake loose leptons that pass the tight selection. Decays of the \Zboson{}~boson to two leptons are used to measure the $\epsilon_{\rm real}$, while the $\epsilon_{\rm fake}$ are measured in control regions which are dominated by contributions from fake leptons. These control regions are defined by requiring low \Etmiss{}, low $\mtw$, or by selecting leptons with high track impact parameter. Contributions from \Wboson+jets and \Zboson+jets production are subtracted in the control regions using simulation~\cite{atlasXsec3}. The resulting multijet background is larger for the \ejets{} channel than it is for the \mujets{} channel.

\section{Reconstructed Event Variables}\label{sec:YieldsAndPlots}

The event yields after the selection described in Sec.~\ref{sec:EventReco} are displayed in Table~\ref{tab:yields}, separately for the \ejets{} and \mujets{} channels, for the data, the simulated \ljets signal from $t\bar{t}$ production, and for the various backgrounds discussed in Sec.~\ref{sec:BackgroundDetermination}. 

\begin{figure*}[htbp]
\centering
\subfigure[]{ \includegraphics[width=0.38\textwidth]{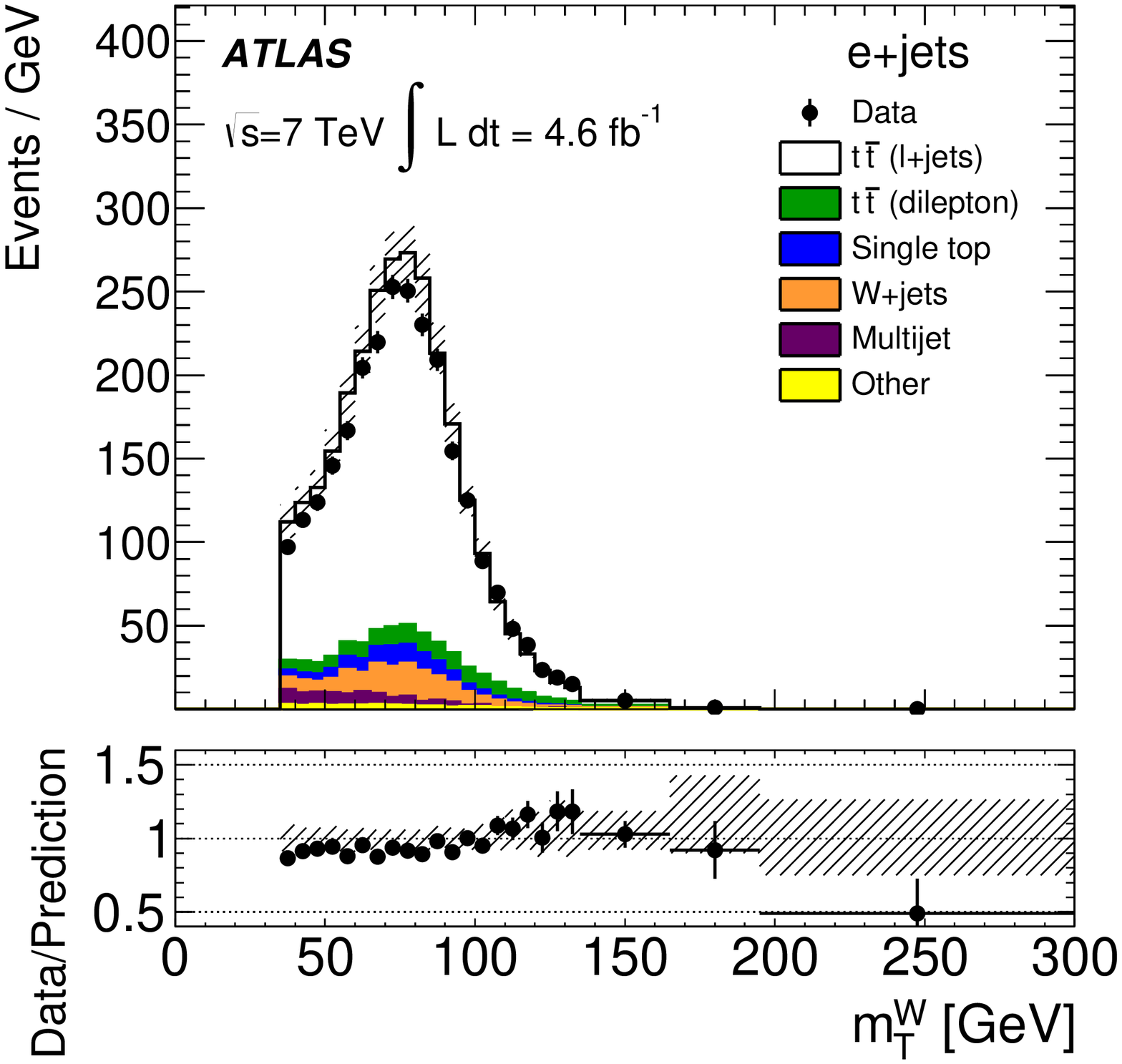}\label{mwt_el}}
\subfigure[]{ \includegraphics[width=0.38\textwidth]{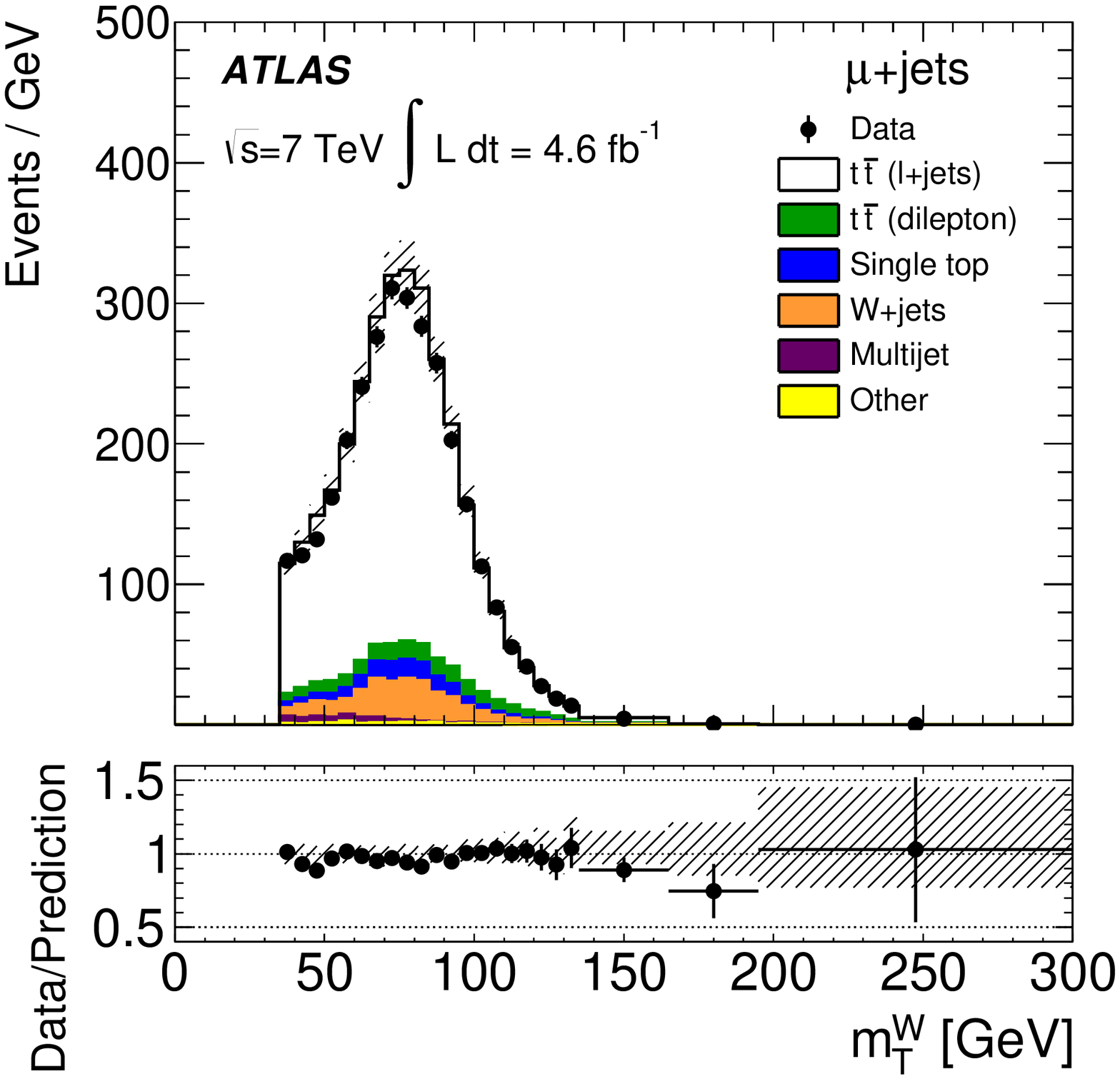}\label{mwt_mu}}
\subfigure[]{ \includegraphics[width=0.38\textwidth]{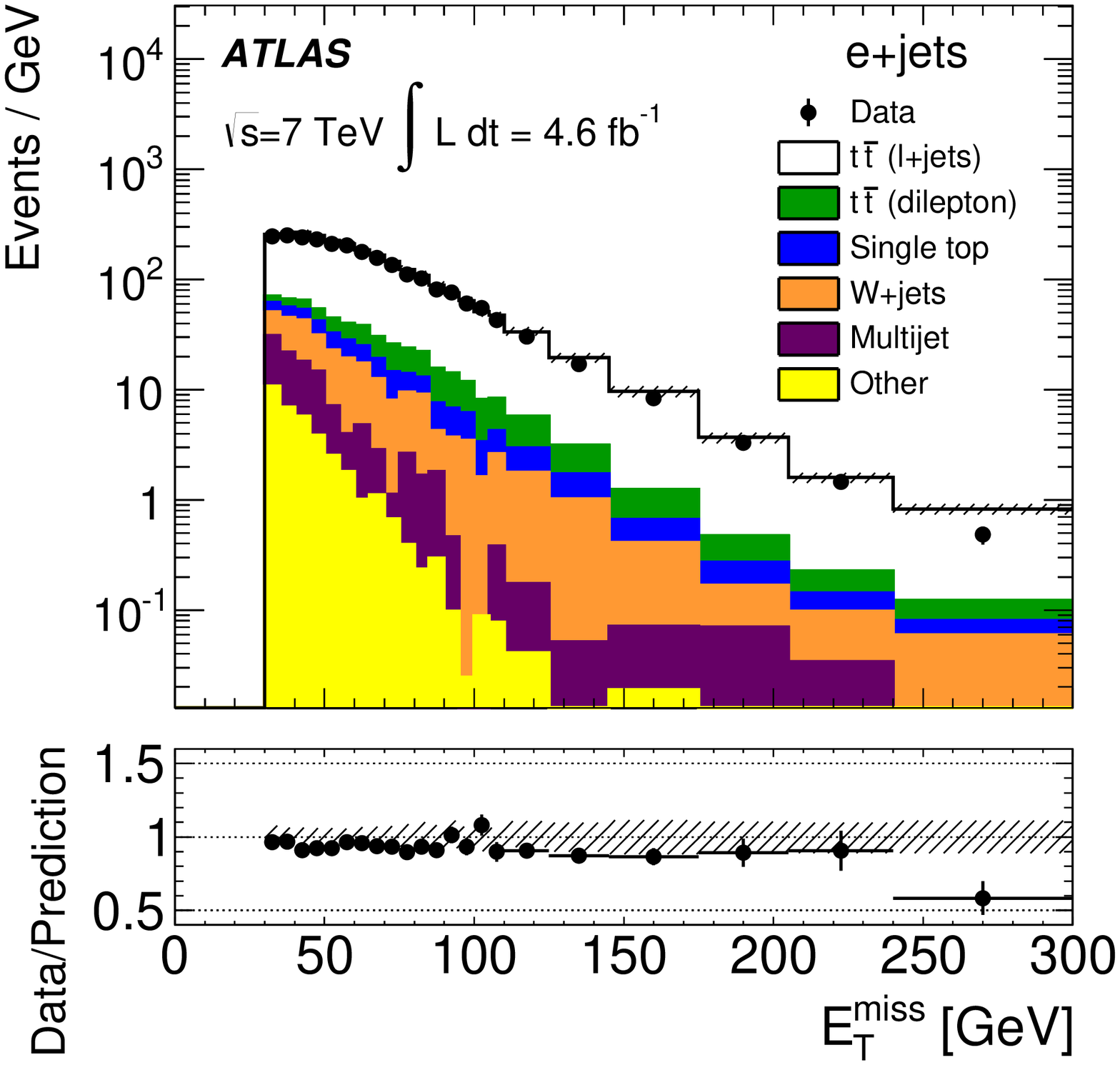}\label{met_el}}
\subfigure[]{ \includegraphics[width=0.38\textwidth]{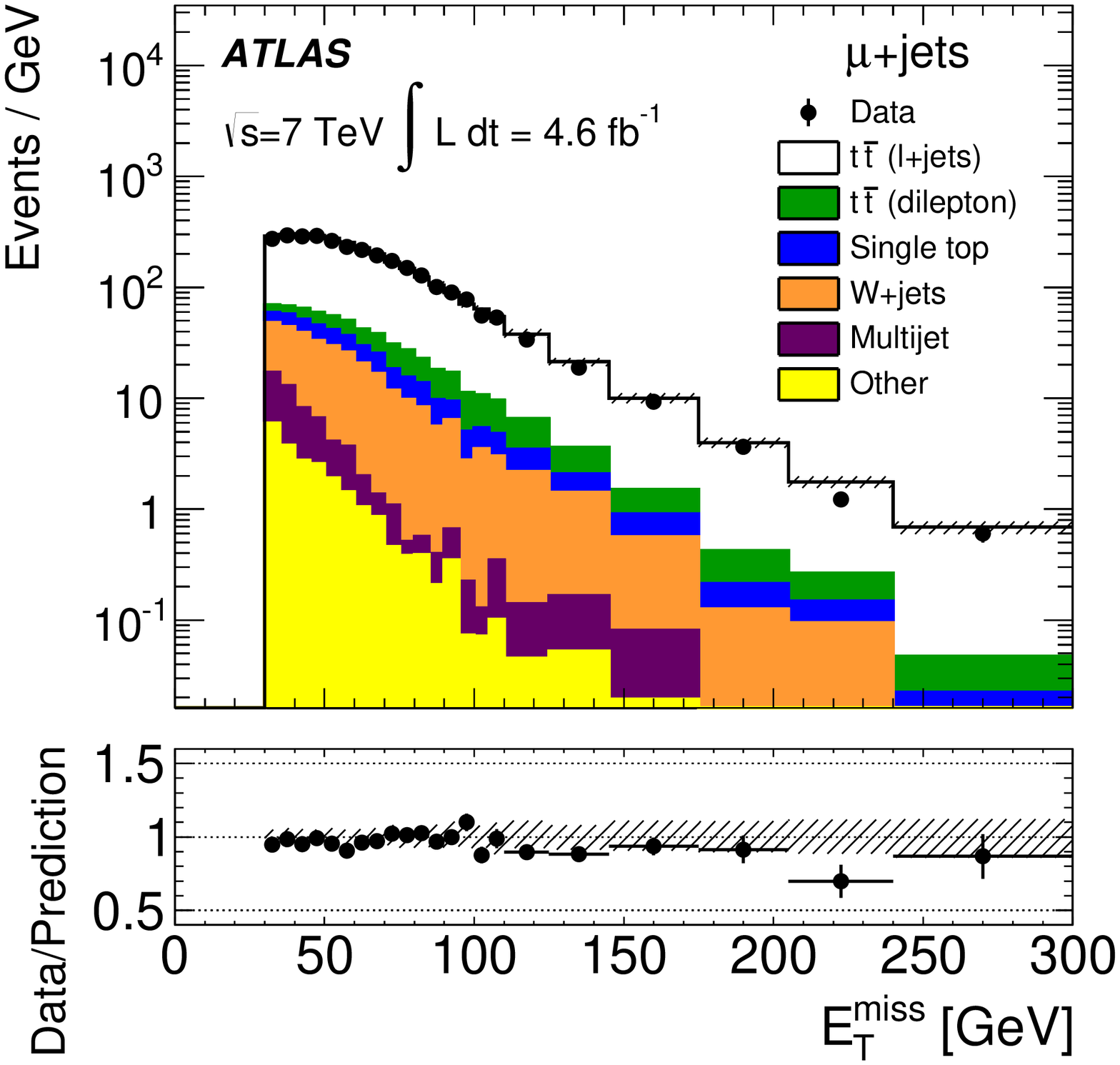}\label{met_mu}}
\subfigure[]{ \includegraphics[width=0.38\textwidth]{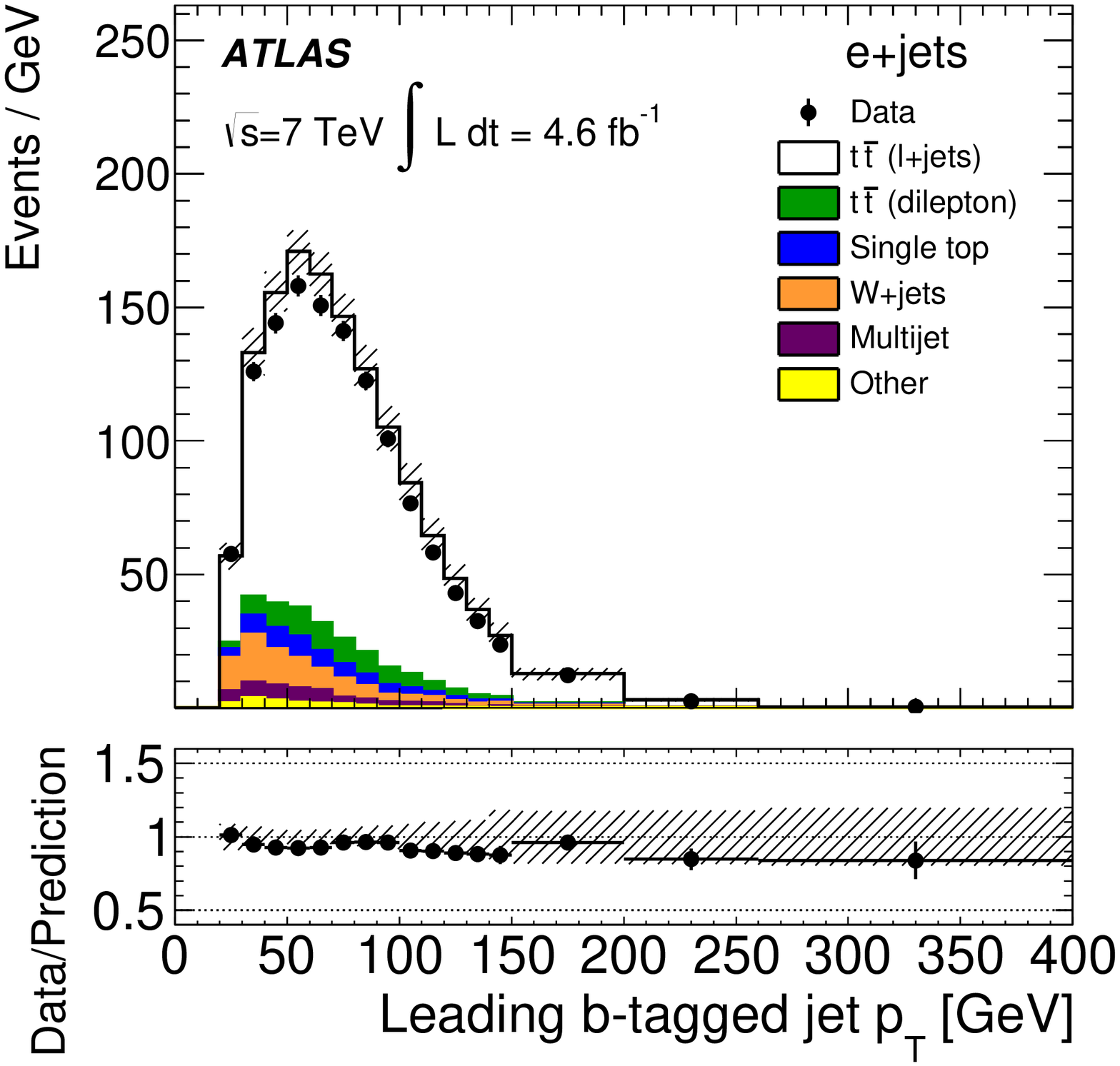}\label{bjet_el}}
\subfigure[]{ \includegraphics[width=0.38\textwidth]{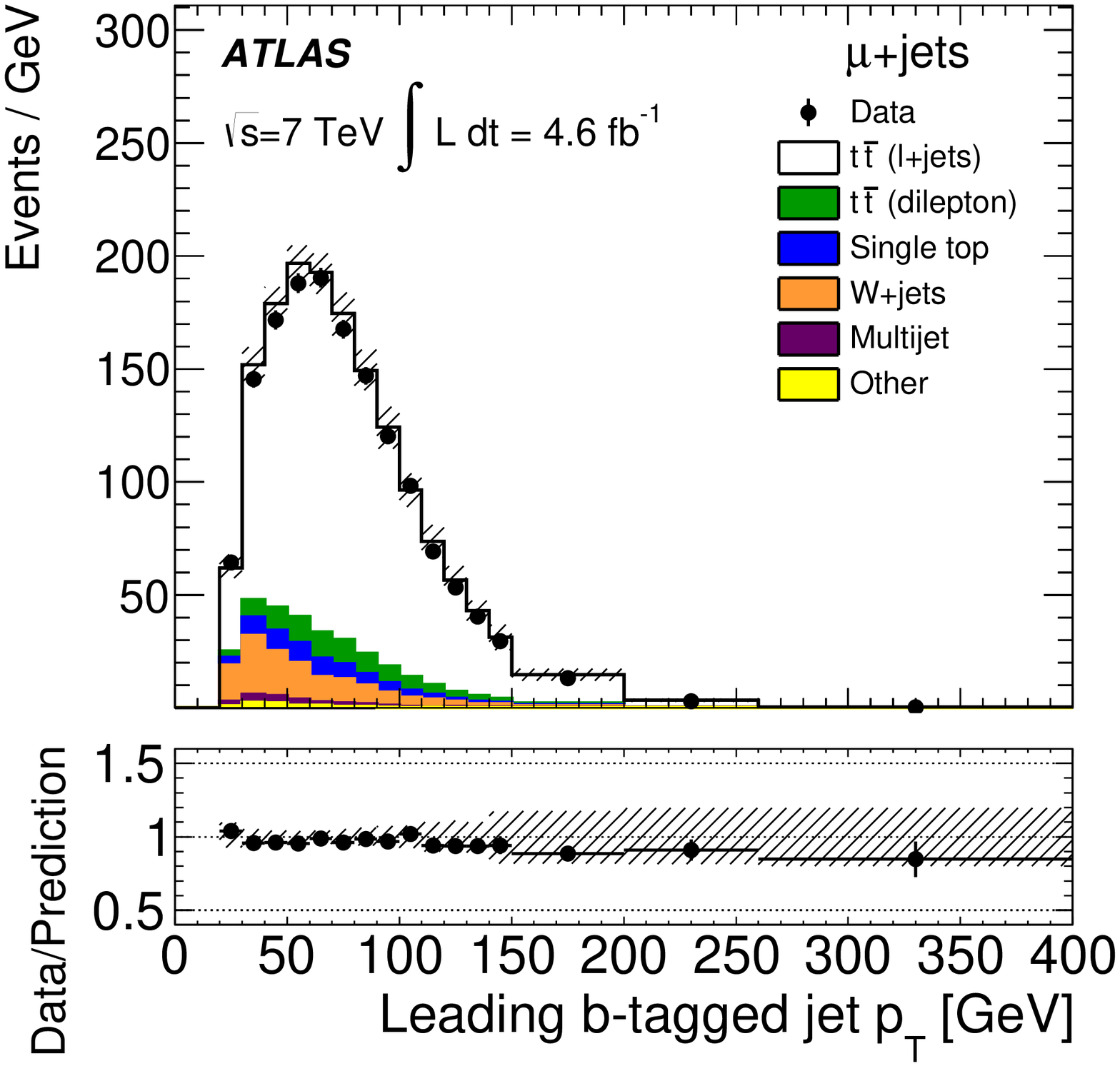}\label{bjet_mu}}
\caption{(Color online) Observables at the reconstruction level: $W$ transverse mass (\mtw{}) in the ~\subref{mwt_el}~\ejets{} and \subref{mwt_mu}~\mujets{} channels, missing transverse momentum (\Etmiss{}) in the \subref{met_el}~\ejets{} and \subref{met_mu}~\mujets{} channels, and leading $b$-tagged jet \pt{} in the \subref{bjet_el}~\ejets{} and \subref{bjet_mu}~\mujets{} channels. Data distributions are compared to predictions, using \Alpgen{}+\Herwig{} as the \ttbar{} signal model. The hashed area indicates the combined statistical and systematic uncertainties in the total prediction, excluding systematic uncertainties related to the modeling of the $\ttbar$ system. Signal and background processes are shown in different colors, with ``Other" including the small backgrounds from diboson and $Z+$jets production. Events beyond the range of the horizontal axis are included in the last bin. 
The lower parts of the figures show the ratios of data to the predictions.}
\label{fig:controls_tagged}
\end{figure*}

A~comparison of the data with the $\ttbar$ signal and background distributions, after all selection criteria are applied, is shown in Fig.~\ref{fig:controls_tagged} as functions of the $W$~boson transverse mass, the missing transverse momentum and the \pt{} of the highest-\pt{} (leading) $b$-tagged jet. Within the uncertainties shown, which cover the experimental and background systematic uncertainties 
but not the $t\bar{t}$ modeling uncertainties (discussed in Sec.~\ref{sec:signalmodeling}), the data and predictions are in agreement.

\begin{table}[!ht]
\begin{center}
  \begin{tabular}{lrr}
      \toprule
  \multicolumn{2}{r}{\ejets{}} & \mujets{} \\
    \hline
    \vspace{.10cm}
 $t\bar{t}$ (\ljets{})${}^{\phantom{1}^{\phantom{1}}}$ 		&  $11200 \pm 1900$ &  $13100 \pm 2000$ \\
 \vspace{.10cm}
 $t\bar{t}$ (dilepton) 	&  $850 \pm 170$ &  $930 \pm 170$ \\
 \vspace{.10cm}
 Single top		& $560 \pm 120$ & $660 \pm 160$ \\
  \vspace{.10cm}
 $W+$jets		&  $920 \pm 240$ &  $1300 \pm 300$ \\
  \vspace{.10cm}
 Multijet		&  $400 \pm 200 $ &  $200 \pm 40$\\
  \vspace{.10cm}
$Z$+jets  & $160 \pm 110$ &    $89 \pm 60$ \\
Diboson   & $22 \pm 13$    &  $25 \pm 14$ \\
\hline
\vspace{.10cm}
 Prediction${}^{\phantom{1}^{\phantom{1}}}$		&  $14100 \pm 1900$ &  $16300 \pm 2000$\\
  \vspace{.10cm}
 Data 		 		& $13167$ & $15752$ \\ 
    \hline \hline
  \end{tabular}
  \end{center}
  \caption{Event yields in the \ejets{} and \mujets{} channels. The signal model, denoted $t\bar{t}$ (\ljets{}) in the table, is generated using \Alpgen{}. Errors indicate the total statistical and systematic uncertainties on each subsample and the uncertainty on the signal includes the generator systematic uncertainty discussed in Sec.~\ref{sec:signalmodeling}. 
}
  \label{tab:yields}
\end{table}

\begin{figure*}[htbp]
\centering
\subfigure[]{ \includegraphics[width=0.45\textwidth]{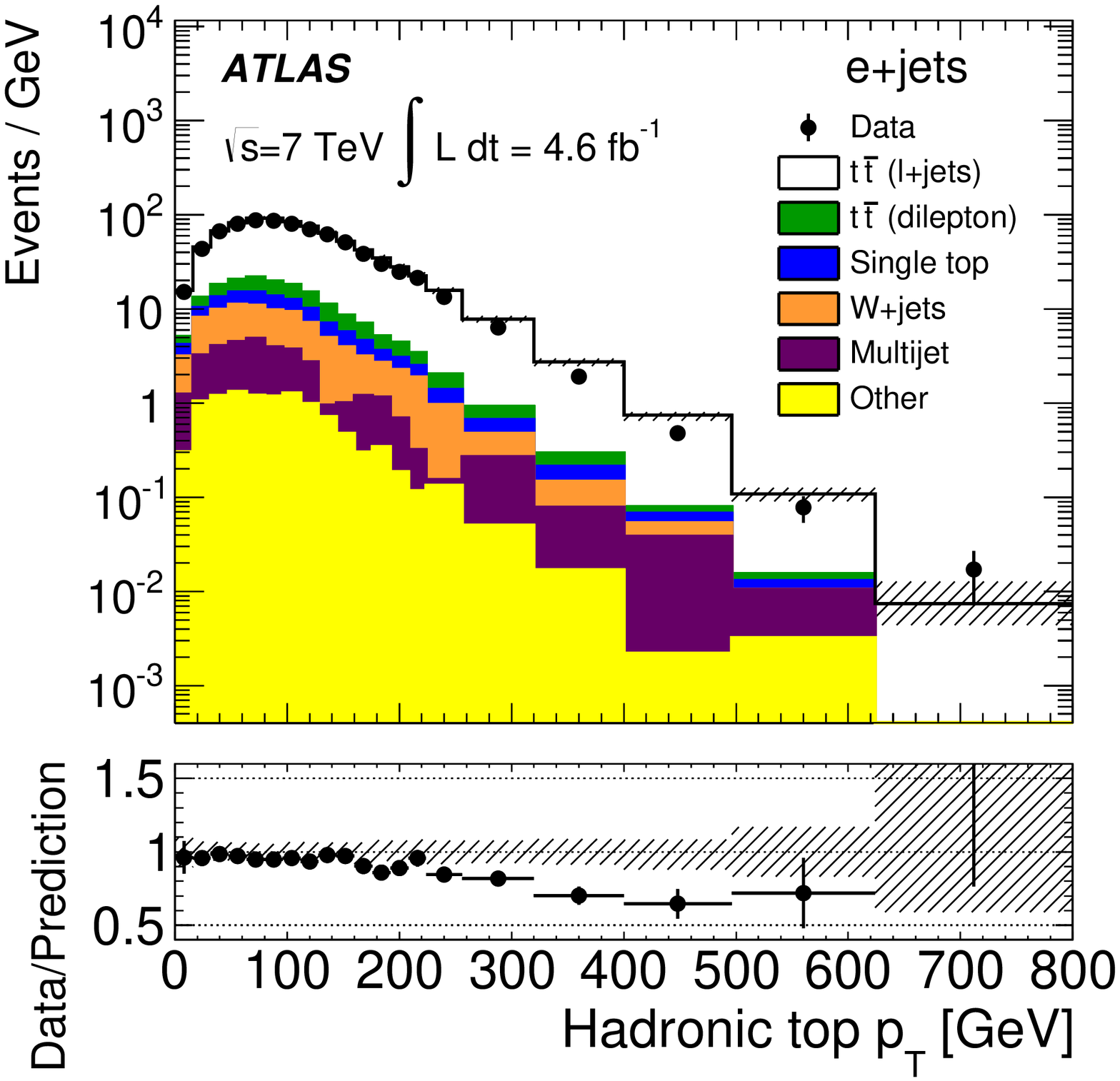}\label{reco_had_top_el}}
\subfigure[]{ \includegraphics[width=0.45\textwidth]{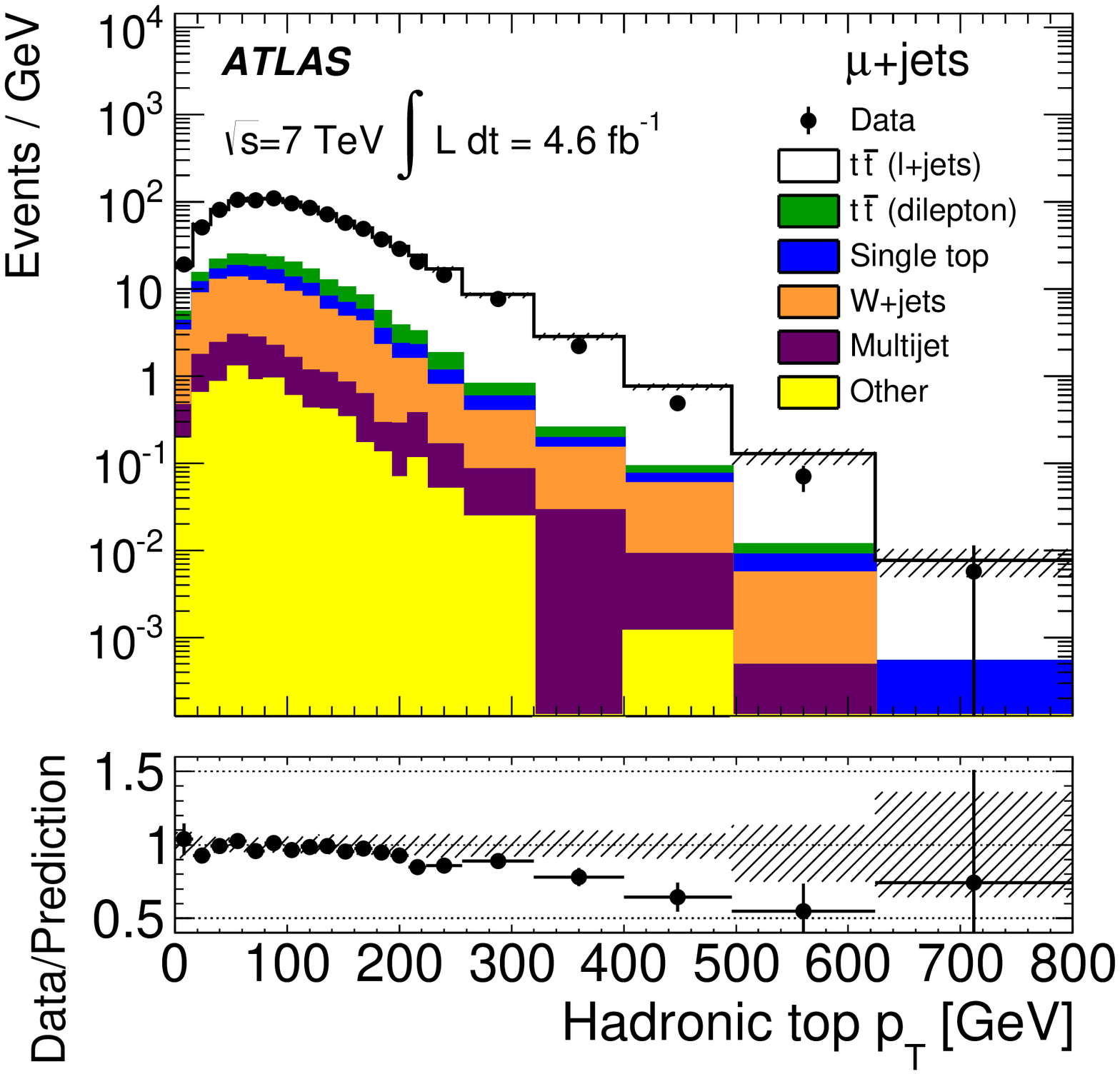}\label{reco_had_top_mu}}
\subfigure[]{ \includegraphics[width=0.45\textwidth]{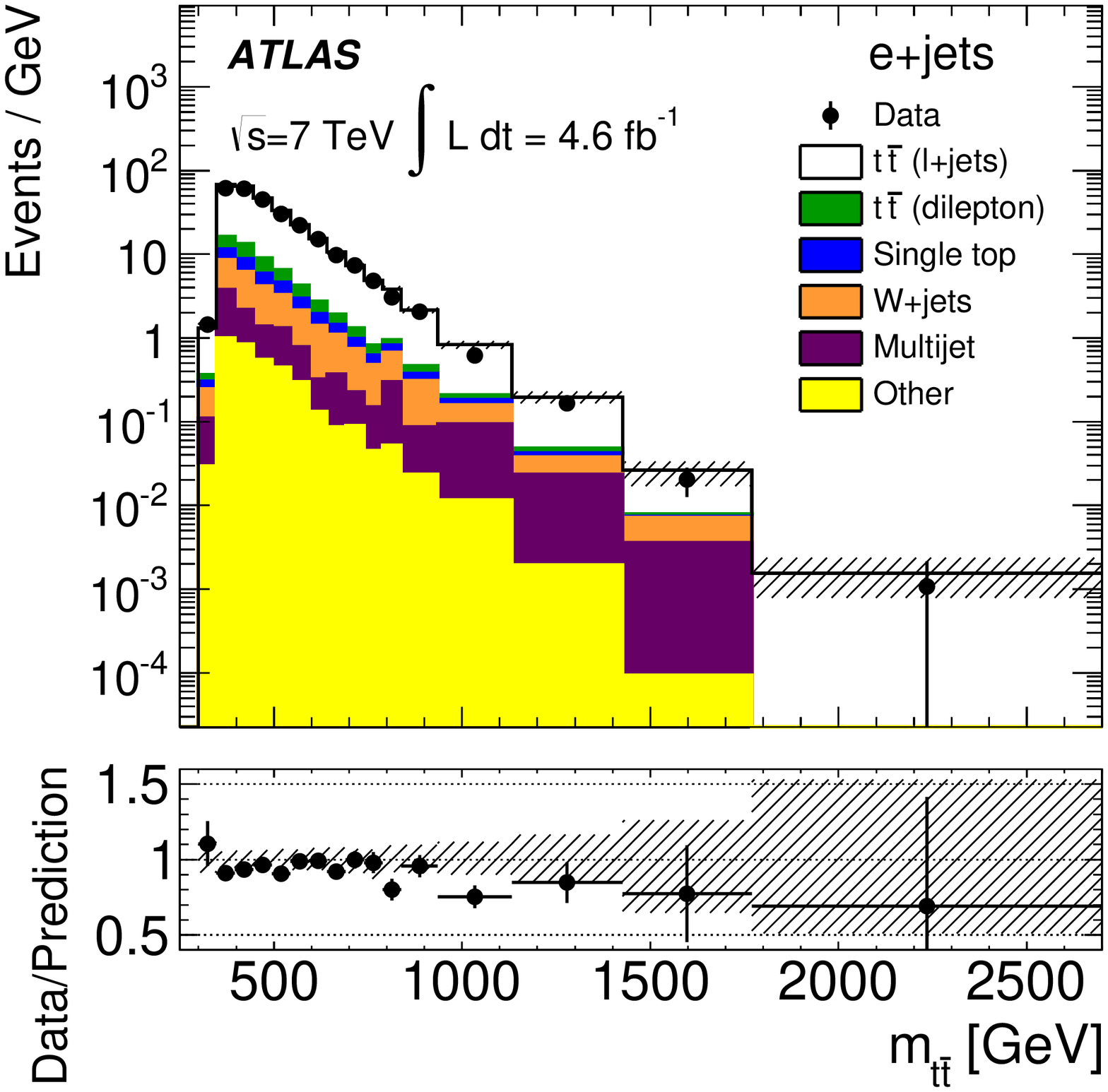}\label{reco_mass_el}}
\subfigure[]{ \includegraphics[width=0.45\textwidth]{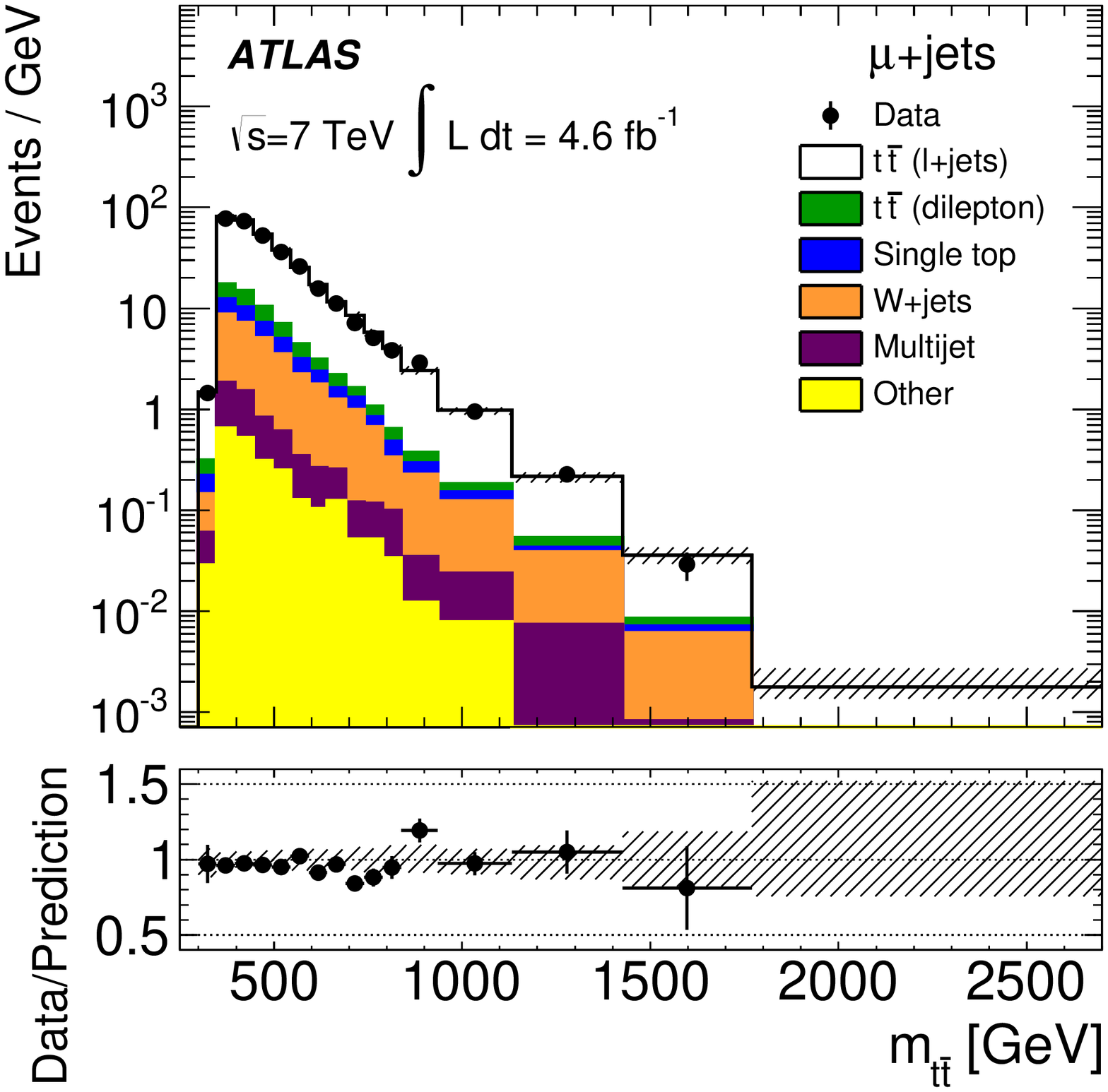}\label{reco_mass_mu}}
\caption{(Color online) Reconstructed distributions for the transverse momentum of the hadronically decaying top quark ($\ptt$) in the \subref{reco_had_top_el}~\ejets{} and \subref{reco_had_top_mu}~\mujets{} channels and for the mass of the \ttbar{} system (\mttbar{}) in the \subref{reco_mass_el}~\ejets{} and \subref{reco_mass_mu}~\mujets{} channels. Data distributions are compared to predictions, using \Alpgen{}+\Herwig{} as the \ttbar{} signal model. 
The hashed area indicates the combined statistical and systematic uncertainties in the total prediction, excluding systematic uncertainties related to the modeling of the $\ttbar$ system. Signal and background processes are shown in different colors, with ``Other" including the small backgrounds from diboson and $Z+$jets production. Events beyond the axis range are included in the last bin. 
The lower parts of the figures show the ratios of data to the predictions.
}
\label{fig:recoTop_tagged}
\end{figure*}

\begin{figure*}[htbp]
\centering
\subfigure[]{ \includegraphics[width=0.45\textwidth]{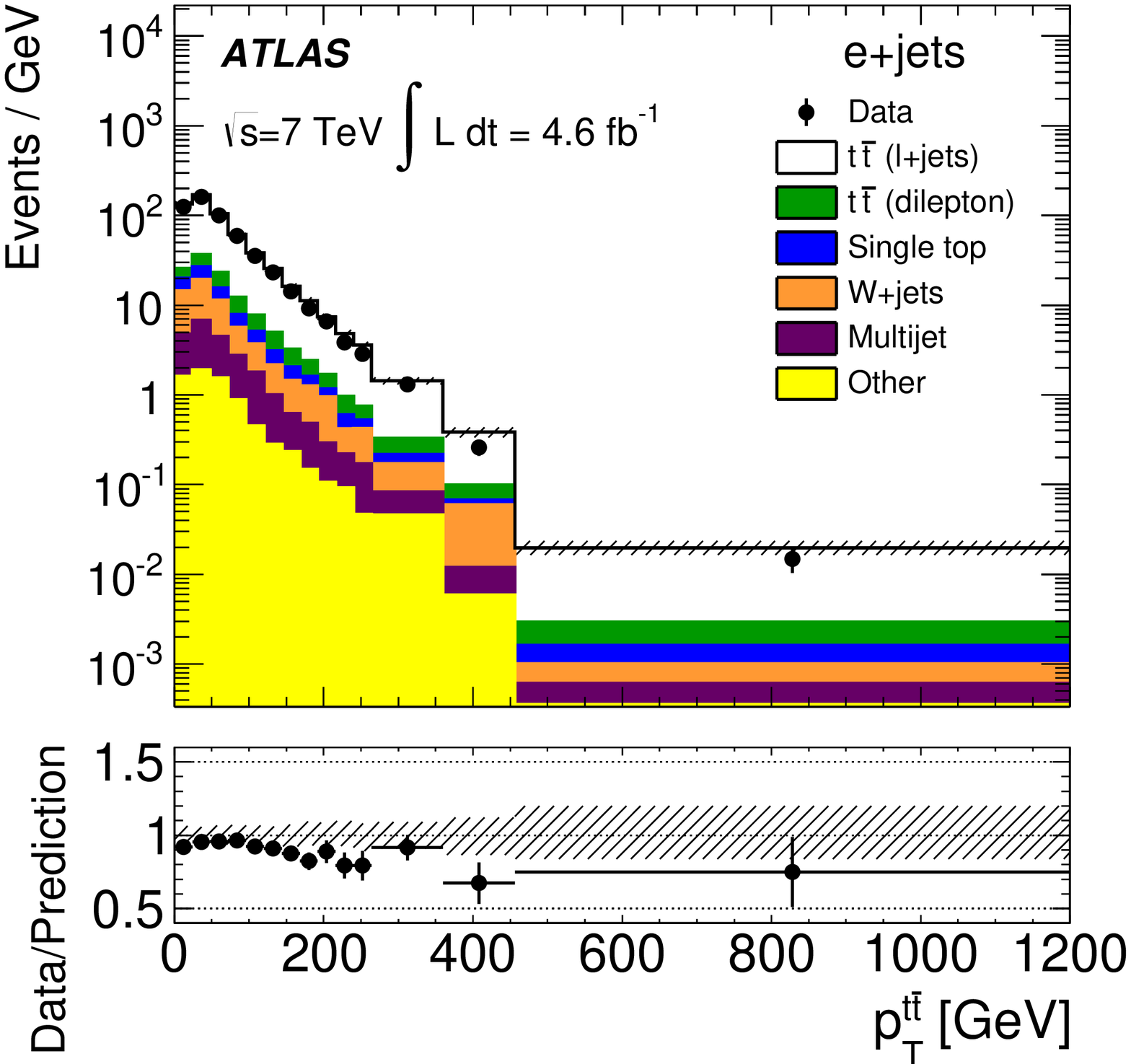}\label{reco_pt_el}}
\subfigure[]{ \includegraphics[width=0.45\textwidth]{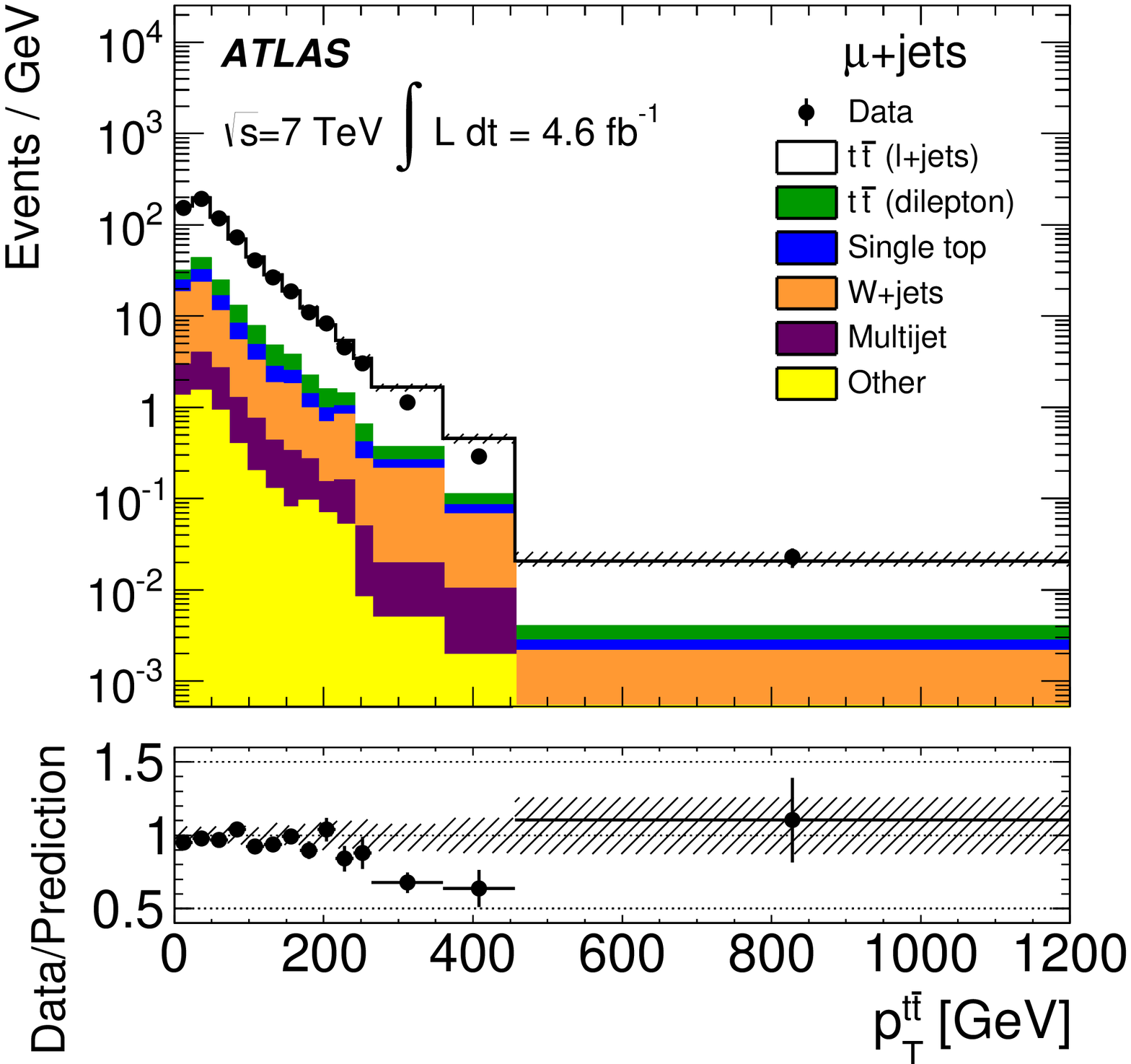}\label{reco_pt_mu}}
\subfigure[]{ \includegraphics[width=0.45\textwidth]{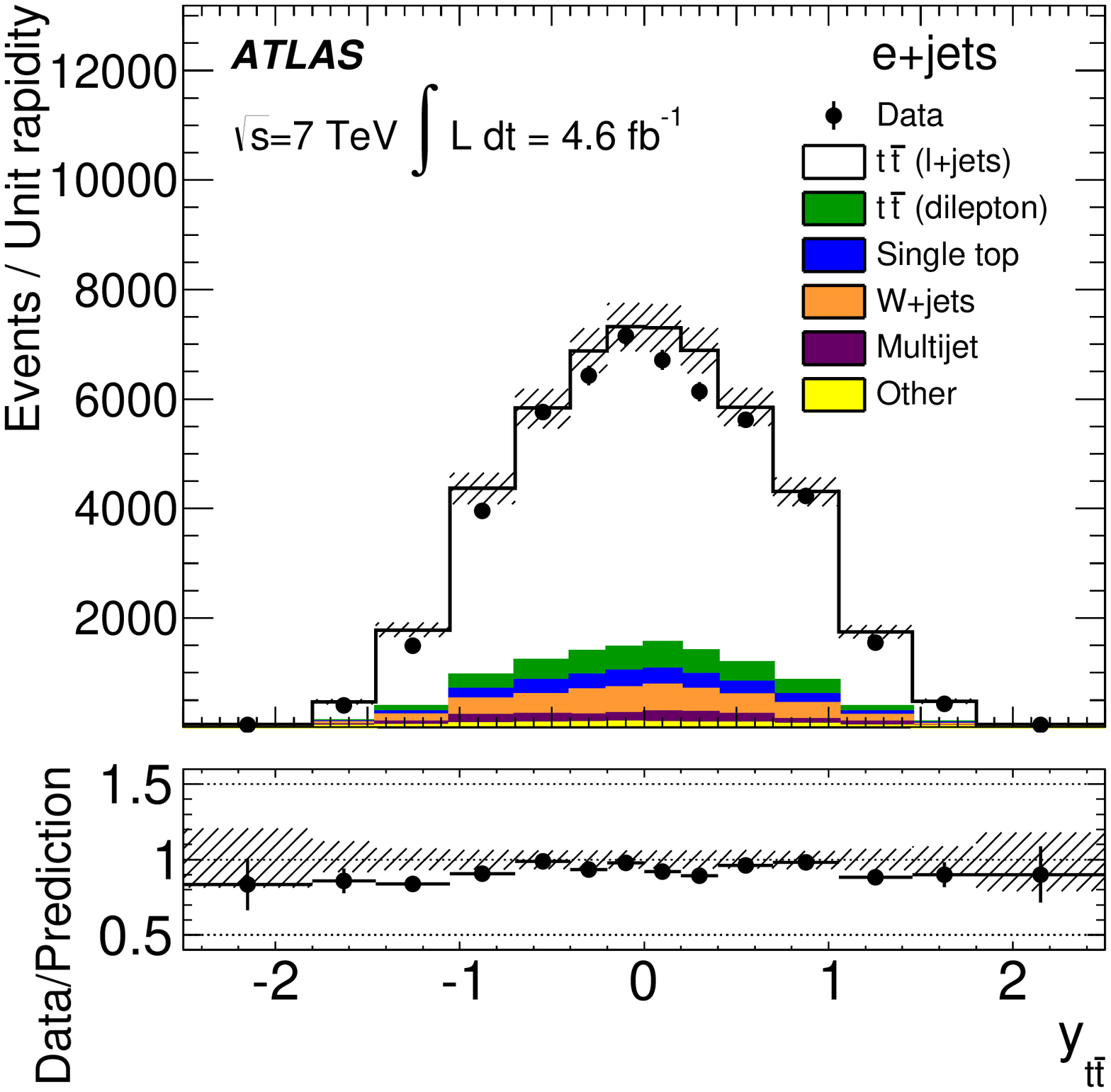}\label{reco_rap_el}}
\subfigure[]{ \includegraphics[width=0.45\textwidth]{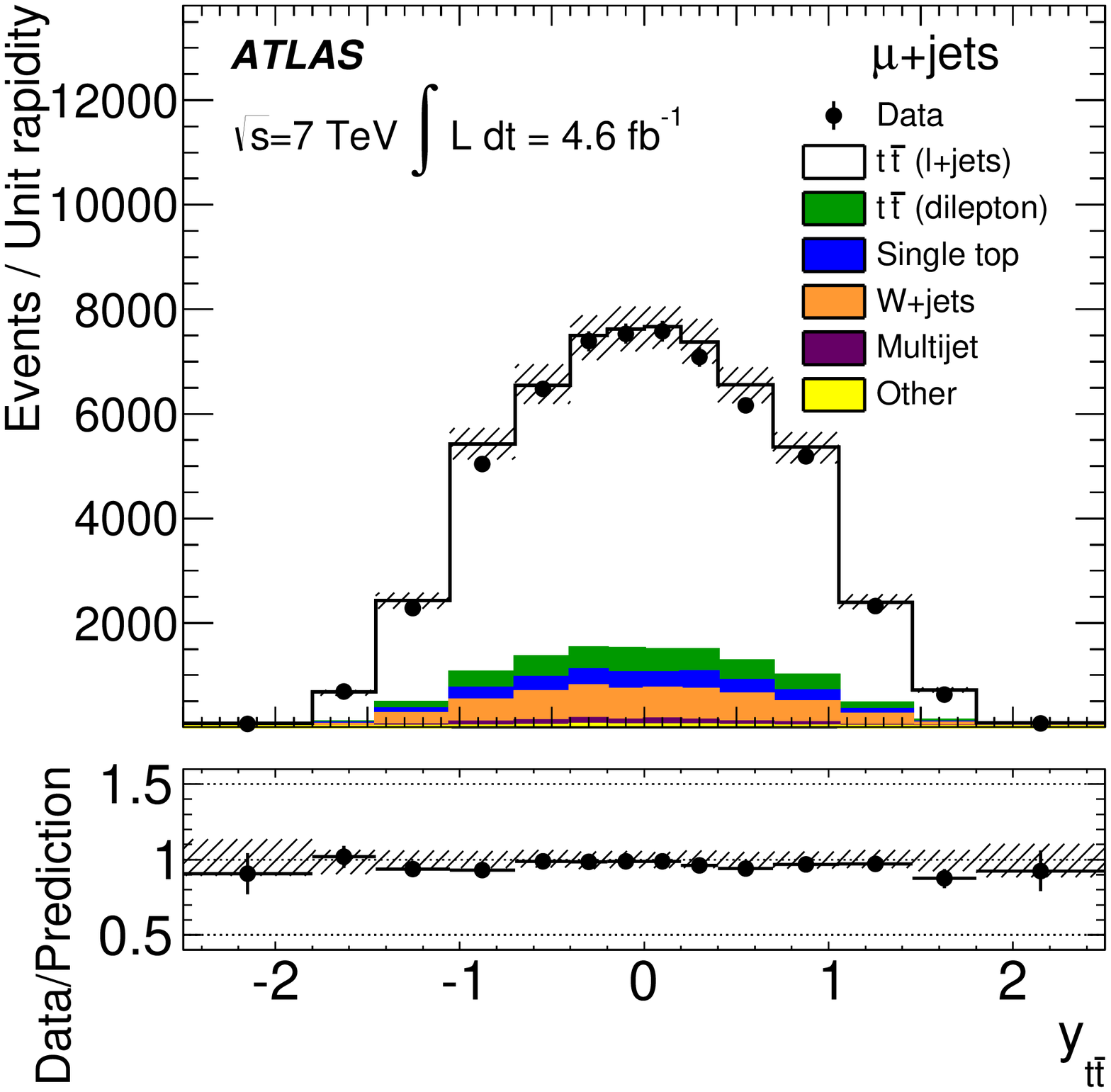}\label{reco_rap_mu}}
\caption{(Color online) Reconstructed distributions for the transverse momentum of the \ttbar{} system (\ptttbar{}) in the \subref{reco_pt_el}~\ejets{} and \subref{reco_pt_mu}~\mujets{} channels and for the rapidity of the \ttbar{} system (\yttbar{}) in the \subref{reco_rap_el}~\ejets{} and \subref{reco_rap_mu}~\mujets{} channels. Data distributions are compared to predictions, using \Alpgen{}+\Herwig{} as the \ttbar{} signal model. The hashed area indicates the combined statistical and systematic uncertainties in the total prediction, excluding systematic uncertainties related to the modeling of the $\ttbar$ system. Signal and background processes are shown in different colors, with ``Other" including the small backgrounds from diboson and $Z+$jets production. Events beyond the axis range are included in the last bin, or in the case of the \yttbar{} spectrum the first and last bin. 
The lower parts of the figures show the ratios of data to the predictions.}
\label{fig:recottbar_tagged}
\end{figure*}

The kinematic spectra corresponding to individual top quarks as well as to the reconstructed $\ttbar$ system are shown in Figs.~\ref{fig:recoTop_tagged} and~\ref{fig:recottbar_tagged}. 
Data and predictions agree within uncertainties with the exception of the high-\pt{} tails of the \ptt{} and \ptttbar{} distributions where data fall below the prediction.

\section{Differential Cross-section Determination}\label{sec:XSDetermination}

The estimated background contributions are subtracted from the measured distributions, which are then corrected for the efficiency to pass the event selection, for the detector resolution, and the branching ratio for the $\ttbar\rightarrow \ell$+jets channel. To facilitate the comparison to theoretical predictions, the cross-section measurements are defined with respect to the top quarks before the decay (parton level) and after QCD radiation~\footnote{Technically, the parton level, used both for the unfolding and for the predictions of the MC generators, is defined as status-code 155 for \Herwig{} and 3 for \Pythia{}.}.

\begin{figure*}[htbp]
\centering
\subfigure[]{ \includegraphics[width=0.45\textwidth]{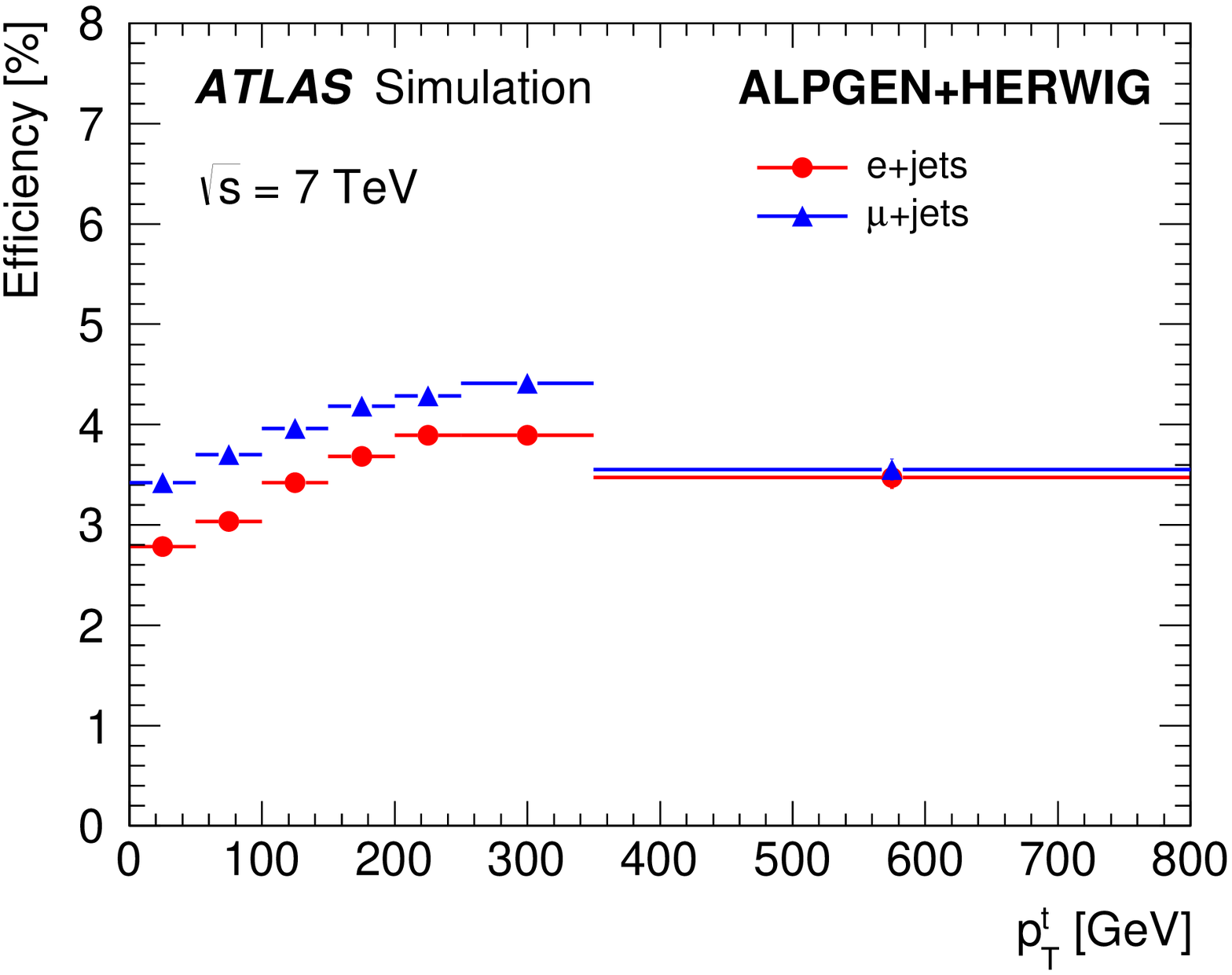}\label{eff_had_top}}
\subfigure[]{ \includegraphics[width=0.45\textwidth]{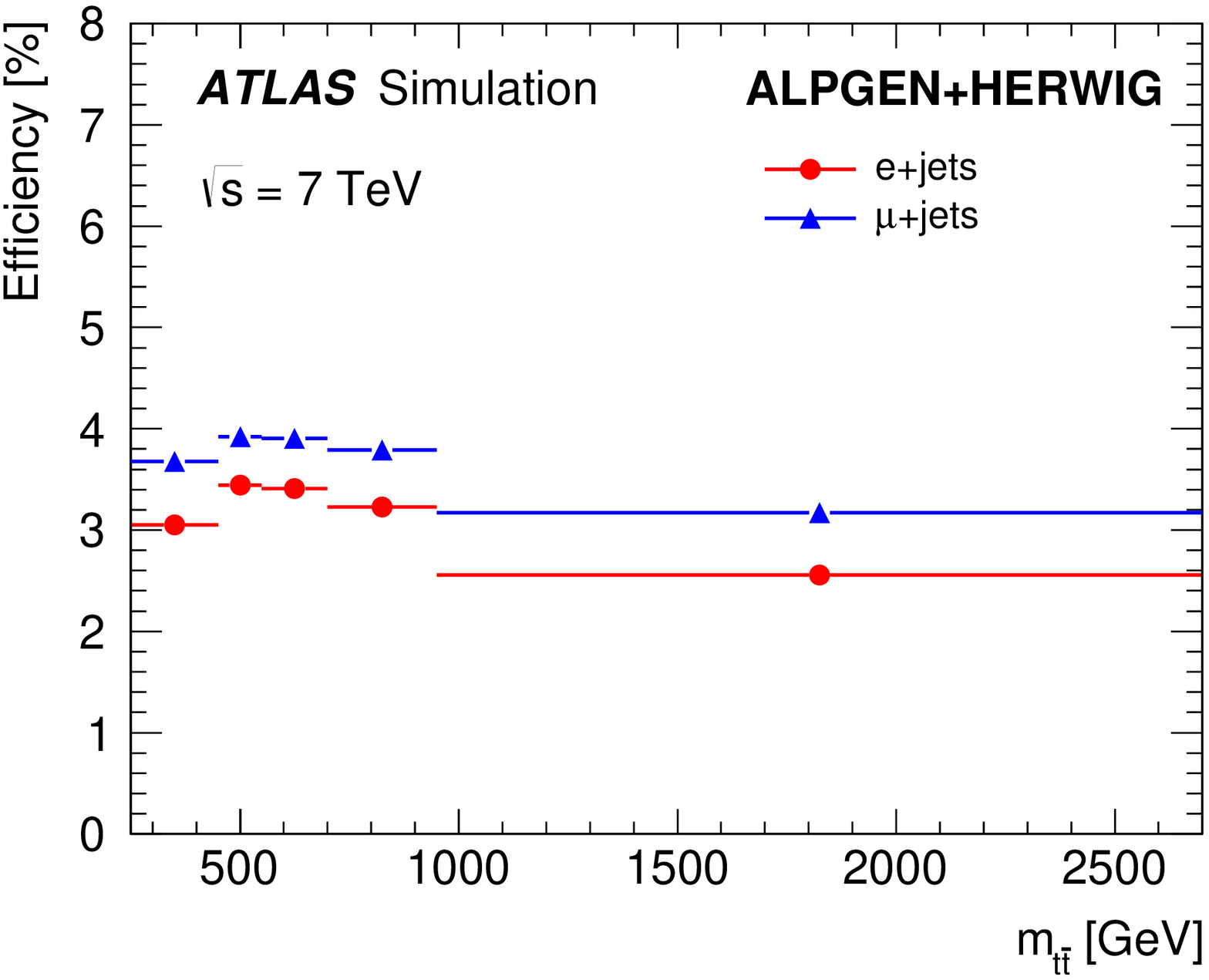}\label{eff_mass}}
\subfigure[]{ \includegraphics[width=0.45\textwidth]{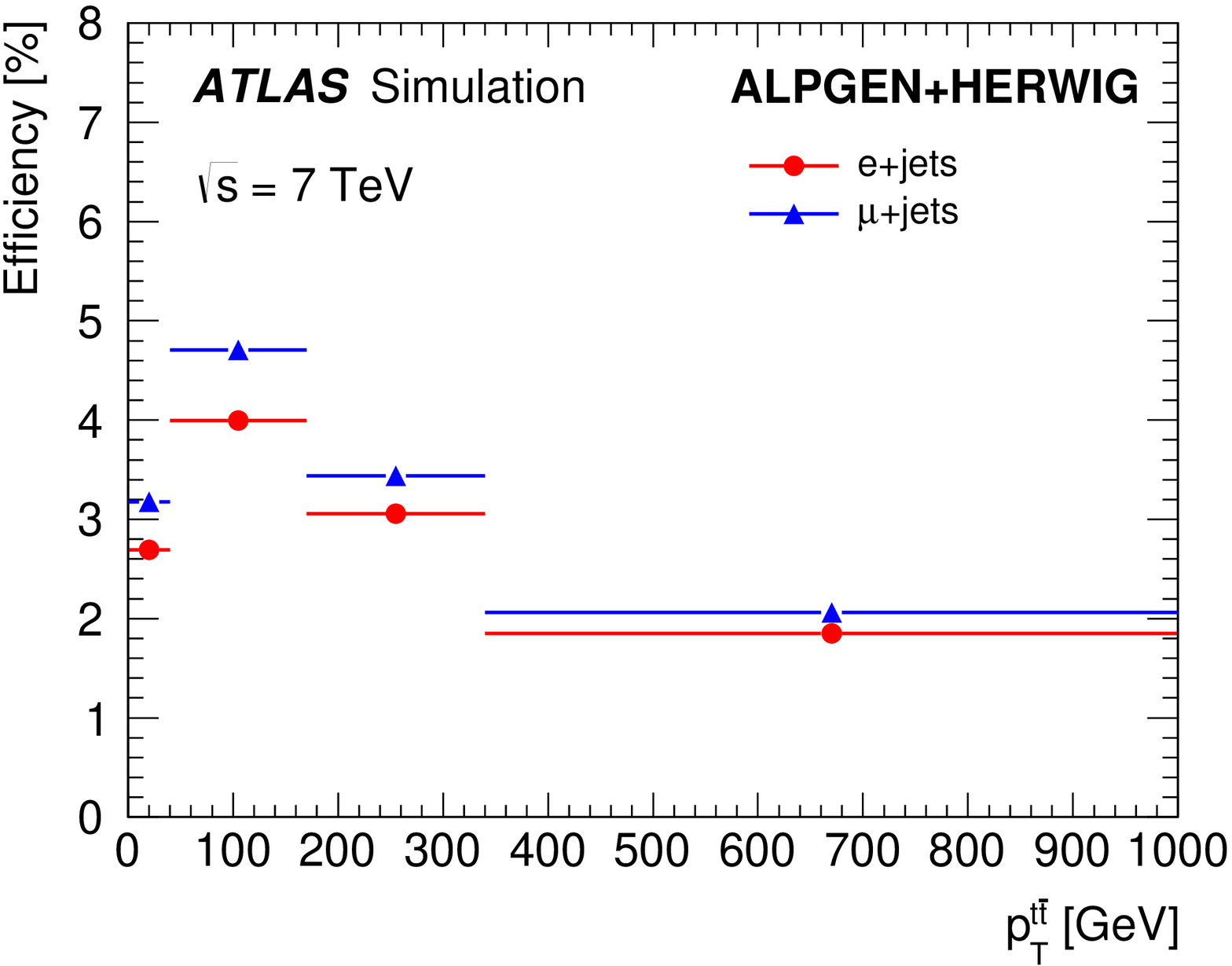}\label{eff_pt}}
\subfigure[]{ \includegraphics[width=0.45\textwidth]{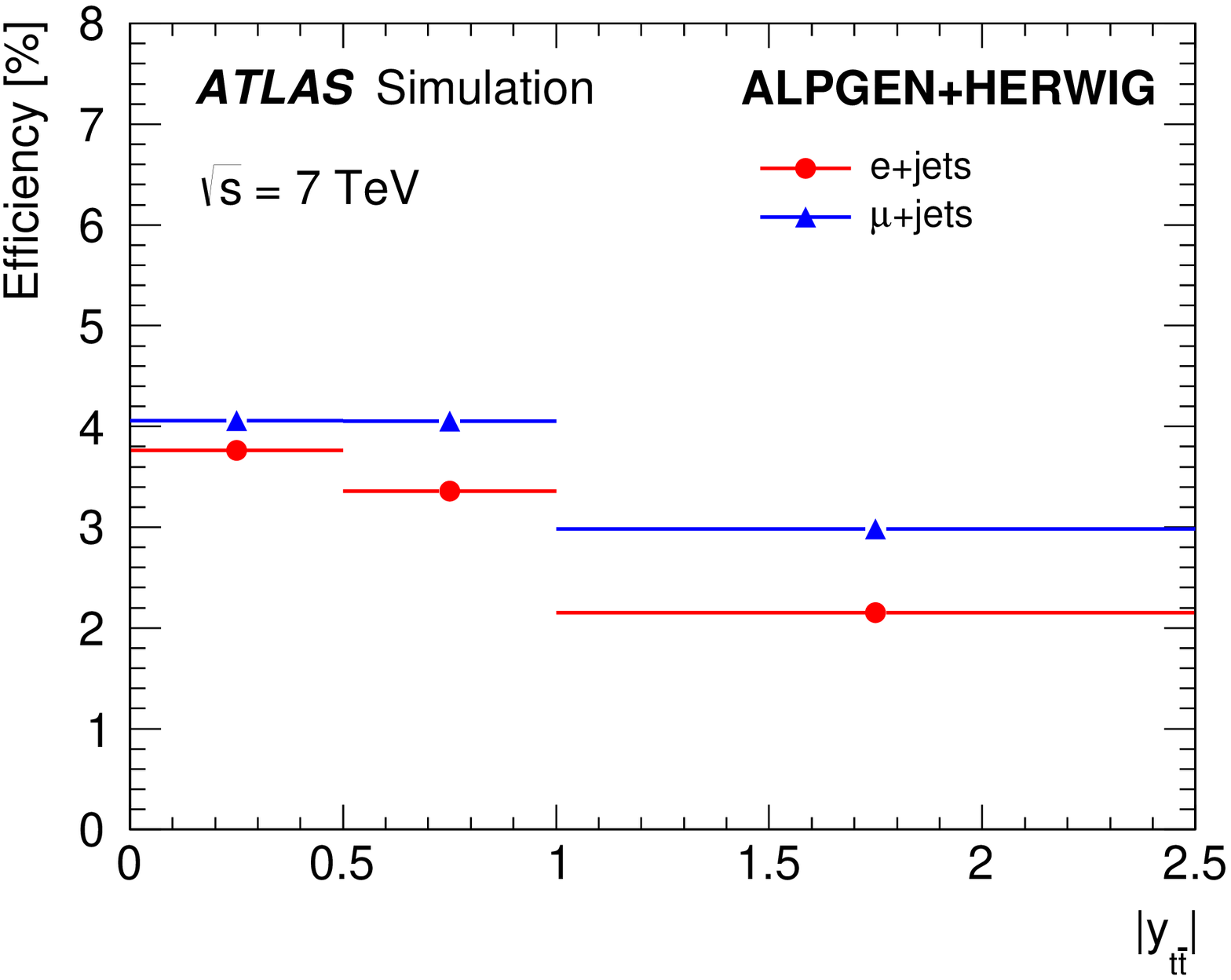}\label{eff_rap}}

\caption{(Color online) The selection efficiencies binned in the~\subref{eff_had_top} transverse momentum of the top-quark ($\ptt$), and the~\subref{eff_mass}~mass ($\mttbar$), \subref{eff_pt} transverse momentum ($\ptttbar$) and the~\subref{eff_rap} absolute value of the rapidity ($\absyttbar$) of the $\ttbar$ system obtained from the \Alpgen{}+\Herwig{} simulation of the \ttbar{} signal. The horizontal axes refers to parton-level variables.}
\label{fig:effs_tagged}
\end{figure*}

The efficiency ($\epsilon_j$) to satisfy the selection criteria in bin $j$ for each variable is evaluated as the ratio of the parton-level spectra before and after implementing the event selection at the reconstruction level. The efficiencies are displayed in Fig.~\ref{fig:effs_tagged} and are typically in the 3--5\% range. The decrease in the efficiencies at high values of \ptt{}, \mttbar{}, and \ptttbar{} is primarily due to the increasingly large fraction of non-isolated leptons and angularly close or merged jets in events with high top-quark \pt{}. There is also a~decrease in the efficiency at high \absyttbar{} due to jets and leptons falling outside of the pseudorapidity range required for the reconstructed lepton and jets. 
The absolute variation of the efficiency as a~function of a~different choice of the top-quark mass is found to be $+0.025 \%/$GeV, independently of the kinematic variable and bin.

The influence of detector resolution is corrected by unfolding. The measured distributions in the \ejets{} and \mujets{} channels are unfolded separately by a~regularized inversion of the migration matrix (symbolized by $\mathcal{M}^{-1}$) described in Sec.~\ref{sec:Unfolding} and then the channels are combined as described in Sec.~\ref{sec:Combination}. The formula used to extract the cross-section in each bin is

\begin{equation}
\frac{\dsigma}{\dX_j} \equiv \frac{1}{\Delta X_j} \cdot \frac{\sum\limits_{i} \mathcal{M}_{ji}^{-1}[D_i - B_i]}{{\rm BR} \,\cdot\, \mathcal{L} \,\cdot\, \epsilon_j},
\end{equation}

\noindent where $\Delta X_j$ is the bin width, $D_i$ ($B_i$) are the data (expected background) yields in each bin $i$ of the reconstructed variable, $\mathcal{L}$ is the integrated luminosity of the data sample, $\epsilon_j$ is the event selection efficiency,  and ${\rm BR}=0.438$ is the branching ratio of $\ttbar\rightarrow \ell$+jets~\cite{PDG}.

The normalized cross-section 
$1/\sigma \, \dsigma/\dX_j$  
is computed by dividing by the measured total cross-section, evaluated by integrating over all bins. 
The normalized distributions have substantially reduced systematic uncertainties since most of the relevant sources of uncertainty (luminosity, jet energy scale, $b$-tagging, and absolute normalization of the data-driven background estimate) have large bin-to-bin correlations.

\subsection{Unfolding Procedure} \label{sec:Unfolding}

The binning for each of the distributions is determined by the experimental resolution of the kinematic variables, and poorly populated bins are combined with neighboring bins to reduce the uncertainty on the final result. Typical values of the fractional resolution for \ptt{} and \mttbar{} are 25\% and 15\%, respectively, while the fractional resolution for \ptttbar{} improves as a~function of \ptttbar{} and is 40\% at $100\,$GeV. For \absyttbar{}, the resolution varies from 0.25 to 0.35, from central to forward rapidities.

\begin{figure*}[htbp]
\centering
\subfigure[]{ \includegraphics[width=0.49\textwidth]{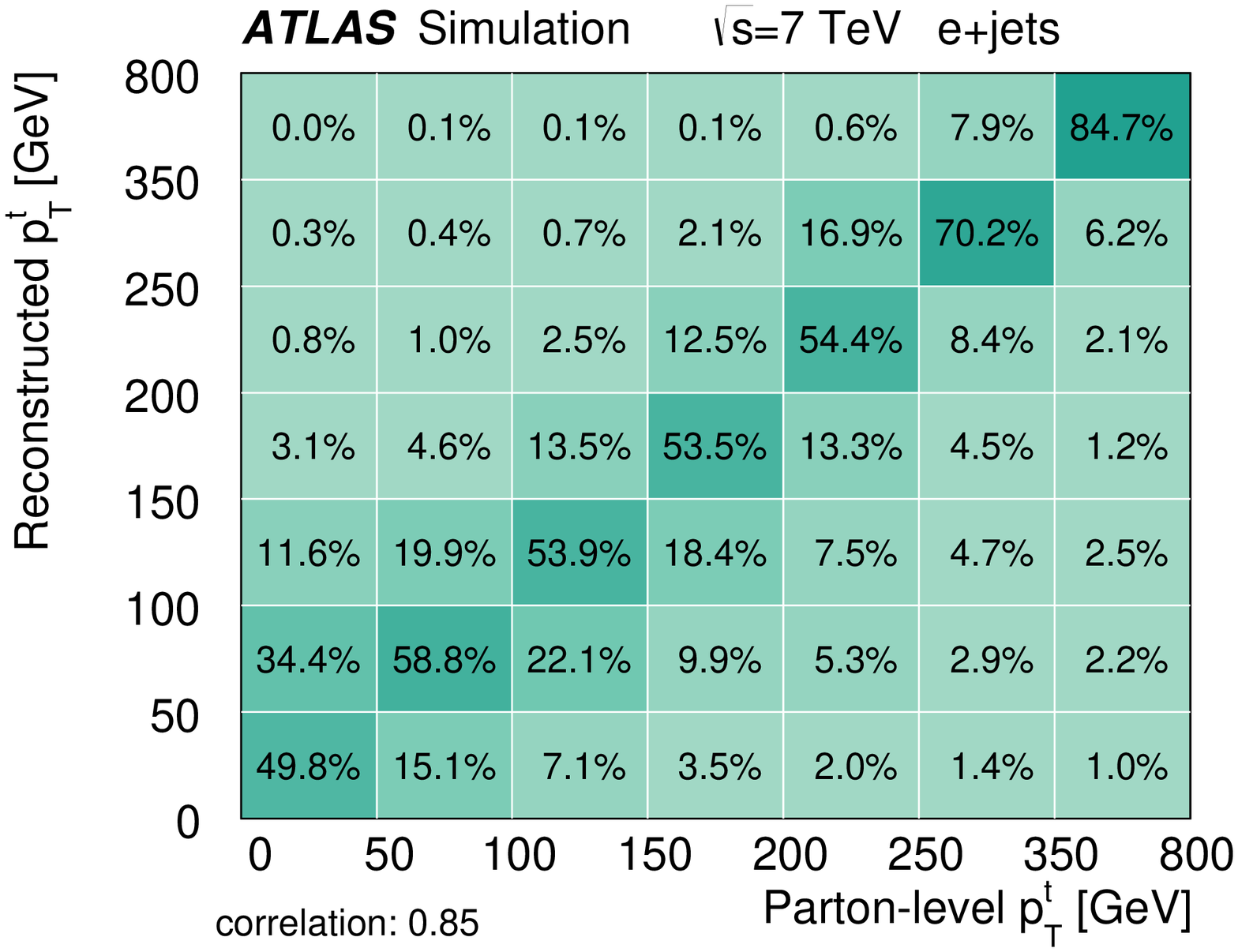}\label{migra_top2_el}}
\subfigure[]{ \includegraphics[width=0.49\textwidth]{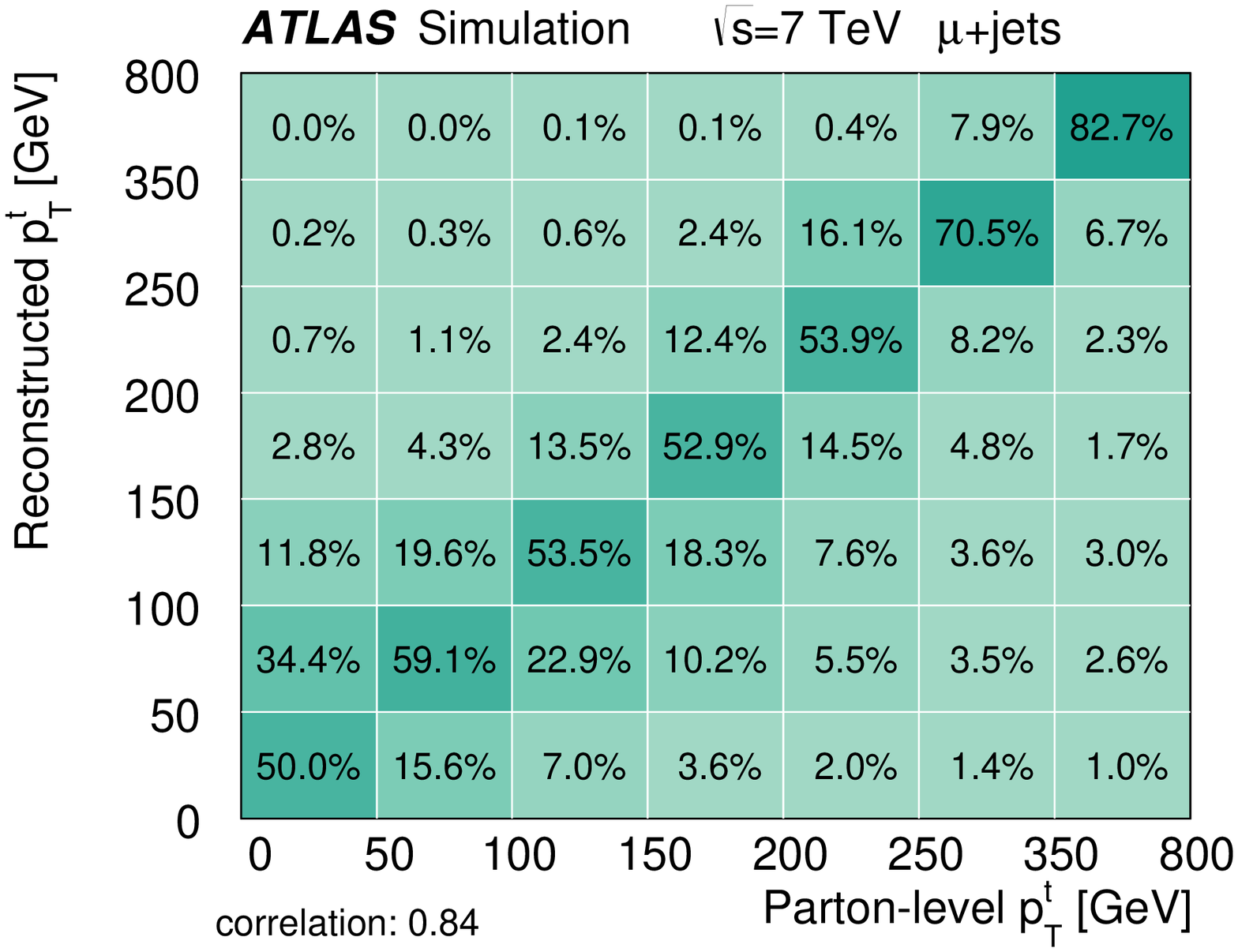}\label{migra_top2_mu}}
\subfigure[]{ \includegraphics[width=0.49\textwidth]{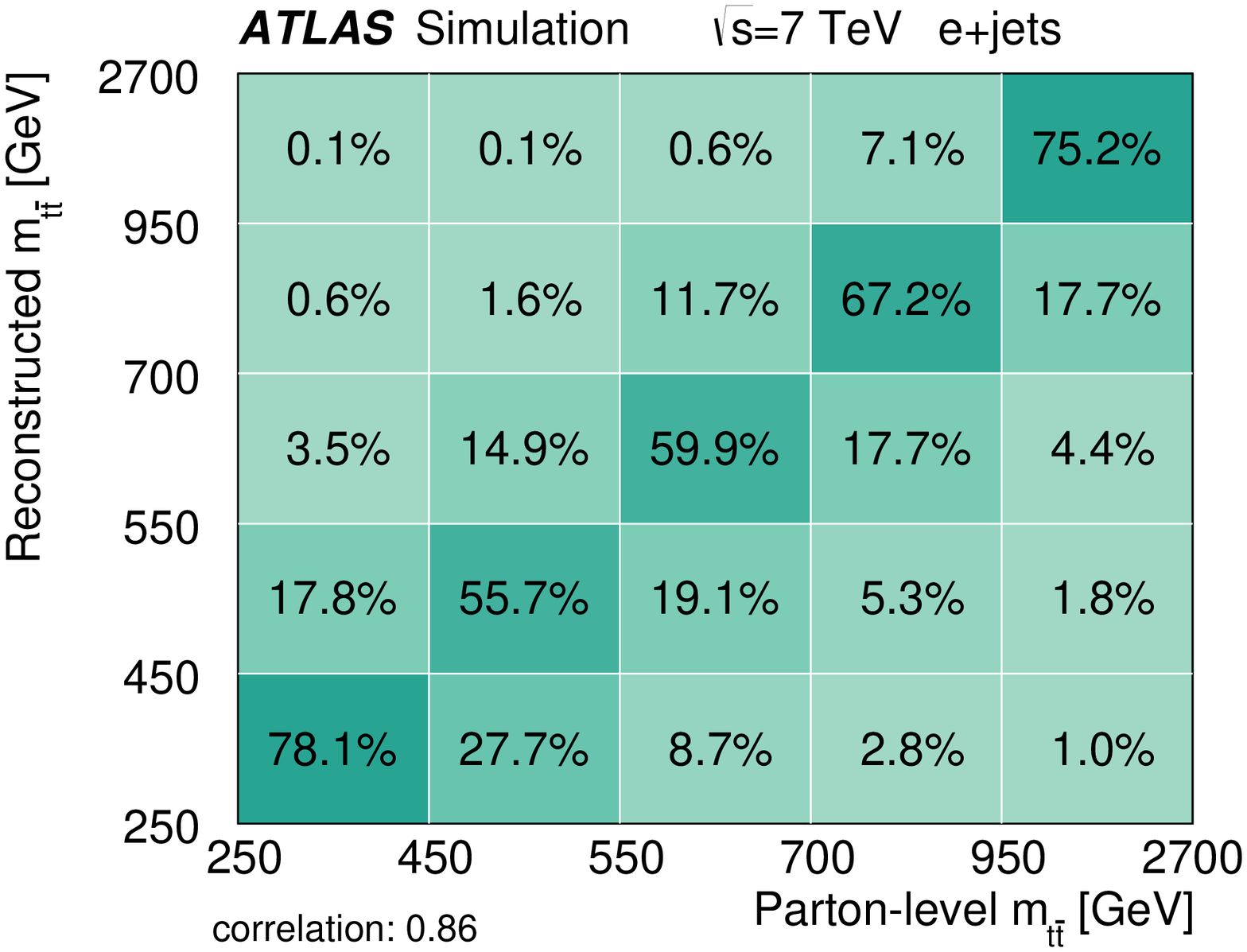}\label{migra_mass_el}}
\subfigure[]{ \includegraphics[width=0.49\textwidth]{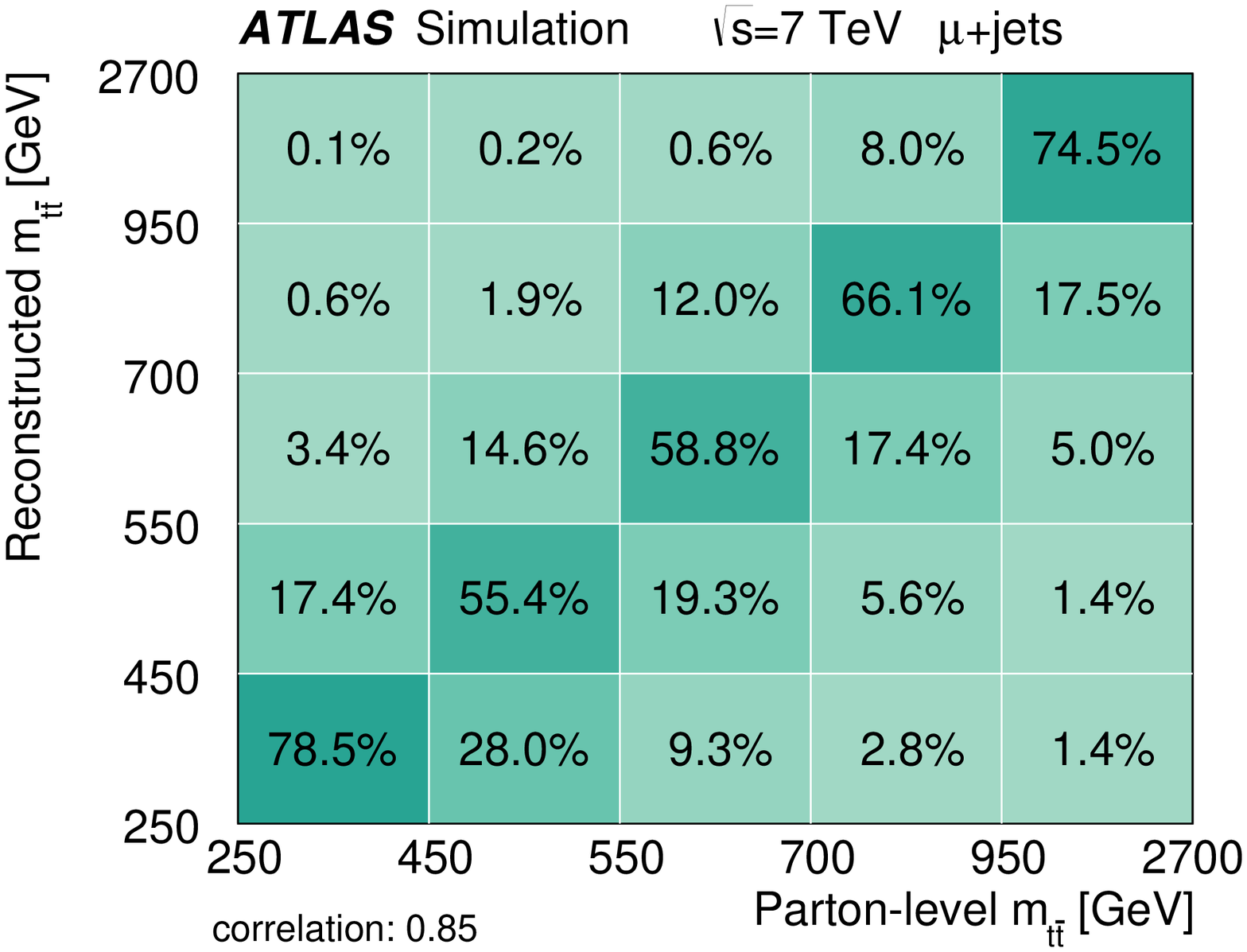}\label{migra_mass_mu}}
\caption{(Color online) The migration matrices obtained from the \Alpgen{}+\Herwig{} simulation, relating the parton and reconstructed levels for the transverse momentum of the hadronically decaying top quark ($\ptt$) in the \subref{migra_top2_el}~\ejets{} and \subref{migra_top2_mu}~\mujets{} channels, and the mass of the $\ttbar$ system ($\mttbar$) in the \subref{migra_mass_el}~\ejets{} and \subref{migra_mass_mu}~\mujets{} channels. The linear correlation coefficient is given below each plot and all columns are normalized to unity (before rounding-off).}
\label{fig:migrations_tagged_1}
\end{figure*}

\begin{figure*}[htbp]
\centering
\subfigure[]{ \includegraphics[width=0.49\textwidth]{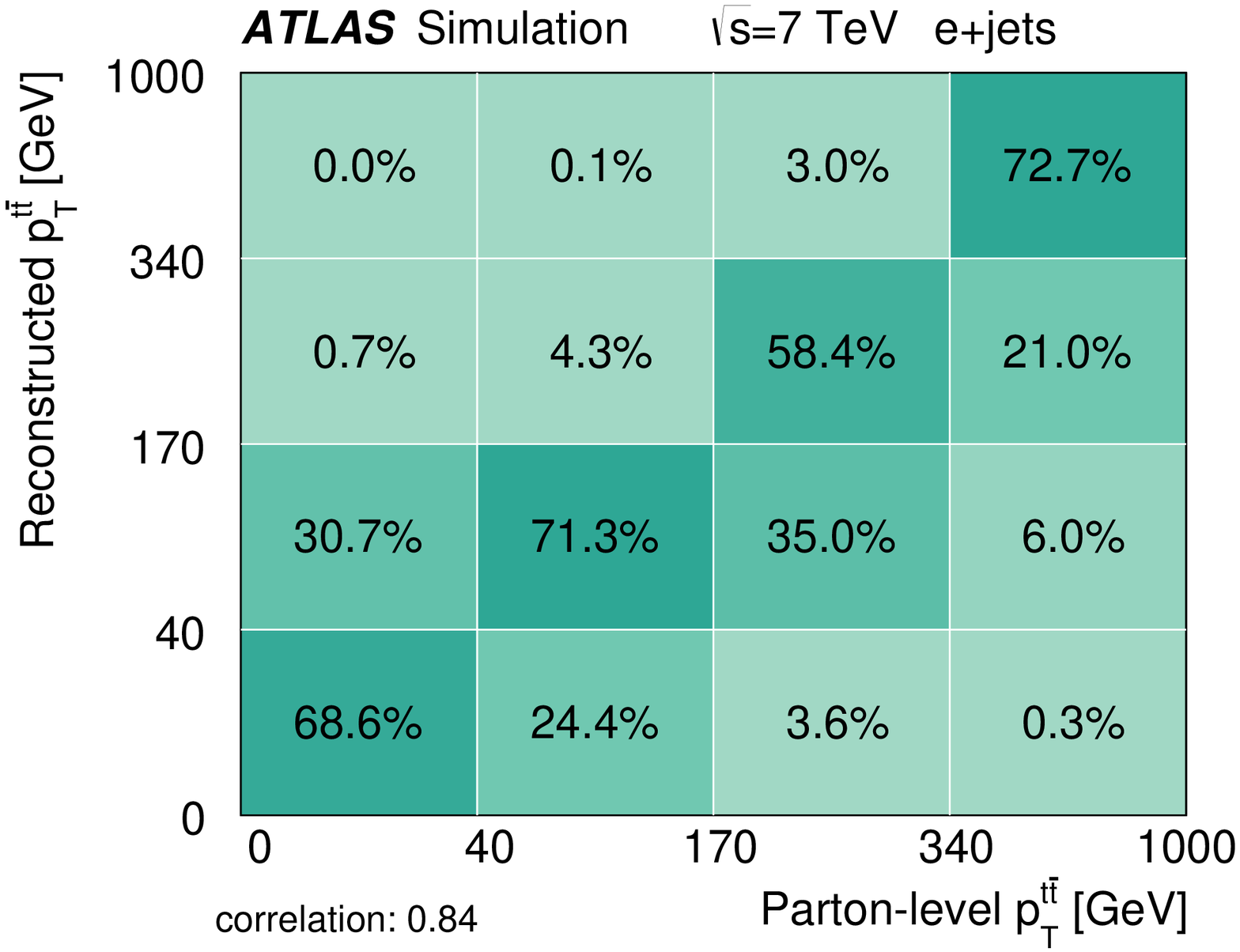}\label{migra_pt_el}}
\subfigure[]{ \includegraphics[width=0.49\textwidth]{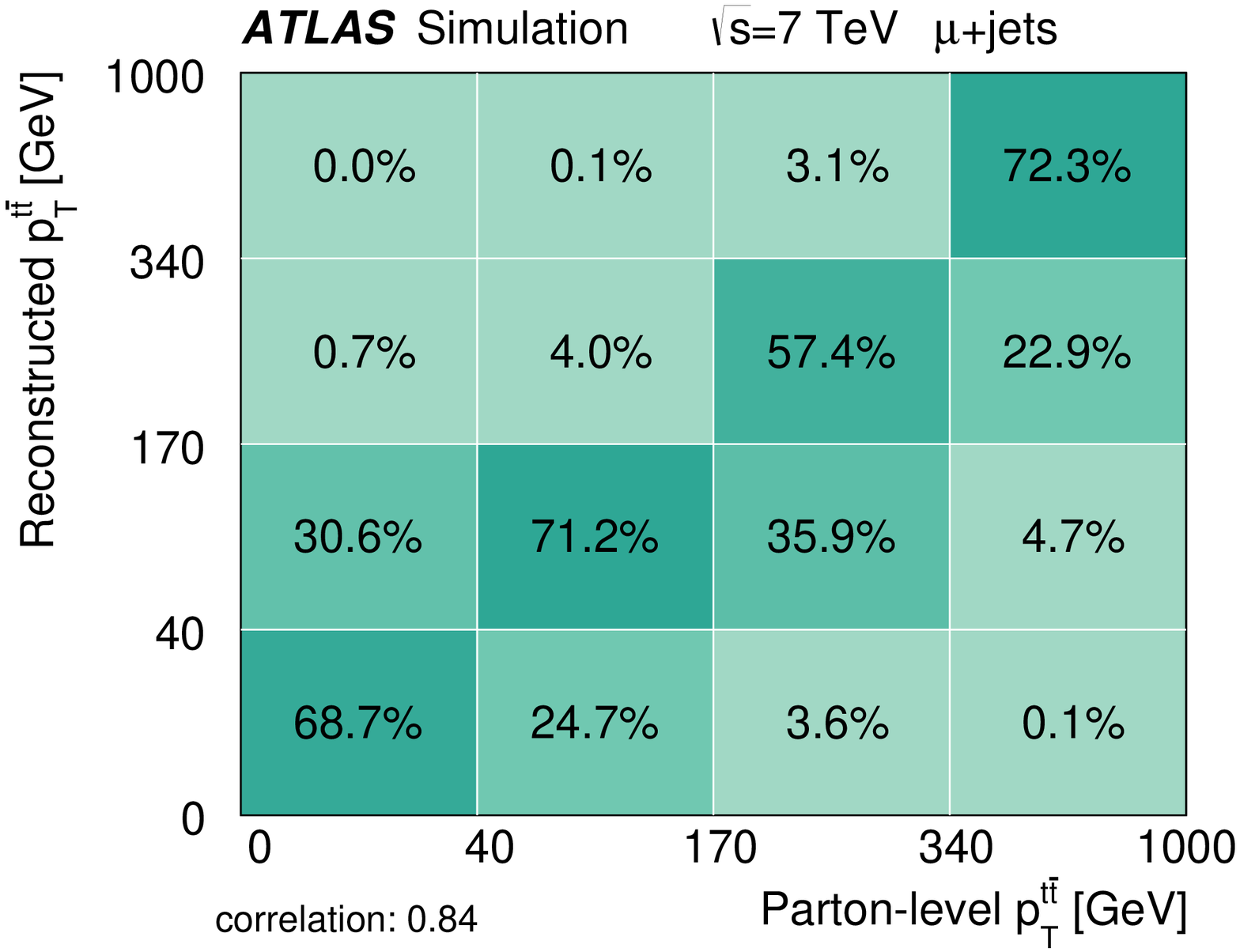}\label{migra_pt_mu}}
\subfigure[]{ \includegraphics[width=0.49\textwidth]{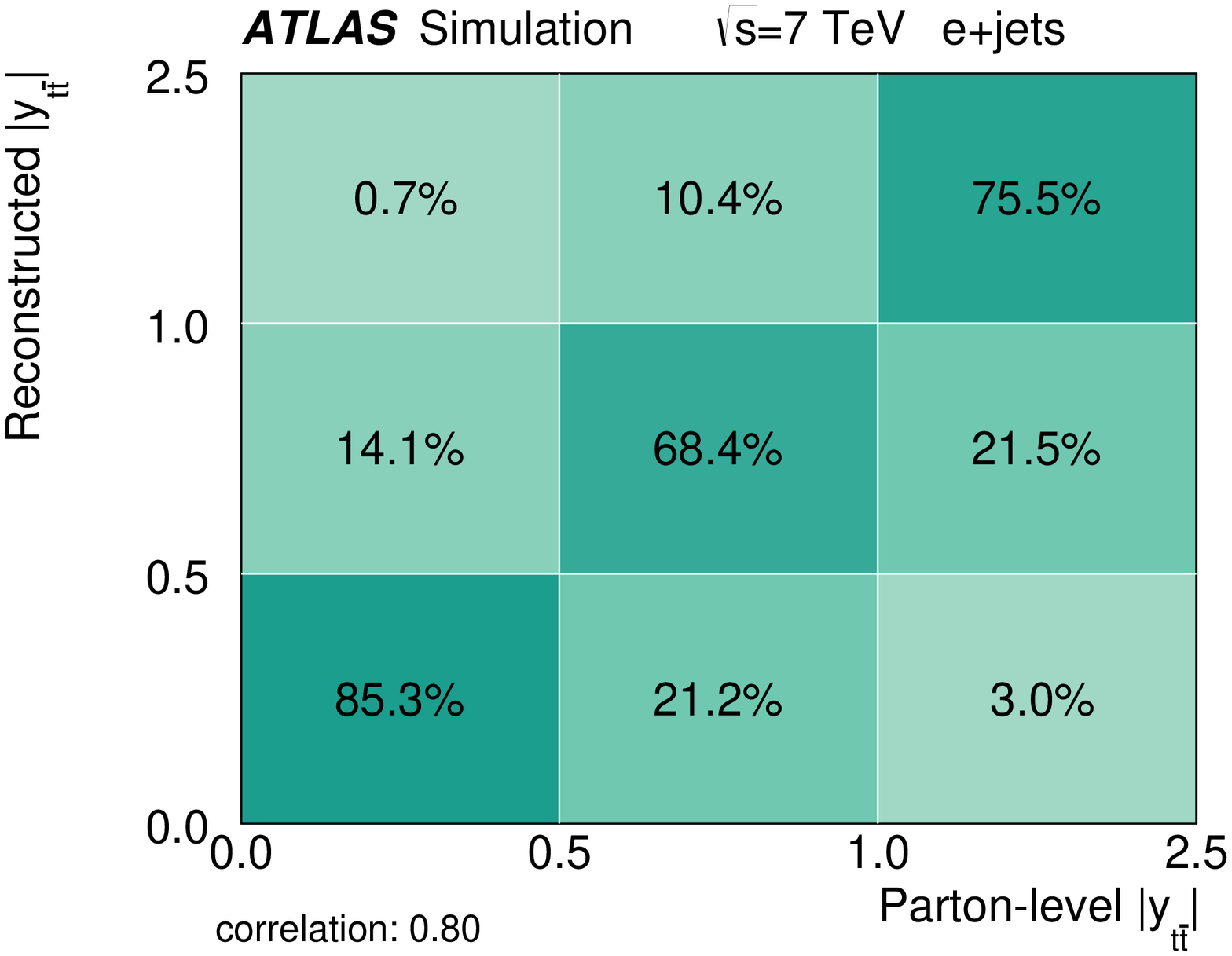}\label{migra_rap_el}}
\subfigure[]{ \includegraphics[width=0.49\textwidth]{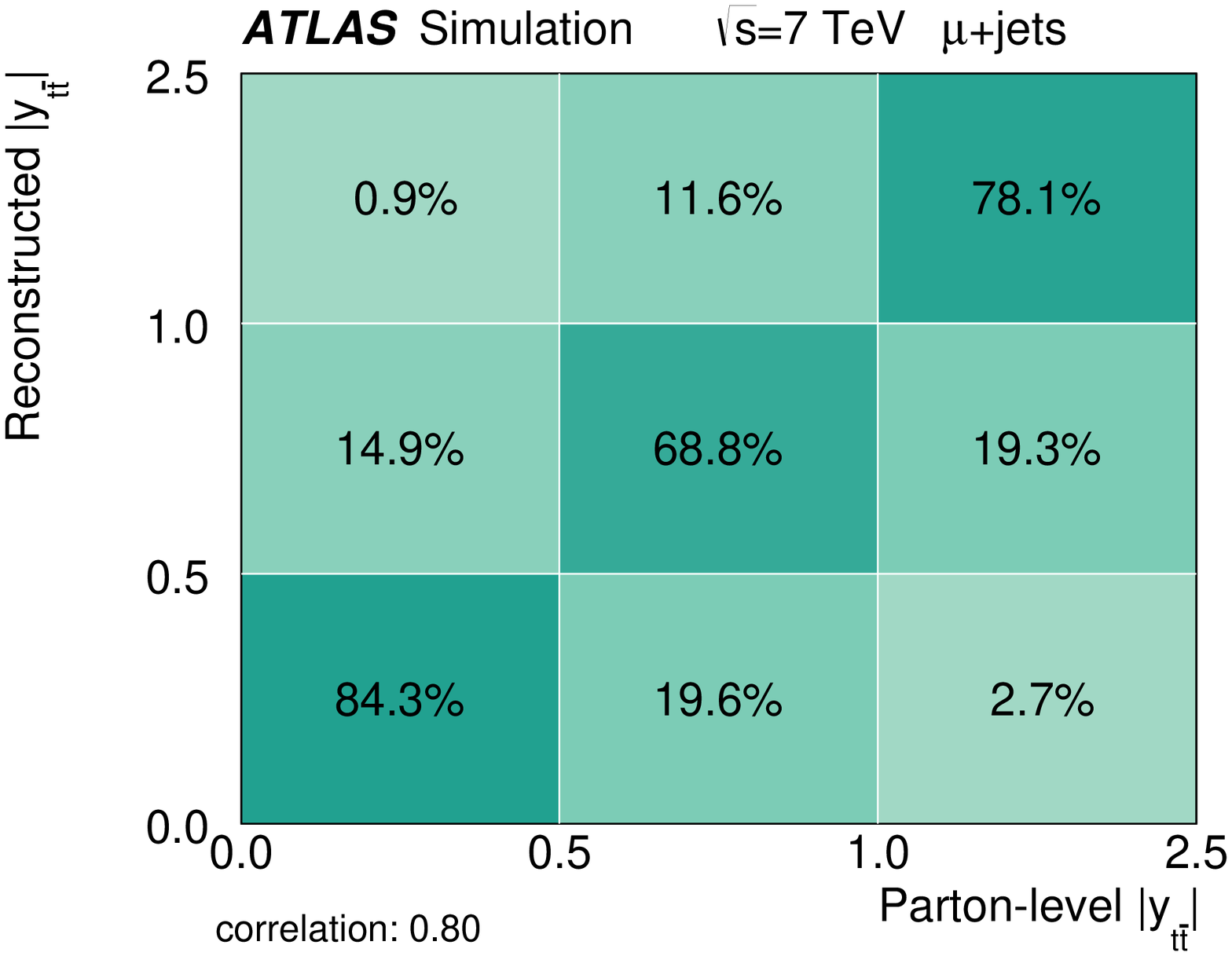}\label{migra_rap_mu}}
\caption{(Color online) The migration matrices obtained from the \Alpgen{}+\Herwig{} simulation, relating the parton and reconstructed levels for the transverse momentum of the $\ttbar$ system ($\ptttbar$) in the \subref{migra_pt_el}~\ejets{} and \subref{migra_pt_mu}~\mujets{} channels, and the absolute value of the rapidity of the $\ttbar$ system ($\absyttbar$) in the \subref{migra_rap_el}~\ejets{} and \subref{migra_rap_mu}~\mujets{} channels. The linear correlation coefficient is given below each plot and all columns are normalized to unity (before rounding-off).}
\label{fig:migrations_tagged_2}
\end{figure*}

The effect of detector resolution is taken into account by constructing the migration matrices, relating the variables of interest at the reconstructed and parton levels, using the \ttbar{} signal simulation.
In Figs.~\ref{fig:migrations_tagged_1} and \ref{fig:migrations_tagged_2}, normalized versions of the migration matrices are presented, where each column is normalized by the number of parton-level events in that bin. The probability for parton-level events to remain in the same bin is therefore shown on the diagonal, and the off-diagonal elements represent the fraction of parton-level events that migrate into other bins. The fraction of events in the diagonal bins is always greater than 50\%, but significant migrations are present in several bins.
The regularized Singular Value Decomposition~\cite{SVD} method is used for the unfolding procedure. A~regularized unfolding technique is chosen in order to prevent large statistical fluctuations that can be introduced when directly inverting the migration matrix. 

To ensure that the results are not biased by the MC generator used for unfolding, the parton-level spectra in simulation are altered by changing the slopes of the \ptt{} and \ptttbar{} distributions by a~factor of two, while for the \mttbar{} distribution the content of one bin ($550$--$700\,$GeV) is increased by a~factor of two to simulate the presence of a~resonance. The shape of the rapidity of the $\ttbar$ system is changed by a~symmetric Gaussian distribution that results in a~reweighting factor of approximately $1.15$ at high $\absyttbar$.

The studies confirm that these altered shapes are indeed recovered within statistical uncertainties by the unfolding based on the nominal migration matrices.
\clearpage

\subsection{Combination of Decay Channels}\label{sec:Combination}

The individual \ejets{} and \mujets{} channels give consistent results: the differences observed in the corresponding bins for all variables of interest are below two standard deviations, taking into account the correlated uncertainties between the two channels.

The Asymmetric BLUE method~\cite{BLUE} is used to combine the cross-sections measured in the \ejets{} and \mujets{} channels, where BLUE refers to the best linear unbiased estimator~\cite{BLUE_paper1}.
The covariance matrix between the two channels is constructed in each kinematic bin by assuming zero or full correlation for channel-specific or common systematic uncertainty sources, respectively. 
The cross-sections are normalized to unity after the combination.
The combined results are compared and found to be in good agreement with the results of unfolding a~merged dataset of both the \ejets{} and \mujets{} channels.

\section{Uncertainties} \label{sec:Uncertainties}

The statistical uncertainty on the data is evaluated with pseudo-experiments by assuming Poisson fluctuations in the data event counts.

The systematic uncertainties are evaluated by varying each source of uncertainty by one standard deviation, propagating this effect through the event selection, unfolding and efficiency corrections, and then considering, for each channel, variable and bin, the variation with respect to the nominal result. This is done separately for the upward and downward variations. For one-sided uncertainties, as in the case of the comparison of two different models, the resulting variation is assumed to be of the same size in both directions and is therefore symmetrized. 
The combined systematic uncertainties are obtained by using the nominal BLUE weights, assigned to each channel in each bin, to linearly combine the systematic uncertainties in the individual channels, and normalizing after the combination.
The total systematic uncertainty in each kinematic bin is computed as the sum in quadrature of individual systematic variations.

The systematic uncertainties and how they affect each of the variables studied are given, grouped into categories, in Tables~\ref{tab:CombSyst_1} and~\ref{tab:CombSyst_2}. The individual systematic uncertainties are listed for completeness in Appendix~\ref{Sec:Appendix:Syst}. The precision of the measurement is dominated by systematic uncertainties. They can be classified into three categories: systematic uncertainties affecting the detector modeling, signal modeling, and background modeling.
\begin{table*} [!htbp]
\footnotesize
\centering

\noindent\makebox[\textwidth]{
\begin{tabular}{l c c c c c c c}
\hline
$\frac{1}{\sigma} \frac{\dsigma}{\dptt}$ Uncertainties [\%] $/$ Bins [GeV] &     0--50 &     50--100 &     100--150 &     150--200 &     200--250 &     250--350 &     350--800 \\ 
\hline
\hline
Jet energy scale & $^{+    3.2 }_{-    2.9}$ & $^{+    1.0 }_{-    1.1}$ & $^{+    1.5 }_{-    1.6}$ & $^{+    2.4 }_{-    2.3}$ & $^{+    2.4 }_{-    2.1}$ & $\pm    2.5$ & $\pm    3.6$ \\ [1mm] 
Jet energy resolution & $\pm    0.4$ & $\pm    0.1$ & $\pm    0.5$ & ${\rm --}$ & $\pm    0.3$ & ${\rm --}$ & $\pm    0.5$ \\ [1mm] 
Jet reconstruction efficiency & ${\rm --}$ & ${\rm --}$ & ${\rm --}$ & ${\rm --}$ & ${\rm --}$ & ${\rm --}$ & $\pm    0.1$ \\ [1mm] 
$b$-quark tagging efficiency & $^{+    1.1 }_{-    1.4}$ & $^{+    0.6 }_{-    0.8}$ & $\pm    0.3$ & $^{+    1.3 }_{-    1.1}$ & $^{+    2.1 }_{-    1.5}$ & $^{+    2.6 }_{-    1.6}$ & $^{+    3.0 }_{-    1.6}$ \\ [1mm] 
$c$-quark tagging efficiency & ${\rm --}$ & ${\rm --}$ & ${\rm --}$ & ${\rm --}$ & $\pm    0.1$ & $\pm    0.1$ & $\pm    0.2$ \\ [1mm] 
Light-jet tagging efficiency & $\pm    0.3$ & ${\rm --}$ & $\pm    0.2$ & ${\rm --}$ & ${\rm --}$ & ${\rm --}$ & $\pm    0.2$ \\ [1mm] 
Lepton selection and momentum scale & $^{+    0.9 }_{-    0.8}$ & $^{+    0.2 }_{-    0.1}$ & $^{+    1.3 }_{-    1.2}$ & $\pm    0.6$ & $\pm    0.9$ & $\pm    1.1$ & $^{+    1.0 }_{-    0.8}$ \\ [1mm] 
\Etmiss{} unassociated cells & $^{+    0.4 }_{-    0.1}$ & ${\rm --}$ & $^{+    0.2 }_{-    0.4}$ & ${\rm --}$ & $^{+    0.3 }_{-    0.2}$ & $^{+    0.3 }_{-    0.4}$ & $^{+    0.3 }_{\rm --}$ \\ [1mm] 
\Etmiss{} pile-up & $^{+    0.6 }_{-    0.1}$ & ${\rm --}$ & $^{+    0.1 }_{-    0.6}$ & $^{\rm --}_{-    0.1}$ & $^{+    0.4 }_{\rm --}$ & $^{+    0.6 }_{\rm --}$ & $^{+    0.8 }_{\rm --}$ \\ [1mm] 
MC generator & $^{+    1.9 }_{-    1.5}$ & $^{+    0.5 }_{-    0.7}$ & $\pm    0.2$ & $^{+    1.5 }_{-    1.9}$ & $\pm    0.1$ & $^{+    3.5 }_{-    2.8}$ & $^{+     11 }_{-    8.6}$ \\ [1mm] 
Fragmentation & $\pm    0.6$ & $\pm    0.7$ & $\pm    0.7$ & $^{+    0.9 }_{-    0.8}$ & $^{+    0.9 }_{-    1.0}$ & $\pm    0.7$ & $\pm    1.9$ \\ [1mm] 
IFSR & $^{+    2.2 }_{-    2.1}$ & $\pm    0.9$ & ${\rm --}$ & $^{+    3.1 }_{-    3.2}$ & $^{+    3.1 }_{-    3.2}$ & $^{+    1.5 }_{-    1.6}$ & ${\rm --}$ \\ [1mm] 
PDF & $\pm    0.1$ & $\pm    0.1$ & ${\rm --}$ & $\pm    0.2$ & $\pm    0.5$ & $\pm    0.8$ & $\pm    0.8$ \\ [1mm] 
MC statistics & $\pm    1.0$ & $\pm    0.4$ & $\pm    0.7$ & $\pm    0.9$ & $\pm    1.1$ & $\pm    1.4$ & $\pm    2.6$ \\ [1mm] 
\Wboson+jets & $\pm    1.7$ & $\pm    0.3$ & $\pm    0.7$ & $^{+    0.9 }_{-    0.8}$ & $^{+    1.0 }_{-    0.9}$ & $^{+    1.4 }_{-    1.3}$ & $\pm    1.4$ \\ [1mm] 
Other backgrounds & $^{+    1.5 }_{-    1.6}$ & $\pm    0.2$ & $^{+    1.0 }_{-    0.9}$ & $^{+    0.7 }_{-    0.5}$ & $^{+    0.6 }_{-    0.4}$ & $\pm    0.8$ & $^{+    0.9 }_{-    1.0}$ \\ [1mm] 
\hline
\hline
 Statistical uncertainty  & $\pm$2.4 & $\pm$1.2 & $\pm$2.5 & $\pm$2.0 & $\pm$2.4 & $\pm$3.5 & $\pm$6.1 \\ [2mm] 
 Total systematic uncertainty  & $^{+    5.3 }_{-    5.0}$ & $^{+    1.8 }_{-    2.0}$ & $^{+    2.6 }_{-    2.7}$ & $\pm    4.8$ & $^{+    4.9 }_{-    4.6}$ & $^{+    5.9 }_{-    5.1}$ & $^{+   12 }_{-   10}$ \\ [2mm] 
\hline
\hline
\end{tabular}}

\vspace{.1 cm}

\noindent\makebox[\textwidth]{
\begin{tabular}{l c c c c c}
\hline
$\frac{1}{\sigma} \frac{\dsigma}{\dmttbar} $ Uncertainties [\%] $/$ Bins [GeV] &     250--450 &     450--550 &     550--700 &     700--950 &     950--2700 \\ 
\hline
\hline
Jet energy scale & $^{+    1.4 }_{-    1.3}$ & $^{+    0.9 }_{-    0.7}$ & $^{+    2.1 }_{-    1.7}$ & $^{+    3.0 }_{-    3.1}$ & $^{+    3.6 }_{-    4.4}$ \\ [1mm] 
Jet energy resolution & $\pm    0.6$ & $\pm    0.9$ & $\pm    0.2$ & $\pm    0.2$ & ${\rm --}$ \\ [1mm] 
Jet reconstruction efficiency & ${\rm --}$ & ${\rm --}$ & ${\rm --}$ & ${\rm --}$ & $\pm    0.2$ \\ [1mm] 
$b$-quark tagging efficiency & $^{+    0.8 }_{-    1.0}$ & $\pm    0.4$ & $^{+    1.6 }_{-    1.3}$ & $^{+    2.0 }_{-    1.3}$ & $^{+    2.2 }_{-    1.2}$ \\ [1mm] 
$c$-quark tagging efficiency & ${\rm --}$ & ${\rm --}$ & $\pm    0.2$ & ${\rm --}$ & $\pm    0.1$ \\ [1mm] 
Light-jet tagging efficiency & ${\rm --}$ & $\pm    0.1$ & ${\rm --}$ & ${\rm --}$ & $\pm    0.1$ \\ [1mm] 
Lepton selection and momentum scale & $\pm    0.5$ & $\pm    0.8$ & $\pm    0.9$ & $\pm    1.7$ & $^{+    1.9 }_{-    1.8}$ \\ [1mm] 
\Etmiss{} unassociated cells & ${\rm --}$ & $^{+    0.1 }_{\rm --}$ & ${\rm --}$ & $^{\rm --}_{-    0.2}$ & $^{+    0.5 }_{-    0.4}$ \\ [1mm] 
\Etmiss{} pile-up & $^{\rm --}_{-    0.1}$ & ${\rm --}$ & $^{+    0.2 }_{\rm --}$ & $^{+    0.2 }_{\rm --}$ & $^{+    0.6 }_{-    0.3}$ \\ [1mm] 
MC generator & $^{+    2.7 }_{-    2.2}$ & $^{+    1.9 }_{-    2.3}$ & $^{+    2.6 }_{-    3.2}$ & $^{+    3.0 }_{-    3.7}$ & $^{+    2.5 }_{-    3.1}$ \\ [1mm] 
Fragmentation & $\pm    0.2$ & $\pm    0.2$ & $\pm    0.5$ & $\pm    1.7$ & $^{+    2.1 }_{-    2.2}$ \\ [1mm] 
IFSR & $^{+    0.6 }_{-    0.5}$ & $\pm    0.2$ & $\pm    0.9$ & $^{+    1.4 }_{-    1.5}$ & $\pm    0.4$ \\ [1mm] 
PDF & ${\rm --}$ & ${\rm --}$ & ${\rm --}$ & $^{+    0.5 }_{-    0.6}$ & $^{+    2.2 }_{-    2.3}$ \\ [1mm] 
MC statistics & $\pm    0.4$ & $\pm    0.4$ & $\pm    0.6$ & $\pm    1.0$ & $\pm    1.6$ \\ [1mm] 
\Wboson+jets & $\pm    0.2$ & $^{+    0.3 }_{-    0.2}$ & $^{+    0.5 }_{-    0.4}$ & $^{+    1.2 }_{-    1.0}$ & $^{+    1.9 }_{-    1.7}$ \\ [1mm] 
Other backgrounds & $\pm    0.3$ & $\pm    0.7$ & $^{+    0.8 }_{-    0.9}$ & $^{+    2.3 }_{-    2.6}$ & $^{+    4.5 }_{-    5.4}$ \\ [1mm] 
\hline
\hline
 Statistical uncertainty  & $\pm$1.2 & $\pm$1.5 & $\pm$2.7 & $\pm$3.2 & $\pm$5.5 \\ [2mm] 
 Total systematic uncertainty  & $^{+    3.4 }_{-    2.9}$ & $^{+    2.6 }_{-    2.9}$ & $^{+    4.1 }_{-    4.3}$ & $^{+    6.1 }_{-    6.5}$ & $^{+    8.0 }_{-    8.9}$ \\ [2mm] 
\hline
\hline
\end{tabular}}

\caption{The individual systematic uncertainties in the normalized differential cross-sections after combining the \ejets{} and \mujets{} channels for \ptt{} and \mttbar{}, grouped into broad categories, and calculated as a~percentage of the cross-section in each bin. ``Other backgrounds" includes the systematic uncertainties in the single top-quark, dilepton, \Zboson{}+jets and QCD multijet backgrounds, and IFSR refers to initial- and final-state radiation. Dashes are used when the estimated relative systematic uncertainty for that bin is below 0.1\%.}
\label{tab:CombSyst_1}
\end{table*}
\begin{table*} [!htbp]
\footnotesize
\centering

\noindent\makebox[\textwidth]{
\begin{tabular}{l c c c c}
\hline
$\frac{1}{\sigma} \frac{\dsigma}{\dptttbar}$ Uncertainties [\%] $/$ Bins [GeV] &     0--40 &     40--170 &     170--340 &     340--1000 \\ 
\hline
\hline
Jet energy scale & $^{+    1.9 }_{-    2.0}$ & $^{+    2.2 }_{-    2.3}$ & $\pm    4.9$ & $^{+    6.2 }_{-    6.5}$ \\ [1mm] 
Jet energy resolution & $^{+    3.4 }_{-    3.5}$ & $^{+    4.2 }_{-    4.1}$ & $^{+    7.2 }_{-    7.1}$ & $^{+    8.2 }_{-    8.0}$ \\ [1mm] 
Jet reconstruction efficiency & ${\rm --}$ & ${\rm --}$ & $\pm    0.1$ & $\pm    0.3$ \\ [1mm] 
$b$-quark tagging efficiency & $^{\rm --}_{-    0.1}$ & $^{+    0.1 }_{\rm --}$ & $^{+    0.4 }_{\rm --}$ & $^{+    1.0 }_{-    0.1}$ \\ [1mm] 
$c$-quark tagging efficiency & ${\rm --}$ & ${\rm --}$ & $\pm    0.2$ & $^{+    0.3 }_{-    0.2}$ \\ [1mm] 
Light-jet tagging efficiency & ${\rm --}$ & ${\rm --}$ & ${\rm --}$ & $^{+    0.1 }_{-    0.2}$ \\ [1mm] 
Lepton selection and momentum scale & $\pm    0.9$ & $^{+    1.3 }_{-    1.2}$ & $\pm    0.8$ & $\pm    1.0$ \\ [1mm] 
\Etmiss{} unassociated cells & $^{+    1.7 }_{-    1.6}$ & $^{+    2.0 }_{-    2.1}$ & $\pm    2.1$ & $\pm    1.8$ \\ [1mm] 
\Etmiss{} pile-up & $^{+    1.0 }_{-    1.2}$ & $^{+    1.5 }_{-    1.3}$ & $^{+    1.6 }_{-    1.4}$ & $^{+    1.5 }_{-    1.6}$ \\ [1mm] 
MC generator & $^{+    4.2 }_{-    3.5}$ & $^{+    4.2 }_{-    5.1}$ & $^{+    8.0 }_{-    9.8}$ & $^{+    1.5 }_{-    1.2}$ \\ [1mm] 
Fragmentation & $\pm    0.6$ & $\pm    0.1$ & $^{+    6.8 }_{-    6.9}$ & $^{+    2.6 }_{-    2.7}$ \\ [1mm] 
IFSR & $^{+    1.2 }_{-    1.3}$ & $\pm    1.0$ & $^{+    6.2 }_{-    5.8}$ & $^{+     10 }_{-    9.5}$ \\ [1mm] 
PDF & ${\rm --}$ & ${\rm --}$ & $\pm    0.2$ & $\pm    1.3$ \\ [1mm] 
MC statistics & $\pm    0.6$ & $\pm    0.8$ & $\pm    1.7$ & $\pm    2.8$ \\ [1mm] 
\Wboson+jets & $^{+    0.6 }_{-    0.8}$ & $^{+    0.7 }_{-    0.9}$ & $^{+    1.8 }_{-    2.4}$ & $^{+    3.1 }_{-    3.7}$ \\ [1mm] 
Other backgrounds & $\pm    0.8$ & $\pm    1.1$ & $\pm    0.9$ & $\pm    1.1$ \\ [1mm] 
\hline
\hline
 Statistical uncertainty  & $\pm$1.5 & $\pm$1.8 & $\pm$4.5 & $\pm$7.7 \\ [2mm] 
 Total systematic uncertainty  & $^{+    6.4 }_{-    6.0}$ & $^{+    7.1 }_{-    7.7}$ & $^{+   15 }_{-   16}$ & $^{+   16 }_{-   15}$ \\ [2mm] 
\hline
\hline
\end{tabular}}

\vspace{.1 cm}

\noindent\makebox[\textwidth]{
\begin{tabular}{l c c c}
\hline
$\frac{1}{\sigma} \frac{\dsigma}{\dabsyttbar}$ Uncertainties [\%] &     0.0--0.5 &     0.5--1.0 &     1.0--2.5 \\ 
\hline
\hline
Jet energy scale & $^{+    0.6 }_{-    0.5}$ & ${\rm --}$ & $^{+    1.1 }_{-    0.9}$ \\ [1mm] 
Jet energy resolution & $\pm    0.1$ & $\pm    0.1$ & $\pm    0.4$ \\ [1mm] 
Jet reconstruction efficiency & ${\rm --}$ & ${\rm --}$ & ${\rm --}$ \\ [1mm] 
$b$-quark tagging efficiency & ${\rm --}$ & ${\rm --}$ & ${\rm --}$ \\ [1mm] 
$c$-quark tagging efficiency & ${\rm --}$ & ${\rm --}$ & ${\rm --}$ \\ [1mm] 
Light-jet tagging efficiency & ${\rm --}$ & ${\rm --}$ & ${\rm --}$ \\ [1mm] 
Lepton selection and momentum scale & $\pm    0.4$ & $\pm    0.1$ & $^{+    0.9 }_{-    0.8}$ \\ [1mm] 
\Etmiss{} unassociated cells & $\pm    0.1$ & ${\rm --}$ & $^{\rm --}_{-    0.2}$ \\ [1mm] 
\Etmiss{} pile-up & ${\rm --}$ & ${\rm --}$ & $^{\rm --}_{-    0.1}$ \\ [1mm] 
MC generator & $^{+    2.5 }_{-    2.0}$ & $^{+    1.5 }_{-    1.2}$ & $^{+    5.0 }_{-    6.2}$ \\ [1mm] 
Fragmentation & $^{+    1.8 }_{-    1.9}$ & $\pm    0.8$ & $^{+    4.3 }_{-    4.1}$ \\ [1mm] 
IFSR & $\pm    0.1$ & ${\rm --}$ & ${\rm --}$ \\ [1mm] 
PDF & $\pm    1.1$ & ${\rm --}$ & $^{+    1.9 }_{-    2.0}$ \\ [1mm] 
MC statistics & $\pm    0.2$ & ${\rm --}$ & $\pm    0.3$ \\ [1mm] 
\Wboson+jets & $\pm    0.3$ & ${\rm --}$ & $^{+    0.5 }_{-    0.4}$ \\ [1mm] 
Other backgrounds & $\pm    0.4$ & $\pm    0.1$ & $\pm    0.9$ \\ [1mm] 
\hline
\hline
 Statistical uncertainty  & $\pm$0.7 & $\pm$0.4 & $\pm$0.9 \\ [2mm] 
 Total systematic uncertainty  & $^{+    3.4 }_{-    3.1}$ & $^{+    1.7 }_{-    1.5}$ & $^{+    7.1 }_{-    7.9}$ \\ [2mm] 
\hline
\hline
\end{tabular}}

\caption{The individual systematic uncertainties in the normalized differential cross-sections after combining the \ejets{} and \mujets{} channels for \ptttbar{} and \absyttbar{}, grouped into broad categories, and calculated as a~percentage of the cross-section in each bin. ``Other backgrounds" includes the systematic uncertainties in the single top-quark, dilepton, \Zboson{}+jets and QCD multijet backgrounds, and IFSR refers to initial- and final-state radiation. Dashes are used when the estimated relative systematic uncertainty for that bin is below 0.1\%.}
\label{tab:CombSyst_2}
\end{table*}

\subsection{Detector Modeling} \label{sec:detectormodeling}

The systematic uncertainties related to the detector modeling induce effects on the reconstruction of the physics objects (leptons, jets and \Etmiss{}) used in the selection and in the reconstruction of the kinematic variables under study. 

The jet energy scale (JES) systematic uncertainty on the signal, acting on both the efficiency and bin migrations, is evaluated using 21 separate components~\cite{jes:2013}, which allow proper treatment of correlations across the kinematic bins. 
The impact of the JES uncertainty on the background is evaluated using the overall JES variation defined as the sum in quadrature of the individual components, and is added to the signal JES systematic uncertainty linearly to account for the correlation between them. The simplified treatment of the JES uncertainty for the background has a~negligible effect on the results.

The uncertainty on the jet energy resolution is modeled by varying the jet energies according to the systematic uncertainties of the resolution measurement performed on data~\cite{jer:2013}. The contribution from this uncertainty is generally small except for the \ptttbar{} distribution.

The uncertainty on the jet reconstruction efficiency is accounted for by 
randomly removing jets, in the simulation, according to the uncertainty 
on the jet reconstruction efficiency measured in data~\cite{jer_2}. 
The effect of this uncertainty is negligible for all the spectra.

The corrections accounting for differences in $b$-tagging efficiencies and mistag rates for $c$-quarks and light-quarks, between data and simulation, are derived from data and parameterized as a~function of $\pt$ and $\eta$~\cite{btag1:2011,btag2:2011}. The uncertainties in these corrections are propagated through the analysis.

Electron and muon trigger, reconstruction, and selection efficiencies are measured in data using $W$ and $Z$~boson decays and are incorporated as appropriate correction factors into the simulation. A~similar procedure is used for the lepton energy and momentum scales and resolutions.
The impact of the uncertainties in all these corrections is at the sub-percent level.

The uncertainties in the energy scale and resolution corrections for jets and high-$\pt$ leptons are propagated to the uncertainty on \Etmiss{}. Other minor systematic uncertainty contributions on the modeling of \Etmiss{} arise from effects due to the pile-up modeling and the uncertainties in the unassociated-cell term~\cite{atlasEtmisPerf}. These contributions are generally at the sub-percent level except for the \ptttbar{} distribution.

The efficiency of the likelihood cut discussed in Sec.~\ref{sec:TopSystemReconstruction} is observed to be $2\pm1\%$ smaller in data than in simulation, but this discrepancy has no kinematic dependence and hence no effect on the unfolded normalized distributions.

\subsection{Signal Modeling} \label{sec:signalmodeling}

The sources of uncertainty for the signal modeling come from the choice of generator used for the simulation of the \ttbar{} process, the parton shower and hadronization model, the model for initial- and final-state QCD radiation (IFSR), and the choice of PDF.

The uncertainties due to the generator choice are evaluated using \McAtNlo{}+\Herwig{} to unfold the data, instead of the nominal \Alpgen{}+\Herwig{}. These uncertainties are larger than those that would result from using \Powheg{}+\Herwig{} as an alternative model for unfolding.
The differences between the fully corrected data distributions obtained in this way and the nominal ones are symmetrized and taken as systematic uncertainties. 

The parton shower and hadronization systematic uncertainties (referred to as fragmentation) are evaluated by comparing the distributions obtained using \Alpgen{}+\Herwig{} and \Alpgen{}+\Pythia{} to unfold the data. The \Alpgen{}+\Pythia{} sample is generated using \Alpgen{} (v2.14) and uses the CTEQ5L PDF~\cite{Lai:1999wy} for the hard process and parton shower.

The effect of IFSR modeling is determined by using two different \Alpgen{}+\Pythia{} samples with varied radiation settings. The distribution of the number of additional partons is changed by varying the renormalization scale associated with $\alpha_{\rm S}$ consistently in the hard matrix element as well as in the parton shower. The parameters controlling the level of radiation via parton showering~\cite{IFSRparameters} were adjusted to encompass the ATLAS measurement of additional jet activity in $\ttbar$ events~\cite{gap_fraction}. These samples are generated with dedicated Perugia 2011 tunes and used to fully correct the data through the unfolding. The IFSR uncertainty is assumed to be half the difference between the two unfolded distributions.

The PDF systematic uncertainty is evaluated by studying the effect on the signal efficiency of using different PDF sets to reweight simulated events at the hard-process level. The PDF sets used are CT10~\cite{CT10}, MSTW2008NLO~\cite{MSTW}, and NNPDF2.3~\cite{NNPDF}. Both the uncertainties within a~given PDF set and the variations between the different PDF sets are taken into account~\cite{PDF4LHC}.

The systematic uncertainties due to the finite size of the simulated samples are evaluated by varying the content of the migration matrix within statistical uncertainties and evaluating the standard deviation of the ensemble of results unfolded with the varied matrices. Simultaneously, the efficiency is re-derived using the parton spectrum projected from the varied migration matrix and therefore accounts for the same statistical fluctuations.

\subsection{Background Modeling} \label{sec:bkgmodeling}

The normalization of the \Wboson+jets background is varied within the uncertainty of the data-driven method, which amounts to 15\% and 13\% for the \ejets{} and \mujets{} channels, respectively. An additional uncertainty of 18\% (\ejets{}) and 21\% (\mujets{}) comes from determining the flavor composition of the sample. This includes the uncertainty on the extrapolation of the flavor composition to jet multiplicities beyond two (the $f^{\rm tag}_{2\rightarrow \geq 4}$ term described in Sec.~\ref{sec:WjetsBackground}).

The multijet background uncertainties are estimated by comparing alternative estimates and their agreement with data in control regions. The resulting normalization uncertainties are 50\% and 20\% for the \ejets{} and \mujets{} channels respectively. 

The statistical uncertainty on the background simulation samples is taken into account by fluctuating the background sum with a~Gaussian distribution in each bin within the uncertainties and propagating the effect to the unfolded distributions.

The uncertainty on the  \Zboson$+$jets background normalization is taken to be 50\% 
in the four-jet bin and the uncertainty on the diboson normalization is taken to be 40\% in the same jet multiplicity bin. The effect of these uncertainties in the final results is negligible. Effects of the uncertainties in the normalizations of the single top and dilepton \ttbar{} backgrounds are also negligible.

\subsection{Main Sources of Systematic Uncertainties}

For \ptt{} and \mttbar{} the largest systematic uncertainties come from JES, signal generator choice, and $b$-quark tagging efficiency.  For \ptttbar{} the uncertainty from IFSR is the largest, followed by signal generator choice, fragmentation and jet energy resolution. Finally, for \yttbar{} the main uncertainties come from the signal generator choice and fragmentation.

\section{Results}
\label{sec:Results}

The unfolded and combined normalized differential cross-sections are shown in Table~\ref{tab:XsectionTable}. The absolute cross-sections, calculated by integrating the spectra before normalization ($160\,{\rm pb}$ for the \ejets{} and \mujets{} channels combined, with a~relative uncertainty of $15\%$), agree with the theoretical calculations within uncertainties. The total uncertainty is dominated by systematic sources as discussed in Sec.~\ref{sec:Uncertainties}.

\begin{table}[!ht]
\begin{center}

\begin{tabular}{r@{  --}rr@{$\pm$}lcc}
\hline
\multicolumn{2}{l}{$\ptt{}$ [GeV]} & \multicolumn{2}{c}{$\frac{1}{\sigma} \frac{\dsigma}{\dptt} \left[\rm 10^{-3} {\rm GeV}^{-1} \, \right]$} & Stat. [\%] & Syst. [\%] \\ 
\hline
   0 &   50 & 3.4 &  0.2 & $\pm$     2.4 & $\pm$     5.1 \\ 
  50 &  100 & 6.7 &  0.2 & $\pm$     1.2 & $\pm$     1.9 \\ 
 100 &  150 & 5.3 &  0.2 & $\pm$     2.5 & $\pm$     2.6 \\ 
 150 &  200 & 2.6 &  0.1 & $\pm$     2.0 & $\pm$     4.8 \\ 
 200 &  250 & 1.12 &  0.06 & $\pm$     2.4 & $\pm$     4.8 \\ 
 250 &  350 & 0.32 &  0.02 & $\pm$     3.5 & $\pm$     5.5 \\ 
 350 &  800 & 0.018 &  0.002 & $\pm$     6.1 & $\pm$    11 \\ 

\multicolumn{2}{c}{}  &  \multicolumn{2}{c}{} & \\
\hline
\multicolumn{2}{l}{$\mttbar{}$ [GeV]} & \multicolumn{2}{c}{$\frac{1}{\sigma} \frac{\dsigma}{\dmttbar} \left[\rm 10^{-3} \, {\rm GeV}^{-1} \right]$} & Stat. [\%] & Syst. [\%] \\ 
\hline
 250 &  450 & 2.52 &  0.09 & $\pm$     1.2 & $\pm$     3.1 \\ 
 450 &  550 & 2.76 &  0.09 & $\pm$     1.5 & $\pm$     2.8 \\ 
 550 &  700 & 1.01 &  0.05 & $\pm$     2.7 & $\pm$     4.2 \\ 
 700 &  950 & 0.23 &  0.02 & $\pm$     3.2 & $\pm$     6.3 \\ 
 950 & 2700 & 0.0071 &  0.0007 & $\pm$     5.5 & $\pm$     8.5 \\ 

\multicolumn{2}{c}{}  &  \multicolumn{2}{c}{} & \\
\hline
\multicolumn{2}{l}{$\ptttbar{}$ [GeV]} & \multicolumn{2}{c}{$\frac{1}{\sigma} \frac{\dsigma}{\dptttbar} \left[\rm 10^{-3} \, {\rm GeV}^{-1} \right]$} & Stat. [\%] & Syst. [\%] \\ 
\hline
   0 &   40 & 14.1 &  0.9 & $\pm$     1.5 & $\pm$     6.2 \\ 
  40 &  170 & 3.0 &  0.2 & $\pm$     1.8 & $\pm$     7.4 \\ 
 170 &  340 & 0.25 &  0.04 & $\pm$     4.5 & $\pm$    16 \\ 
 340 & 1000 & 0.008 &  0.001 & $\pm$     7.7 & $\pm$    16 \\ 

\multicolumn{2}{c}{}  &  \multicolumn{2}{c}{} & \\
\hline
\multicolumn{2}{c}{$\absyttbar{}$ } & \multicolumn{2}{l}{$\ \ \ \ \ \frac{1}{\sigma} \frac{\dsigma}{\dabsyttbar}$} & Stat. [\%] & Syst. [\%] \\ 
\hline
 0.0 &  0.5 & 0.86 & 0.03 & $\pm$     0.7 & $\pm$     3.2 \\ 
 0.5 &  1.0 & 0.64 & 0.01 & $\pm$     0.4 & $\pm$     1.6 \\ 
 1.0 &  2.5 & 0.17 & 0.01 & $\pm$     0.9 & $\pm$     7.5 \\ 
\hline
\end{tabular}

\caption{Normalized differential cross-sections for the different variables considered. The cross-section in each bin is given as the integral of the normalized differential cross-section over the bin width, divided by the bin width. The calculation of the cross-sections in the last bins includes events falling outside of the bin edges, and the normalization is done within the quoted bin width. The reported total uncertainty in the second column is obtained by adding the statistical and systematic uncertainties in quadrature.}
\label{tab:XsectionTable}
\end{center}
\end{table}

The unfolded distributions are also shown compared to different MC generators in Fig.~\ref{fig:combined_results_with_MC}. \Alpgen{} and \McAtNlo{} use \Herwig{} for parton shower and hadronization, while the PDFs are different as mentioned in Sec.~\ref{sec:Simulation}, and \Powheg{} is shown interfaced with both \Herwig{} and \Pythia{}

The covariance matrices for the normalized unfolded spectra due to the statistical and systematic uncertainties are displayed in~Table~\ref{tab:Cov}. They are obtained by 
evaluating the covariance between the kinematic bins using pseudo-experiments simultaneously in both the \ejets{} and \mujets{} 
channels and combining them as described in Sec.~\ref{sec:Combination}. 

The correlations due to statistical fluctuations are shown in Appendix~\ref{Sec:Appendix:CorrVars}.
They are evaluated by varying the data event counts independently in every bin before unfolding, propagating the statistical uncertainties through the unfolding 
separately for the \ejets{} and \mujets{} channels, and then performing the
combination of the two channels.
Large off-diagonal correlations come from 
the normalization constraint for the spectra and the regularization in
the unfolding procedure.
The statistical correlations between bins of different variables have also been 
evaluated and are presented in Appendix~\ref{Sec:Appendix:CorrVars}. 

\begin{table*} [!ht]
\footnotesize
\begin{center}


\noindent\makebox[\textwidth]{
\begin{tabular}{r| c c c c c c c }
\hline
{$\ptt{}$ [GeV]}  & \multicolumn{1}{|c}{0--50} & \multicolumn{1}{|c}{50--100} & \multicolumn{1}{|c}{100--150} & \multicolumn{1}{|c}{150--200} & \multicolumn{1}{|c}{200--250} & \multicolumn{1}{|c}{250--350} & \multicolumn{1}{|c}{350--800}\\ 
\hline
    0--50  & $\phantom{-}4.34\cdot 10^{-2}$  & $\phantom{-}1.04\cdot 10^{-2}$  & $-2.13\cdot 10^{-2}$  & $-2.23\cdot 10^{-2}$  & $-8.16\cdot 10^{-3}$  & $-1.49\cdot 10^{-3}$  & $\phantom{-}1.06\cdot 10^{-4}$  \\ 
    50--100  & $\phantom{-}1.04\cdot 10^{-2}$  & $\phantom{-}2.97\cdot 10^{-2}$  & $-1.39\cdot 10^{-2}$  & $-1.36\cdot 10^{-2}$  & $-7.13\cdot 10^{-3}$  & $-2.10\cdot 10^{-3}$  & $-1.43\cdot 10^{-4}$  \\ 
    100--150  & $-2.13\cdot 10^{-2}$  & $-1.39\cdot 10^{-2}$  & $\phantom{-}3.25\cdot 10^{-2}$  & $\phantom{-}3.70\cdot 10^{-3}$  & $-2.39\cdot 10^{-5}$  & $-2.73\cdot 10^{-4}$  & $-4.08\cdot 10^{-5}$  \\ 
    150--200  & $-2.23\cdot 10^{-2}$  & $-1.36\cdot 10^{-2}$  & $\phantom{-}3.70\cdot 10^{-3}$  & $\phantom{-}2.06\cdot 10^{-2}$  & $\phantom{-}8.48\cdot 10^{-3}$  & $\phantom{-}1.68\cdot 10^{-3}$  & $-2.64\cdot 10^{-5}$  \\ 
    200--250  & $-8.16\cdot 10^{-3}$  & $-7.13\cdot 10^{-3}$  & $-2.39\cdot 10^{-5}$  & $\phantom{-}8.48\cdot 10^{-3}$  & $\phantom{-}4.44\cdot 10^{-3}$  & $\phantom{-}1.09\cdot 10^{-3}$  & $\phantom{-}2.44\cdot 10^{-5}$  \\ 
    250--350  & $-1.49\cdot 10^{-3}$  & $-2.10\cdot 10^{-3}$  & $-2.73\cdot 10^{-4}$  & $\phantom{-}1.68\cdot 10^{-3}$  & $\phantom{-}1.09\cdot 10^{-3}$  & $\phantom{-}4.44\cdot 10^{-4}$  & $\phantom{-}2.33\cdot 10^{-5}$  \\ 
    350--800  & $\phantom{-}1.06\cdot 10^{-4}$  & $-1.43\cdot 10^{-4}$  & $-4.08\cdot 10^{-5}$  & $-2.64\cdot 10^{-5}$  & $\phantom{-}2.44\cdot 10^{-5}$  & $\phantom{-}2.33\cdot 10^{-5}$  & $\phantom{-}3.78\cdot 10^{-6}$  \\ 
\hline
\end{tabular}}

\noindent\makebox[\textwidth]{
\begin{tabular}{r| c c c c c }
\hline
{$\mttbar{}$ [GeV]}  & \multicolumn{1}{|c}{250--450} & \multicolumn{1}{|c}{450--550} & \multicolumn{1}{|c}{550--700} & \multicolumn{1}{|c}{700--950} & \multicolumn{1}{|c}{950--2700}\\ 
\hline
    250--450  & $\phantom{-}7.28\cdot 10^{-3}$  & $-6.76\cdot 10^{-3}$  & $-3.66\cdot 10^{-3}$  & $-7.62\cdot 10^{-4}$  & $-2.29\cdot 10^{-5}$  \\ 
    450--550  & $-6.76\cdot 10^{-3}$  & $\phantom{-}8.20\cdot 10^{-3}$  & $\phantom{-}3.06\cdot 10^{-3}$  & $\phantom{-}2.77\cdot 10^{-4}$  & $\phantom{-}1.99\cdot 10^{-6}$  \\ 
    550--700  & $-3.66\cdot 10^{-3}$  & $\phantom{-}3.06\cdot 10^{-3}$  & $\phantom{-}2.43\cdot 10^{-3}$  & $\phantom{-}2.21\cdot 10^{-4}$  & $\phantom{-}3.25\cdot 10^{-6}$  \\ 
    700--950  & $-7.62\cdot 10^{-4}$  & $\phantom{-}2.77\cdot 10^{-4}$  & $\phantom{-}2.21\cdot 10^{-4}$  & $\phantom{-}2.85\cdot 10^{-4}$  & $\phantom{-}1.16\cdot 10^{-5}$  \\ 
    950--2700  & $-2.29\cdot 10^{-5}$  & $\phantom{-}1.99\cdot 10^{-6}$  & $\phantom{-}3.25\cdot 10^{-6}$  & $\phantom{-}1.16\cdot 10^{-5}$  & $\phantom{-}5.60\cdot 10^{-7}$  \\ 
\hline
\end{tabular}}

\noindent\makebox[\textwidth]{
\begin{tabular}{r| c c c c }
\hline
{$\ptttbar{}$ [GeV]}  & \multicolumn{1}{|c}{$[0, 40]$} & \multicolumn{1}{|c}{$[40, 170]$} & \multicolumn{1}{|c}{$[170, 340]$} & \multicolumn{1}{|c}{$[340, 1000]$}\\ 
\hline
    $[ 0,     40 ]$  & $\phantom{-}7.70\cdot 10^{-1}$  & $-1.92\cdot 10^{-1}$  & $-3.16\cdot 10^{-2}$  & $-6.19\cdot 10^{-4}$  \\ 
    $[ 40,     170 ]$  & $-1.92\cdot 10^{-1}$  & $\phantom{-}4.89\cdot 10^{-2}$  & $\phantom{-}7.34\cdot 10^{-3}$  & $\phantom{-}1.31\cdot 10^{-4}$  \\ 
    $[ 170,     340 ]$  & $-3.16\cdot 10^{-2}$  & $\phantom{-}7.34\cdot 10^{-3}$  & $\phantom{-}1.68\cdot 10^{-3}$  & $\phantom{-}3.82\cdot 10^{-5}$  \\ 
    $[ 340,     1000 ]$  & $-6.19\cdot 10^{-4}$  & $\phantom{-}1.31\cdot 10^{-4}$  & $\phantom{-}3.82\cdot 10^{-5}$  & $\phantom{-}1.78\cdot 10^{-6}$  \\ 
\hline
\end{tabular}}

\noindent\makebox[\textwidth]{
\begin{tabular}{r| c c c }
\hline
{$\absyttbar{}$ }  & \multicolumn{1}{|c}{0.0--0.5} & \multicolumn{1}{|c}{0.5--1.0} & \multicolumn{1}{|c}{1.0--2.5}\\ 
\hline
    0.0--0.5   & $\phantom{-}6.35\cdot 10^{-4}$  & $\phantom{-}1.72\cdot 10^{-4}$  & $-2.69\cdot 10^{-4}$  \\ 
    0.5--1.0   & $\phantom{-}1.72\cdot 10^{-4}$  & $\phantom{-}9.56\cdot 10^{-5}$  & $-8.90\cdot 10^{-5}$  \\ 
    1.0--2.5   & $-2.69\cdot 10^{-4}$  & $-8.90\cdot 10^{-5}$  & $\phantom{-}1.19\cdot 10^{-4}$  \\ 
\hline
\end{tabular}}

\caption{Bin-wise full covariance matrices for the normalized
  differential cross-sections. From top to bottom: top-quark $\pt$; and mass, transverse momentum and absolute value of the rapidity of the $\ttbar$ system. The elements of the covariance matrices are in units of $10^{-6} \, {\rm GeV}^{-2}$ for all the spectra except for $\absyttbar$.}
\label{tab:Cov}
\end{center}
\end{table*}

\section{Interpretation}
\label{sec:Interpretation}

The level of agreement between the measured distributions, simulations with different MC generators and 
theoretical predictions was quantified 
by calculating $\chi^2$ values, employing the full covariance matrices, evaluated as
described in Sec.~\ref{sec:Results}, and 
inferring $p$-values (probabilities that the $\chi^2$ is larger than or equal to the observed value) 
from the $\chi^2$ and the number of degrees of freedom (NDF).
The normalization constraint used to derive the normalized differential cross-sections
lowers by one unit the NDF and the rank of the $N_{\rm b} \times N_{\rm b}$ covariance matrix, where $N_{\rm b}$
is the number of bins of the spectrum under consideration. In order to evaluate the 
$\chi^2$ the following relation was used: 
\begin{equation}
\chi^2 = V_{N_{\rm b}-1}^{\rm T} \cdot {\rm Cov}_{N_{\rm b}-1}^{-1} \cdot V_{N_{\rm b}-1}
\end{equation} where $V_{N_{\rm b}-1}$ is the vector of differences between data and predictions obtained
discarding one of the $N_{\rm b}$ elements and ${\rm Cov}_{N_{\rm b}-1}$ is the $(N_{\rm b}-1) \times (N_{\rm b}-1)$ sub-matrix
derived from the full covariance matrix discarding the corresponding row and column. 
The sub-matrix obtained in this way is invertible and allows the $\chi^2$ to be computed. 
The $\chi^2$ value does not depend on the choice of the element discarded for the vector $V_{N_{\rm b}-1}$ and
the corresponding sub-matrix ${\rm Cov}_{N_{\rm b}-1}$. 

\begin{figure*}[!ht]
\centering
\subfigure[]{ \includegraphics[width=0.45\textwidth]{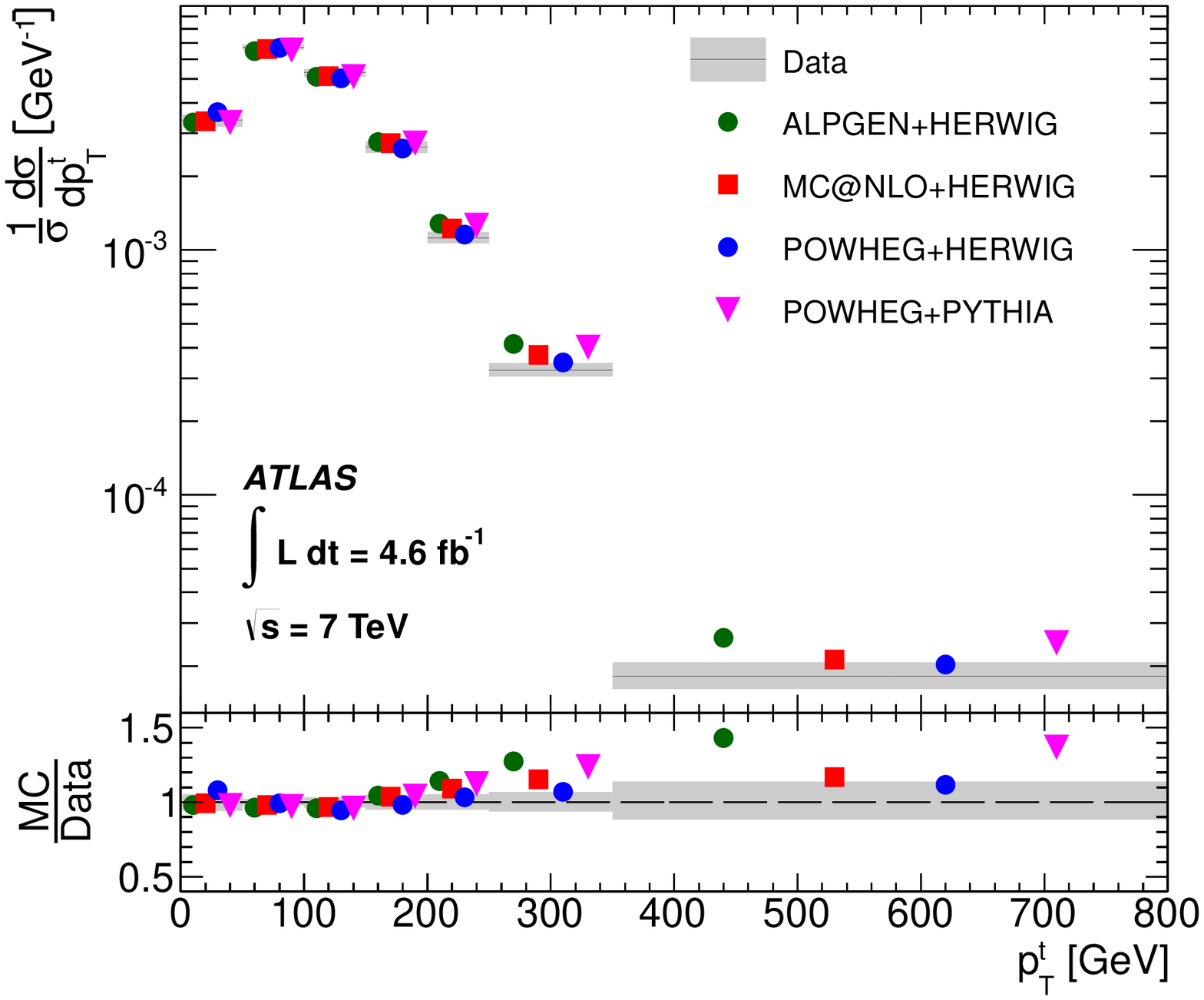}\label{unf_top2_MC}}
\subfigure[]{ \includegraphics[width=0.45\textwidth]{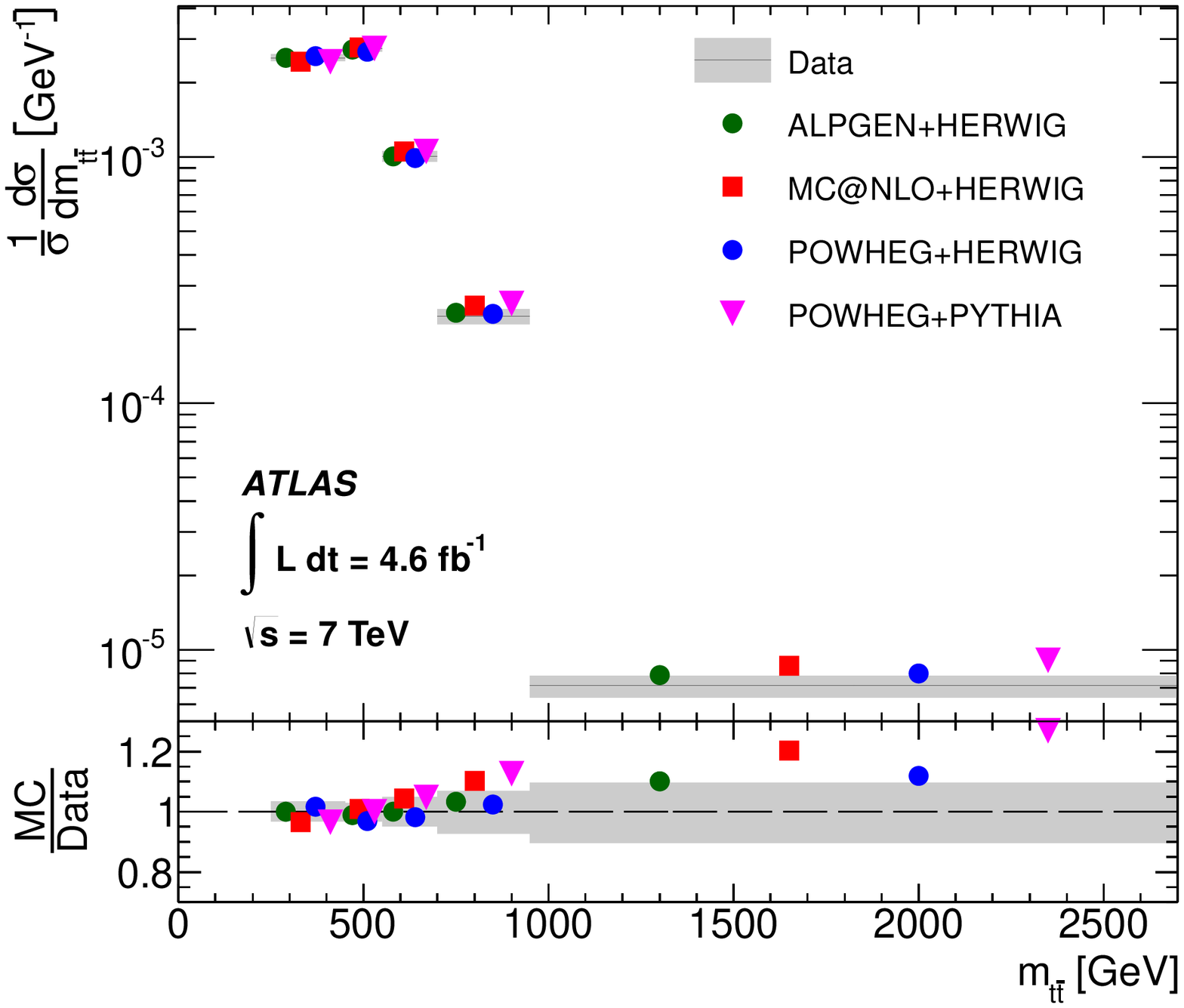}\label{unf_mass_MC}}
\subfigure[]{ \includegraphics[width=0.45\textwidth]{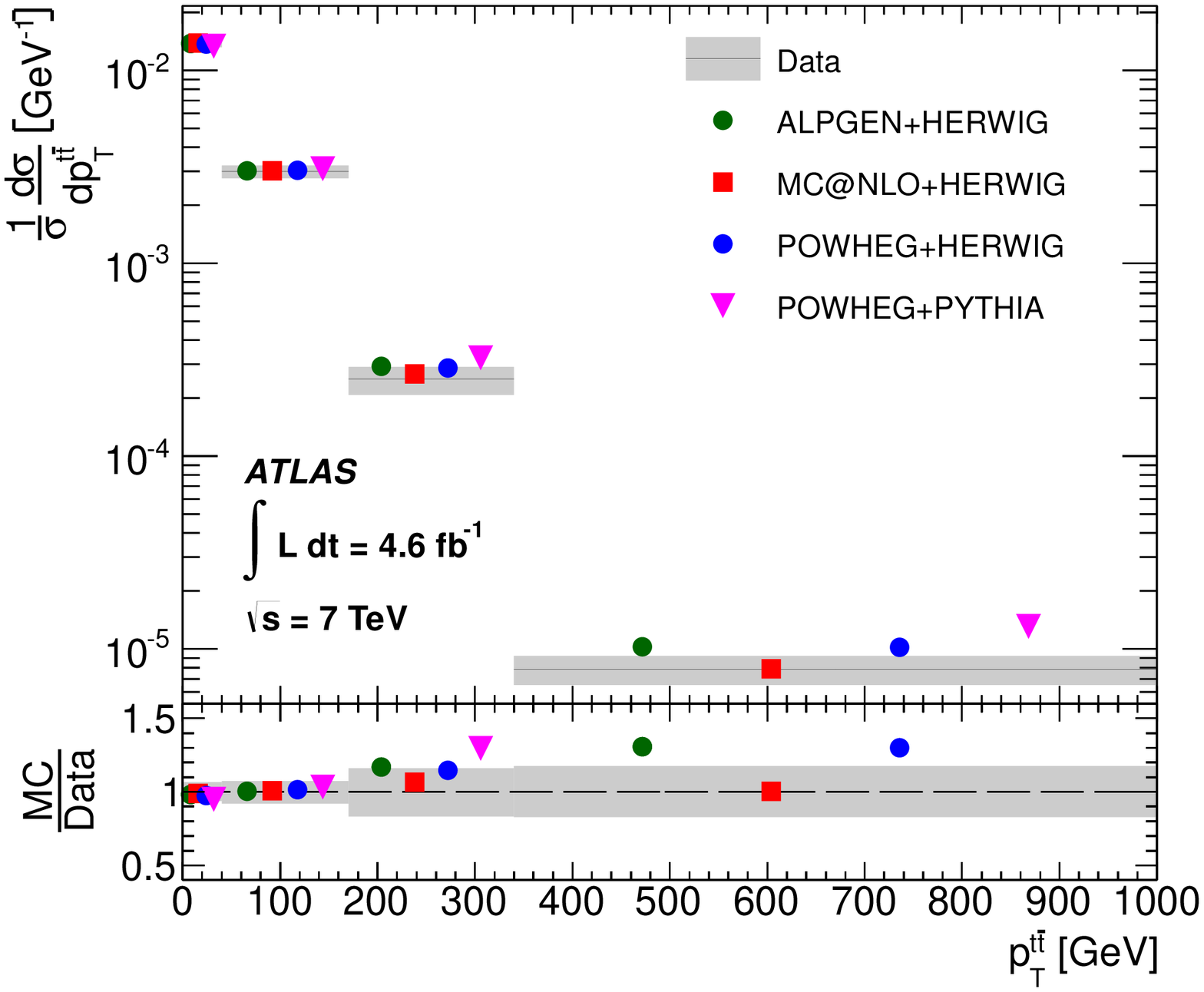}\label{unf_pt_MC}}
\subfigure[]{ \includegraphics[width=0.45\textwidth]{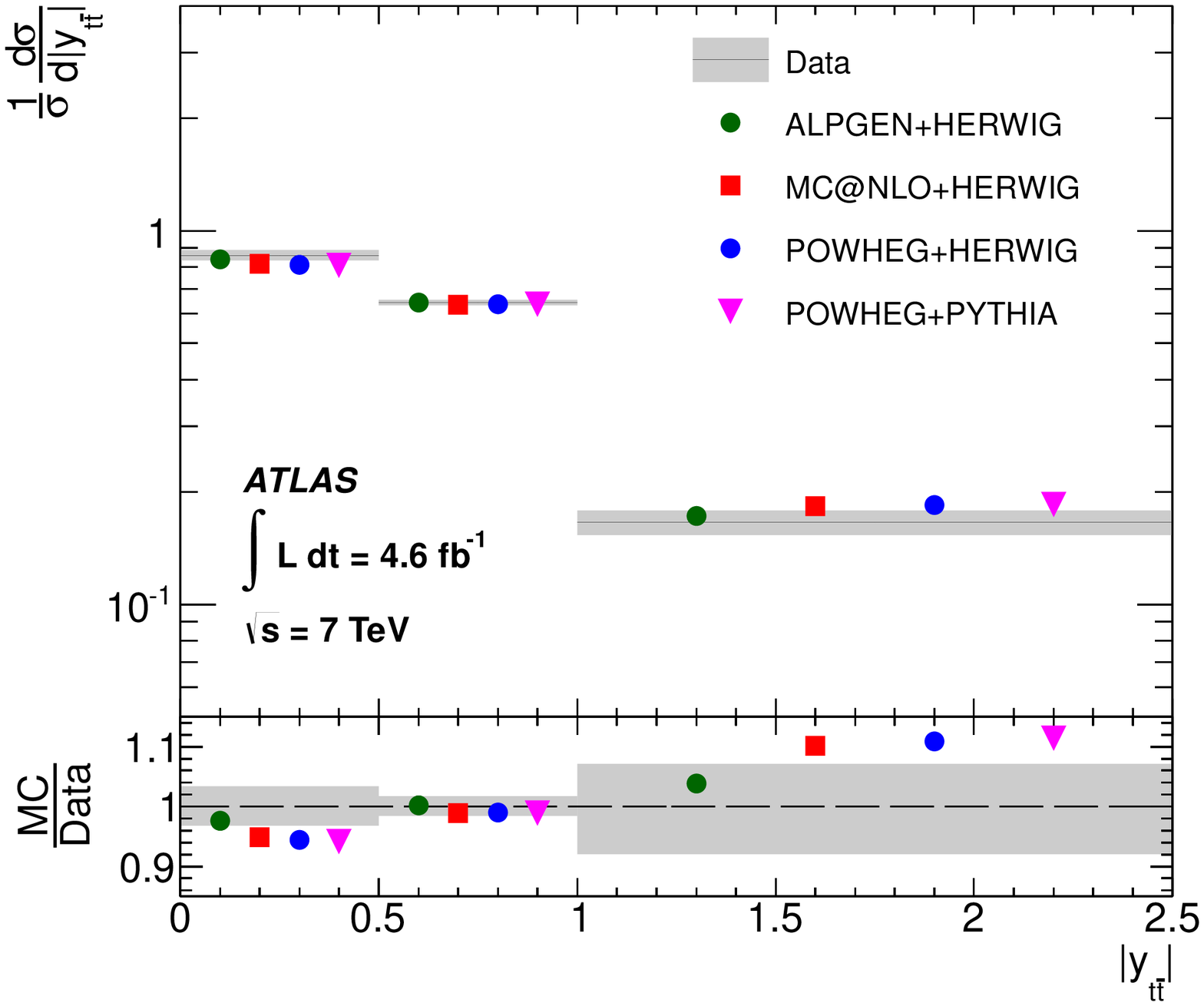}\label{unf_rap_MC}}
\caption{(Color online) Normalized differential cross-sections for the \subref{unf_top2_MC}~transverse momentum of the hadronically decaying top quark  ($\ptt$), and the \subref{unf_mass_MC} mass ($\mttbar$), \subref{unf_pt_MC}~transverse momentum ($\ptttbar$) and the \subref{unf_rap_MC}~absolute value of the rapidity ($\absyttbar$) of the \ttbar{} system. Generator predictions are shown as markers for \Alpgen{}+\Herwig{} (circles), \McAtNlo{}+\Herwig{} (squares), \Powheg{}+\Herwig{} (triangles) and \Powheg{}+\Pythia{} (inverted triangles). The markers are offset within each bin to allow for better visibility. 
The gray bands indicate the total uncertainty on the data in each bin. The lower part of each figure shows the ratio of the generator predictions to data. For \ptttbar{} the \Powheg{}+\Pythia{} marker cannot be seen in the last bin of the ratio plot because it falls beyond the axis range.
The cross-section in each bin is given as the integral of the differential cross-section over the bin width, divided by the bin width. 
The calculation of the cross-sections in the last bins includes events falling outside of the bin edges, and the normalization is done within the quoted bin width. 
The bin ranges along the horizontal axis (and not the position of the markers) can be associated with the normalized differential cross-section values along the vertical axis.
}
\label{fig:combined_results_with_MC}
\end{figure*}

The predictions from MC generators do not include theoretical uncertainties and were evaluated using a~specific set of tuned parameters.
The $p$-values comparing the measured spectra to the predictions of MC generators shown in Fig.~\ref{fig:combined_results_with_MC} are listed in Table~\ref{tab:pvalues_combined_norm}. No single generator performs best for all the kinematic variables; however, the difference in $\chi^2$ between generators demonstrates that the data have sufficient precision to probe the predictions.  For \ptt{} the agreement with \Alpgen{}+\Herwig{} and \Powheg{}+\Pythia{} is particularly bad due to a~significant discrepancy in the tail of the distribution. \McAtNlo{}+\Herwig{} and \Powheg{}+\Herwig{} predict shapes closer to the measured distribution. As can be seen in Fig.~\ref{fig:combined_results_with_MC}, there is a~general trend of data being softer in \ptt{} above $200\,$GeV compared to all generators. The shape of the \mttbar{} distribution is best described by \Alpgen{}+\Herwig{} and \Powheg{}+\Herwig{}. The \ptttbar{} shape is described best by \McAtNlo{}+\Herwig{} and particularly badly by \Powheg{}+\Pythia{} while the \yttbar{} shape is described best by \Alpgen{}+\Herwig{}.

\begin{table*} [p]
\footnotesize
\centering
\noindent\makebox[\textwidth]{
\begin{tabular}{ c | r@{/}l c | r@{/}l c | r@{/}l c | r@{/}l c  | r@{/}l c | r@{/}l c }
\hline
Variable  & \multicolumn{3}{|c|}{\Alpgen{}+\Herwig{}} & \multicolumn{3}{|c|}{MC@NLO+\Herwig{}} & \multicolumn{3}{|c|}{\Powheg{}+\Herwig{}} & \multicolumn{3}{|c|}{\Powheg{}+\Pythia{}} & \multicolumn{3}{|c|}{NLO QCD} & \multicolumn{3}{|c}{NLO+NNLL} \\ 
                   & \multicolumn{2}{|c}{$\chi^{2}$/NDF} &  ~$p$-value &  \multicolumn{2}{|c}{$\chi^{2}$/NDF} &  ~$p$-value &  \multicolumn{2}{|c}{$\chi^{2}$/NDF} &  ~$p$-value & \multicolumn{2}{|c}{$\chi^{2}$/NDF} &  ~$p$-value & \multicolumn{2}{|c}{$\chi^{2}$/NDF} &  ~$p$-value & \multicolumn{2}{|c}{$\chi^{2}$/NDF} &  ~$p$-value   \\ 
\hline
\hline
        $\ptt{}$ & {\ }    24.0 & 6 & 0.00 & {\ } 8.0 & 6 & 0.24 & {\ } 4.8 & 6 & 0.57 & {\ } 18.9 & 6 & 0.00 & {\ } 9.5 & 6 & 0.15 & {\ } 7.6 & 6 & 0.27 \\
        $\mttbar{}$ & {\ }   2.6 & 4 & 0.63 & {\ } 6.9 & 4 & 0.14 & {\ } 5.5 & 4 & 0.24 & {\ } 12.9 & 4 & 0.01 & {\ } 5.5 & 4 & 0.24 & {\ } 5.9 & 4 & 0.20  \\
        $\ptttbar{}$ & {\ }  4.2 & 3 & 0.25 & {\ } 0.5 & 3 & 0.93 & {\ } 3.5 & 3 & 0.32 & {\ } 17.8 & 3 & 0.00 & {\ } 14.4 & 3 & 0.00 & {\ } 8.6 & 3 & 0.02 \\
        $\absyttbar{}$ & {\ }  1.6 & 2 & 0.45 & {\ } 3.4 & 2 & 0.18 & {\ } 4.3 & 2 & 0.11 & {\ } 4.8 & 2 & 0.09 & {\ } 3.7 & 2 & 0.16   \\
\hline
\end{tabular}}

\caption{Comparison between the measured normalized differential cross-sections and the predictions from several MC generators and theoretical calculations. For each variable and prediction a $\chi^2$ and a $p$-value are calculated using the covariance matrix of each measured spectrum. 
The number of degrees of freedom (NDF) is equal to $N_{\rm b}-1$ where $N_{\rm b}$ is the number of bins in the distribution.
In the last column  \ptt{}, \mttbar{} and \ptttbar{} are compared to NLO+NNLL predictions \cite{NNLO_calc} and \cite{nnloMtt,PhysRevLett.110.082001,PhysRevD.88.074004}.}
\label{tab:pvalues_combined_norm}
\end{table*}

\begin{figure*}[!ht]
\centering
\subfigure[]{ \includegraphics[width=0.45\textwidth]{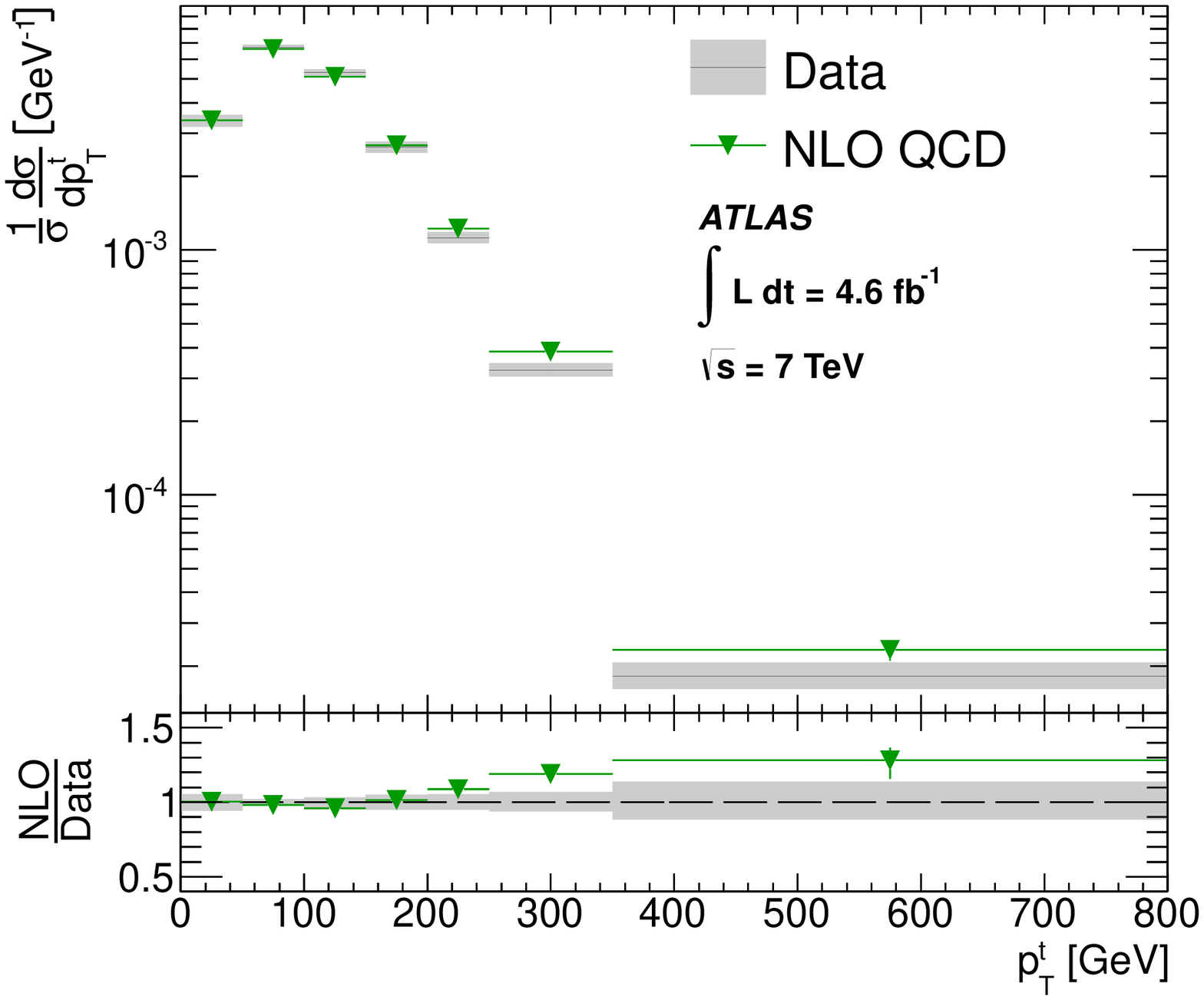}\label{unf_top2_NLO}}
\subfigure[]{ \includegraphics[width=0.45\textwidth]{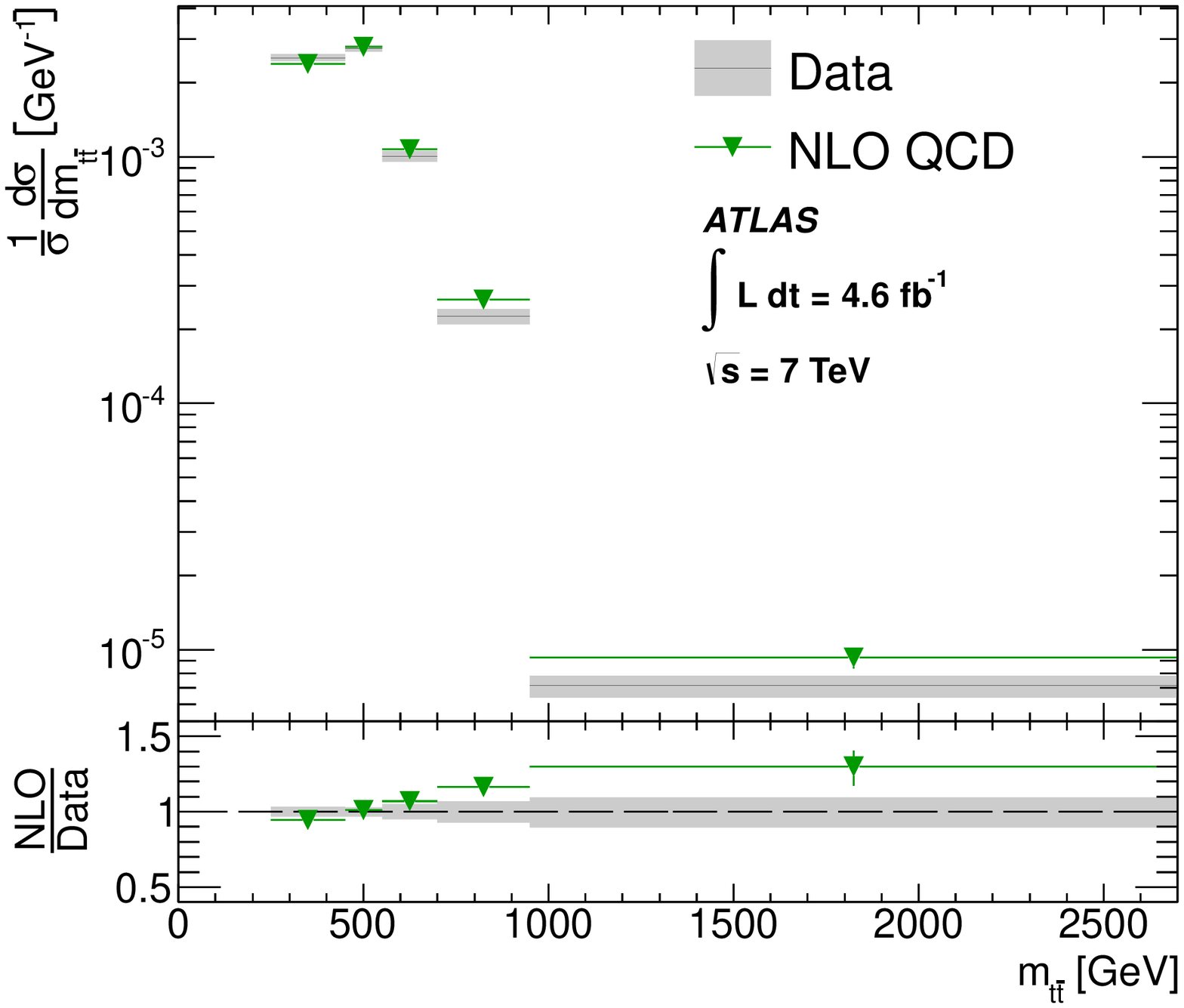}\label{unf_mass_NLO}}
\subfigure[]{ \includegraphics[width=0.45\textwidth]{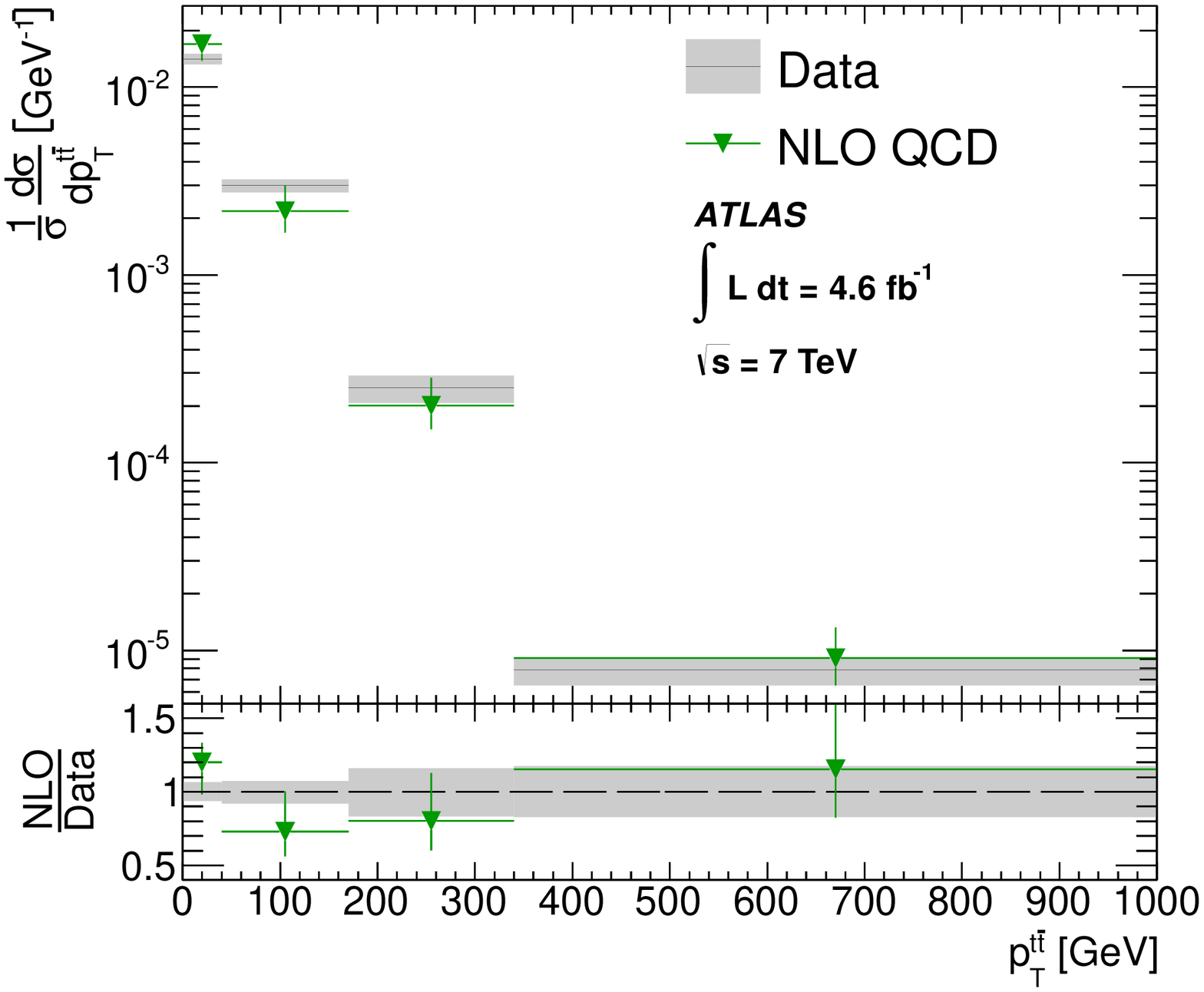}\label{unf_pt_NLO}}
\subfigure[]{ \includegraphics[width=0.45\textwidth]{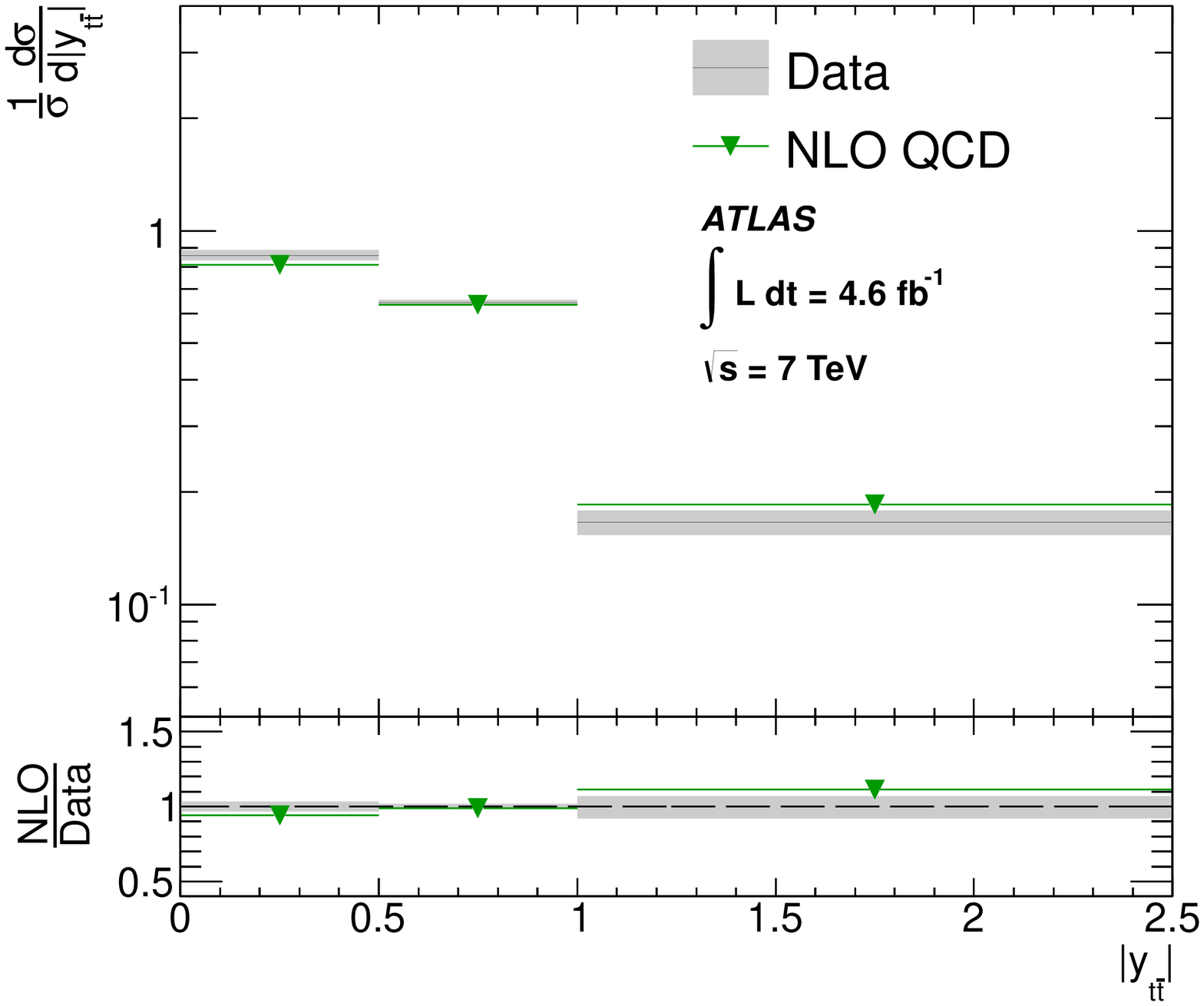}\label{unf_rap_NLO}}
\caption{(Color online) Normalized differential cross-sections for the \subref{unf_top2_NLO}~transverse momentum of the hadronically decaying top-quark ($\ptt$), and the \subref{unf_mass_NLO} mass ($\mttbar$), \subref{unf_pt_NLO}~transverse momentum ($\ptttbar$) and the \subref{unf_rap_NLO}~absolute value of the rapidity ($\absyttbar$) of the \ttbar{} system. The distributions are compared to NLO QCD predictions (based on \MCFM~\cite{MCFM} with the CT10 PDF). The bin ranges along the horizontal axis (and not the position of the markers) can be associated with the normalized differential cross-section values along the vertical axis. The error bars correspond to the PDF and fixed scale uncertainties in the theoretical prediction. The gray bands indicate the total uncertainty on the data in each bin. The lower part of each figure shows the ratio of the NLO QCD predictions to data.
The cross-section in each bin is given as the integral of the differential cross-section over the bin width, divided by the bin width. 
The calculation of the cross-sections in the last bins includes events falling outside of the bin edges, and the normalization is done within the quoted bin width. }
\label{fig:combined_results_with_NLO}
\end{figure*}

\begin{figure*}[!ht]
\centering
\subfigure[]{ \includegraphics[width=0.45\textwidth]{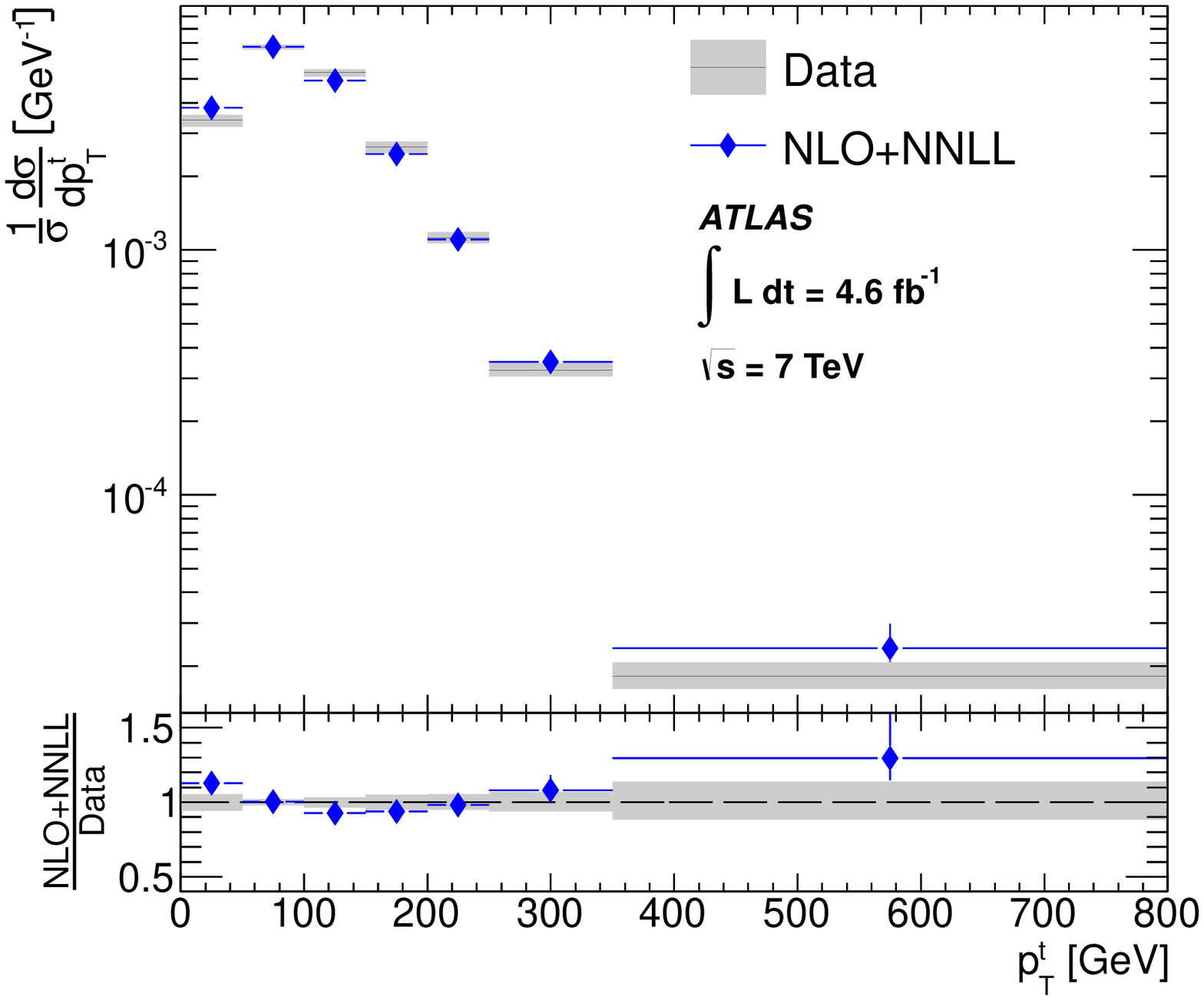}\label{unf_top2_NNLL}}
\subfigure[]{ \includegraphics[width=0.45\textwidth]{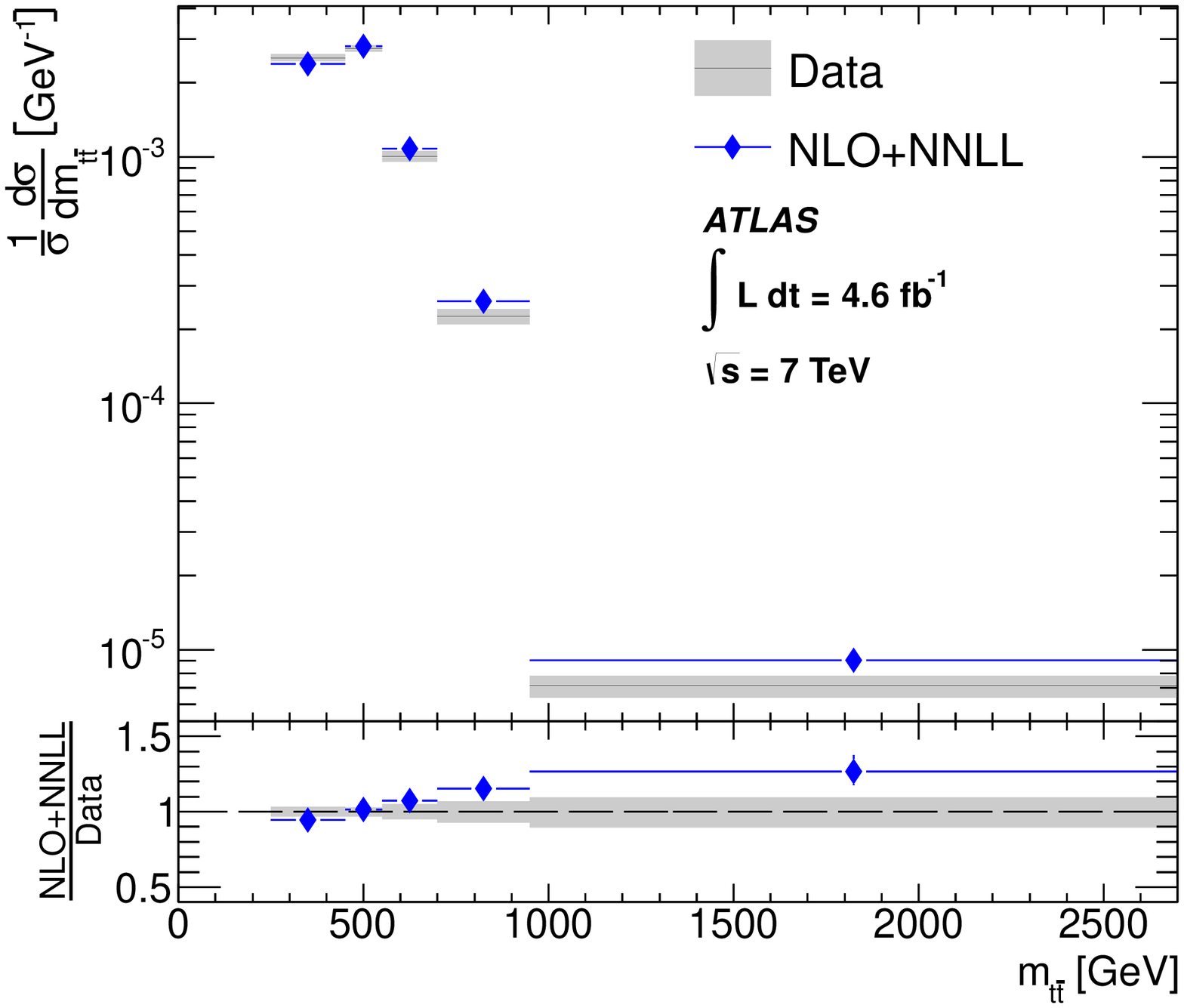}\label{unf_mass_NNLL}}
\subfigure[]{ \includegraphics[width=0.45\textwidth]{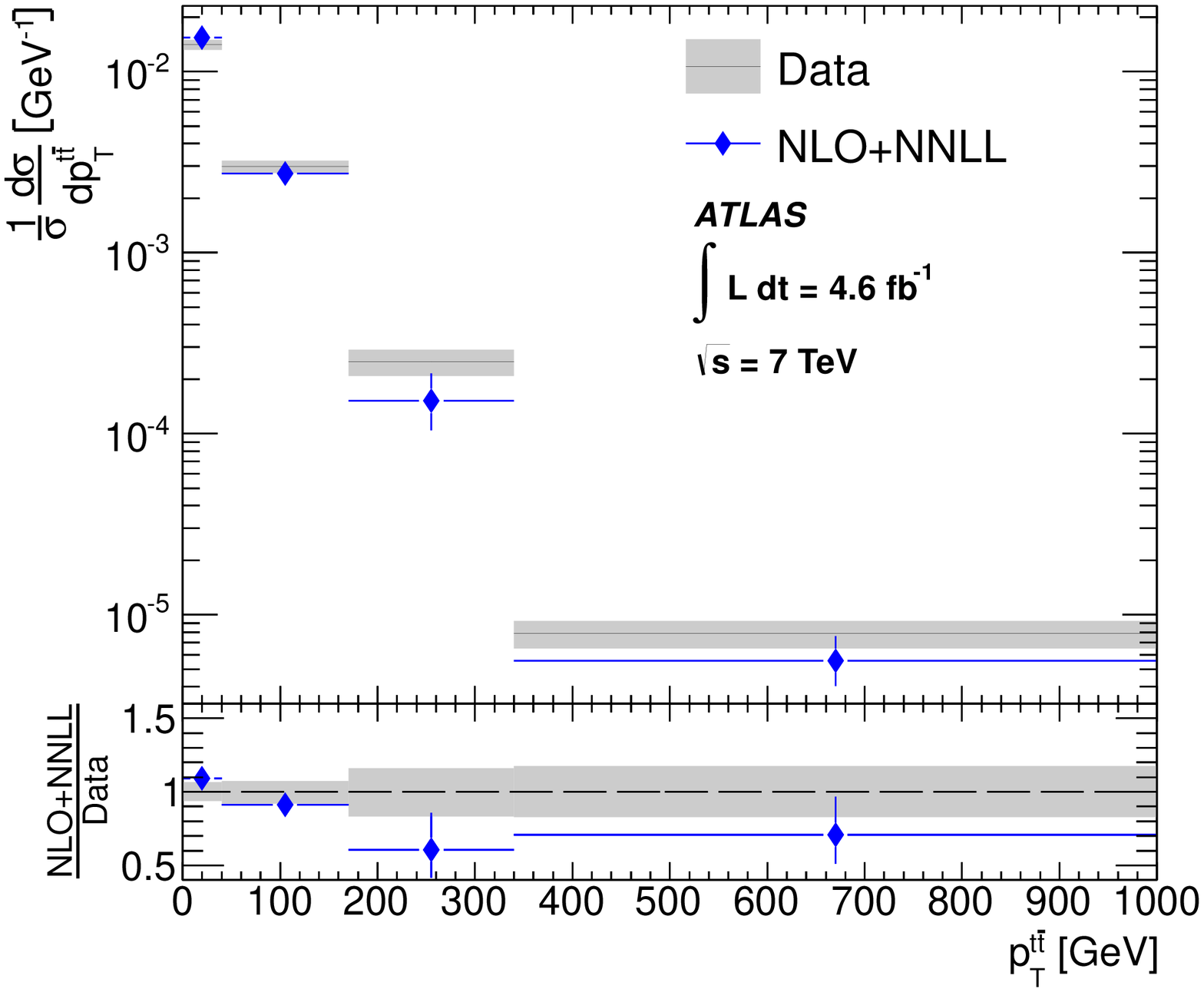}\label{unf_pTtt_NNLL}}
\caption{(Color online) Normalized differential cross-sections for the \subref{unf_top2_NNLL} transverse momentum of the hadronically decaying top-quark ($\ptt$), the \subref{unf_mass_NNLL}~mass of the $\ttbar$ system ($\mttbar$), and the \subref{unf_pTtt_NNLL}~transverse momentum of the $\ttbar$ system ($\ptttbar$) .
The distributions are compared to the predictions from 
NLO+NNLL calculations for $\ptt$~\cite{NNLO_calc}, $\mttbar$~\cite{nnloMtt} and $\ptttbar$~\cite{PhysRevLett.110.082001,PhysRevD.88.074004},
all using the MSTW2008NNLO PDF. The bin ranges along the horizontal axis (and not the position of the markers) can be associated with the normalized differential cross-section values along the vertical axis. The error bars correspond to the fixed (and dynamic in the case of \ptt{}) scale uncertainties in the theoretical prediction. The gray bands indicate the total uncertainty on the data in each bin. The lower part of each figure shows the ratio of the NLO+NNLL calculations to data.
The cross-section in each bin is given as the integral of the differential cross-section over the bin width, divided by the bin width. 
The calculation of the cross-sections in the last bins includes events falling outside of the bin edges, and the normalization is done within the quoted bin width. }
\label{fig:combined_results_with_NNLL}
\end{figure*}

The distributions are also shown compared to QCD calculations at NLO (based on \MCFM~\cite{MCFM} version 6.5 with the CT10 PDF) in 
Fig.~\ref{fig:combined_results_with_NLO} and to NLO+NNLL calculations for \ptt{}~\cite{NNLO_calc}, \mttbar{}~\cite{nnloMtt} and \ptttbar{}~\cite{PhysRevLett.110.082001,PhysRevD.88.074004}, all using the MSTW2008NNLO~\cite{MSTW} PDF,
in Fig.~\ref{fig:combined_results_with_NNLL}. The $p$-values for these comparisons are shown in 
Table~\ref{tab:pvalues_combined_norm}. 

The uncertainties in the NLO predictions due to the parton 
distribution functions were evaluated at the 68$\%$ confidence level (CL) using the CT10 PDF error-sets.
Another source of uncertainty considered is the one related to the factorization and renormalization scales. 
The nominal value was assumed to be $\mu = \mt$ for both scales, and is varied simultaneously 
up and down from $2 m_t$ to $m_t/2$. 
The full covariance matrix, including the bin-wise correlations induced by the uncertainties in the scale and in
the different PDF components, was used for the $\chi^2$ evaluation.

For the NLO+NNLL predictions of \mttbar{} and \ptttbar{} spectra, the calculation is performed using the mass of the \ttbar{} system as the dynamic scale of the process. The uncertainties come from doubling and halving this scale and
from the PDF uncertainty evaluated at the 68$\%$ CL using the MSTW2008NNLO PDF error-sets.
For the NLO+NNLL prediction of the \ptt{} spectrum, besides the fixed scale uncertainty, the contribution of the 
alternative dynamic scale $\mu = \sqrt{m_t^2 + {\ptt}^2}$ is also included; in this case the PDF uncertainty is not provided. 
For both the above theoretical calculations the bin-wise correlations were taken into account in evaluating the $\chi^2$s and $p$-values, which are shown in Table~\ref{tab:pvalues_combined_norm}.

The data are softer than both the NLO and NLO+NNLL QCD calculations in the 
tail of the \ptt{} distribution. 
The measured \mttbar{} spectrum also falls more quickly than either the 
NLO or NLO+NNLL predictions. The \ptttbar{} spectrum agrees poorly with both the NLO and NLO+NNLL predictions.
No electroweak corrections are included in these predictions, and these were 
shown in Ref.~\cite{Manohar:2012rs,Bernreuther:2008aw,epj.c51.37,kuhn:ttp13.015} to have non-negligible effects in the \ptt{} and \mttbar{} distributions. 

\begin{figure*}[!ht]
\centering
\subfigure[]{ \includegraphics[width=0.45\textwidth]{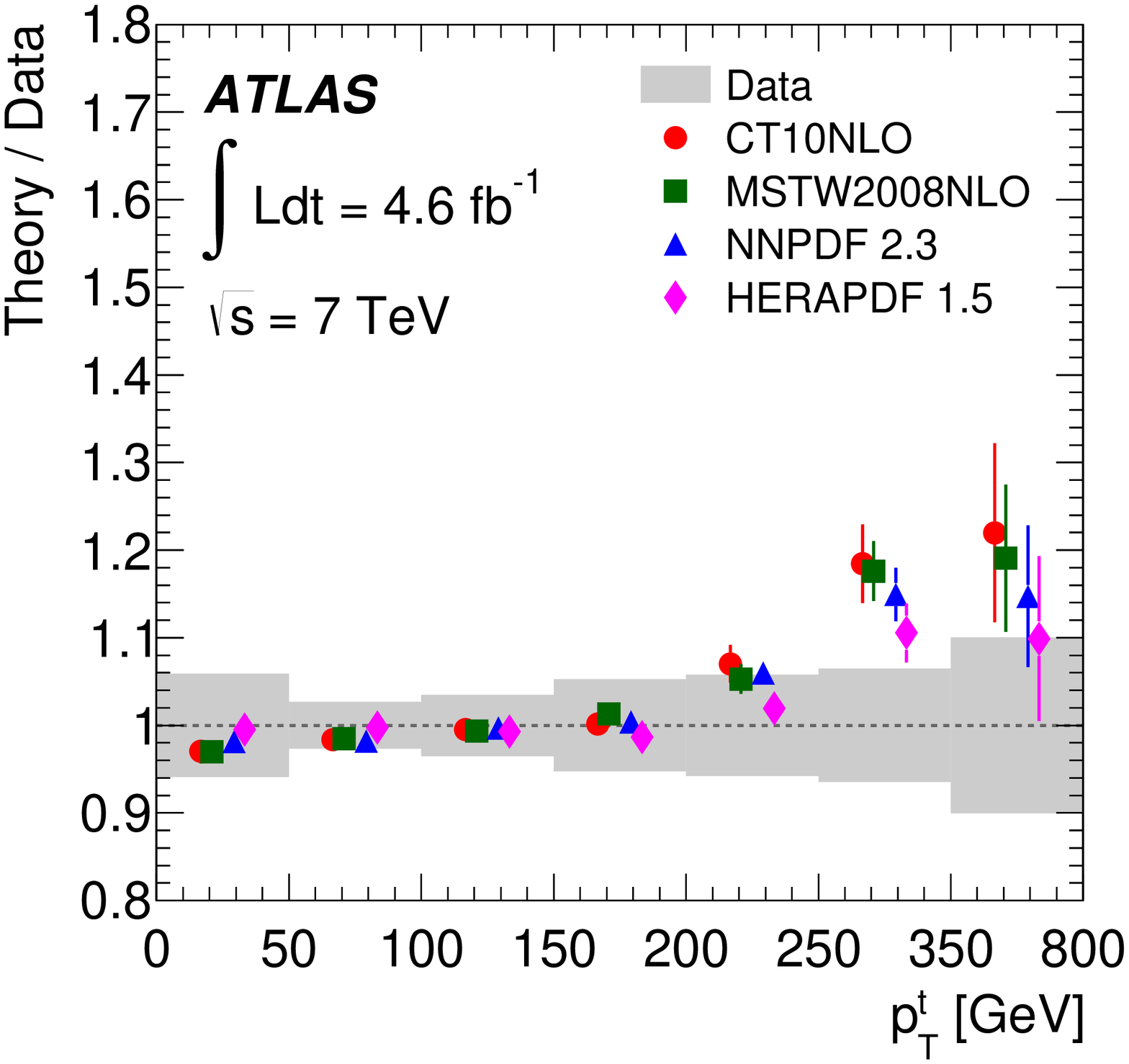} \label{pdf_pt}}
\subfigure[]{ \includegraphics[width=0.45\textwidth]{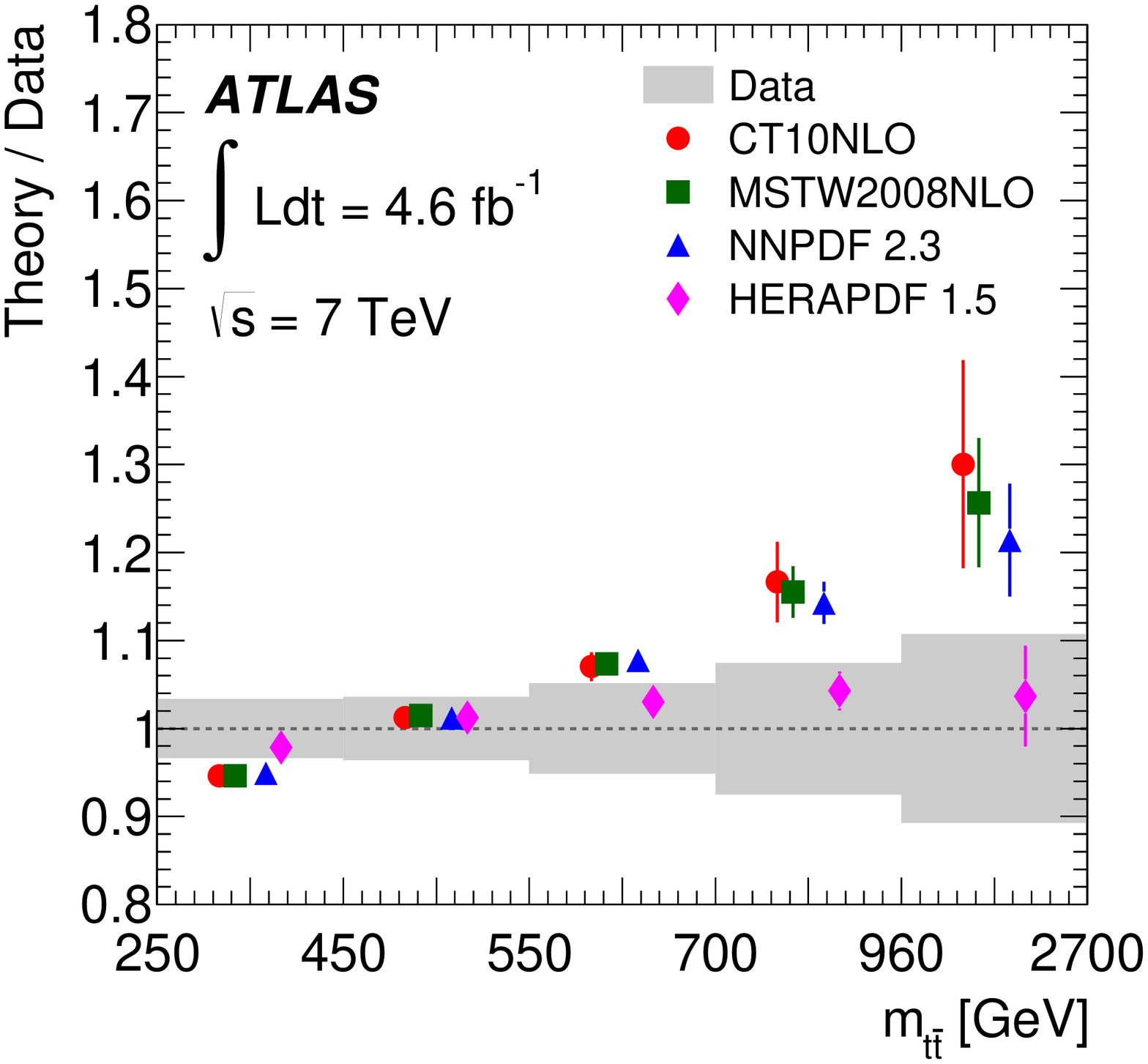}  \label{pdf_mass} }
\subfigure[]{ \includegraphics[width=0.45\textwidth]{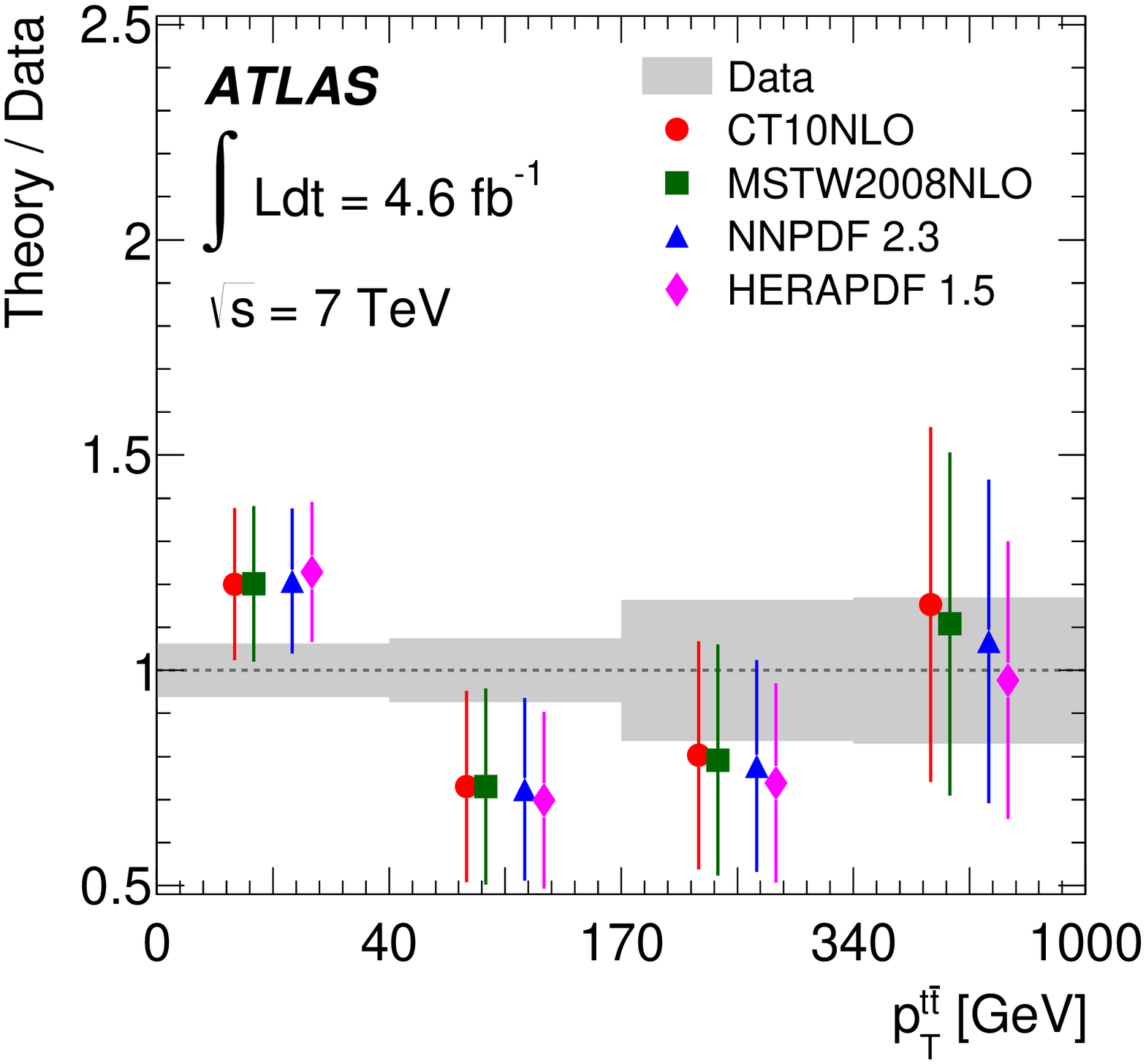}  \label{pdf_pttt}}
\subfigure[]{ \includegraphics[width=0.45\textwidth]{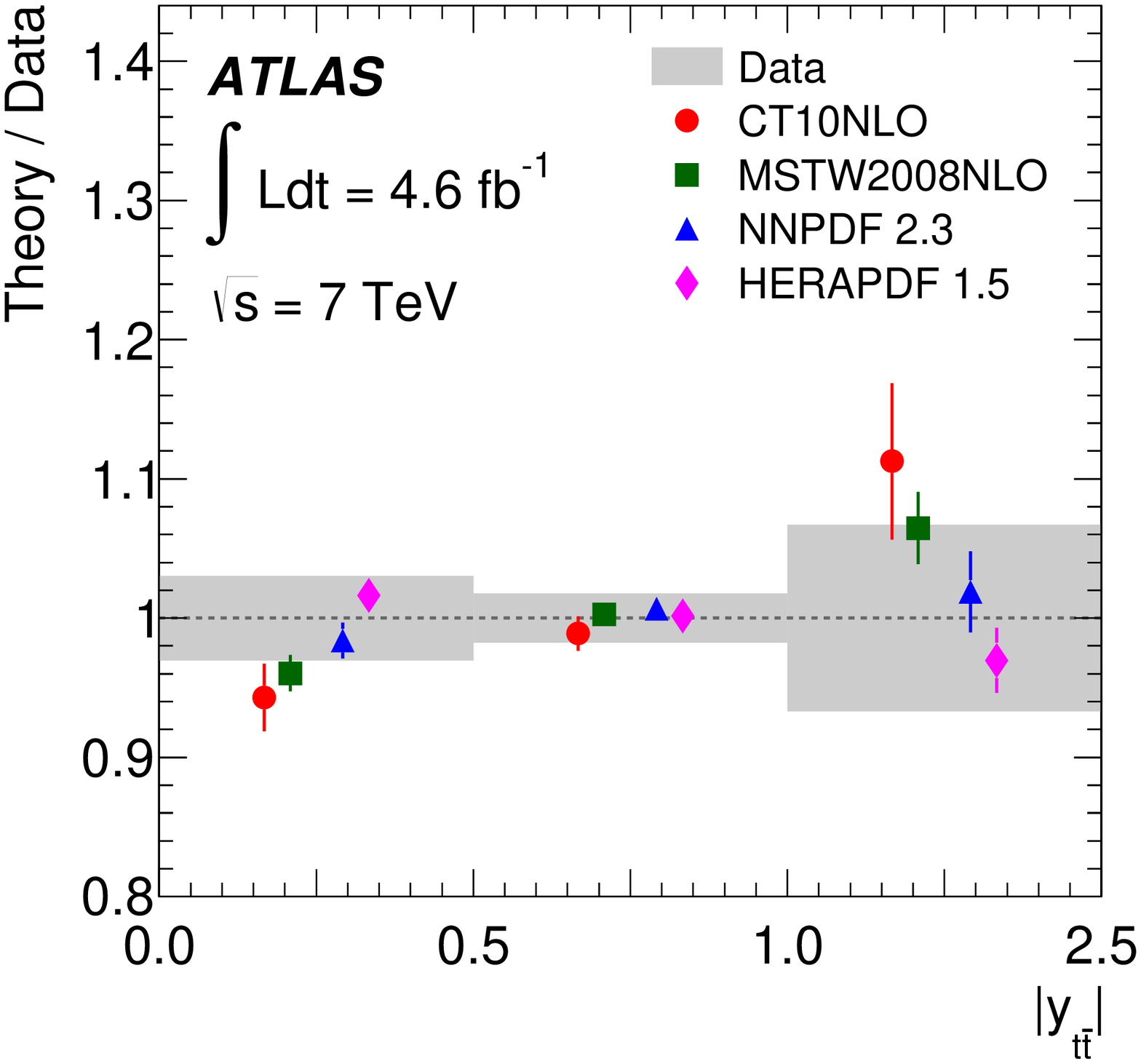}  \label{pdf_aytt}}
\caption{(Color online) Ratios of the NLO QCD predictions~\cite{MCFM} to the measured normalized differential cross-sections for different PDF sets (CT10~\cite{CT10}, MSTW2008NLO~\cite{MSTW}, NNPDF2.3~\cite{NNPDF} and HERAPDF1.5~\cite{HERA}) (markers) for the \subref{pdf_pt}~transverse momentum of the hadronically decaying top-quark ($\ptt$), and the \subref{pdf_mass} mass ($\mttbar$), the \subref{pdf_pttt} transverse momentum ($\ptttbar$), and the \subref{pdf_aytt}~absolute value of the rapidity ($\absyttbar$) of the $\ttbar$ system. The markers are offset in each bin and the bins are of equal size to allow for better visibility. The gray bands indicate the total uncertainty on the data in each bin, while the error bars denote the uncertainties in the predictions, which include the internal PDF set variations and also fixed scale uncertainties.}
\label{fig:pdf_ratios}
\end{figure*}

The predictions of various NLO PDF sets are evaluated using \MCFM, interfaced to four different PDF sets: CT10~\cite{CT10}, 
MSTW2008NLO~\cite{MSTW}, NNPDF2.3~\cite{NNPDF} and HERAPDF1.5~\cite{HERA}. The uncertainties in the predictions 
include the PDF uncertainties~\footnote{For HERAPDF1.5, only the 21 member PDFs accounting for experimental uncertainties are taken into account.} and the fixed scale uncertainties already described. 
The comparisons between data and the 
different predictions are presented in Fig.~\ref{fig:pdf_ratios} for the normalized differential cross-sections and 
the $p$-values for these comparisons are shown in Table~\ref{tab:PvalsPDF}.
The significant changes in $\chi^2$ between the different PDF sets for the \ptt{}, \mttbar{} and \yttbar{} distributions indicate that the data can be used to improve the precision of future PDF fits.

\begin{table*} [p]
\footnotesize
\centering

\begin{tabular}{ c |r@{/}l c  | r@{/}l c | r@{/}l c | r@{/}l c }
\hline
Variable & \multicolumn{3}{|c|}{CT10} & \multicolumn{3}{|c|}{MSTW2008NLO} & \multicolumn{3}{|c|}{NNPDF 2.3} & \multicolumn{3}{|c}{HERAPDF 1.5} \\
         &  \multicolumn{2}{|c}{$\chi^{2}$/NDF} &  ~$p$-value &  \multicolumn{2}{|c}{$\chi^{2}$/NDF} &  ~$p$-value & \multicolumn{2}{|c}{$\chi^{2}$/NDF} &  ~$p$-value & \multicolumn{2}{|c}{$\chi^{2}$/NDF} &  ~$p$-value    \\
\hline
\hline
       \ptt{}  &    9.5 & 6 & 0.15 & {\ } 9.8 & 6 & 0.14 & {\ } 8.2 & 6 & 0.22 & {\ } 5.5 & 6 & 0.49 \\
       \mttbar{} &     	5.5 & 4 & 0.24 & {\ } 6.0 & 4 & 0.20 & {\ } 5.2 & 4 & 0.27 & {\ } 0.63 & 4 & 0.96  \\
       \ptttbar{} & 14.4 & 3 & 0.00 & {\ } 13.0 & 3 & 0.01 & {\ } 12.4 & 3 & 0.01 & {\ } 9.1 & 3 & 0.03   \\	
       \absyttbar{} & 3.7 & 2 & 0.16 & {\ } 4.0 & 2 & 0.13 & {\ } 1.3 & 2 & 0.52 & {\ } 0.44 & 2 & 0.80  \\	
\hline
\end{tabular}

\caption{ Comparison between the measured normalized differential cross-sections and the NLO predictions (MCFM) for different parton distribution functions. 
For each kinematic variable and each parton distribution function, a $\chi^{2}$ and a $p$-value are calculated using the covariance matrix of each measured spectrum as well as the theory PDF and scale covariance matrix. 
The number of degrees of freedom (NDF) is equal to $N_{\rm b}-1$ where $N_{\rm b}$ is the number of bins in the distribution.
}
\label{tab:PvalsPDF}
\end{table*}
 
As can be seen in Fig.~\ref{fig:pdf_ratios}, a~certain tension between data and all predictions is observed in the case of the top-quark $\pt$ distribution at high $\pt$ values. For the \mttbar{} distribution, the agreement with HERAPDF1.5 is better than that with the other PDF predictions. For the \ptttbar{} distribution, one should note that \MCFM{} is effectively only a~leading-order calculation and resummation effects are expected to play an important role at low \ptttbar{}. Finally, for the \absyttbar{} distribution, the NNPDF2.3 and especially HERAPDF1.5 sets are in better agreement with the data. 


\clearpage
\section{Conclusion}
\label{sec:Conclusion}
 
Kinematic distributions of the top quarks in \ttbar{} events, selected in the \ljets channel, were measured using data from $7\,{\rm TeV}$ proton--proton collisions collected by the ATLAS detector at the CERN Large Hadron Collider. This dataset corresponds to an integrated luminosity of \lumitot. Normalized differential cross-sections have been measured as a~function of the top-quark transverse momentum and as a~function of the mass, transverse momentum, and rapidity of the \ttbar{} system. These results agree with the previous ATLAS measurements and supersede them with a~larger dataset, smaller uncertainties, and an additional variable. 

In general the Monte Carlo predictions and the QCD calculations agree with data in a~wide kinematic region. However, data are softer than all predictions in the tail of the \ptt{} spectrum, particularly in the case of the \Alpgen{}+\Herwig{} and \Powheg{}+\Pythia{} generators. 
The same trend is observed for the NLO+NNLL predictions of the \mttbar{} and \ptt{}
spectra which tend to be above the data in the tail of the
distributions. Nevertheless the overall agreement is still found to be
reasonable for these two variables while it is worst for \ptttbar{}.
The distributions show some preference for HERAPDF1.5 when used in conjunction with a~fixed-order NLO QCD calculation. More precise conclusions about PDFs will be possible from the comparison of these measurements to future calculations at NNLO+NNLL in QCD and after including electroweak effects. 

\section*{Acknowledgments}

We thank CERN for the very successful operation of the LHC, as well as the support staff from our institutions without whom ATLAS could not be operated efficiently.

We acknowledge the support of ANPCyT, Argentina; YerPhI, Armenia; ARC,
Australia; BMWF and FWF, Austria; ANAS, Azerbaijan; SSTC, Belarus; CNPq and FAPESP,
Brazil; NSERC, NRC and CFI, Canada; CERN; CONICYT, Chile; CAS, MOST and NSFC,
China; COLCIENCIAS, Colombia; MSMT CR, MPO CR and VSC CR, Czech Republic;
DNRF, DNSRC and Lundbeck Foundation, Denmark; EPLANET, ERC and NSRF, European Union;
IN2P3-CNRS, CEA-DSM/IRFU, France; GNSF, Georgia; BMBF, DFG, HGF, MPG and AvH
Foundation, Germany; GSRT and NSRF, Greece; ISF, MINERVA, GIF, DIP and Benoziyo Center,
Israel; INFN, Italy; MEXT and JSPS, Japan; CNRST, Morocco; FOM and NWO,
Netherlands; BRF and RCN, Norway; MNiSW, Poland; GRICES and FCT, Portugal; MERYS
(MECTS), Romania; MES of Russia and ROSATOM, Russian Federation; JINR; MSTD,
Serbia; MSSR, Slovakia; ARRS and MIZ\v{S}, Slovenia; DST/NRF, South Africa;
MICINN, Spain; SRC and Wallenberg Foundation, Sweden; SER, SNSF and Cantons of
Bern and Geneva, Switzerland; NSC, Taiwan; TAEK, Turkey; STFC, the Royal
Society and Leverhulme Trust, United Kingdom; DOE and NSF, United States of
America.

The crucial computing support from all WLCG partners is acknowledged
gratefully, in particular from CERN and the ATLAS Tier-1 facilities at
TRIUMF (Canada), NDGF (Denmark, Norway, Sweden), CC-IN2P3 (France),
KIT/GridKA (Germany), INFN-CNAF (Italy), NL-T1 (Netherlands), PIC (Spain),
ASGC (Taiwan), RAL (UK) and BNL (USA) and in the Tier-2 facilities
worldwide.

\clearpage

\onecolumngrid
\newpage
\appendix

\section{Additional Tables of Systematic Uncertainties}
\label{Sec:Appendix:Syst}

Appendix~\ref{Sec:Appendix:Syst} includes
Tables~\ref{tab:CombSyst_1_shape},~\ref{tab:CombSyst_2_shape},~\ref{tab:CombSyst_3_shape} and \ref{tab:CombSyst_4_shape} 
with the contribution of each individual source of systematic uncertainty calculated as a~percentage of the normalized differential cross-section in each bin for each variable. Those that contribute to the uncertainty on the jet energy scale are denoted (JES) and are described in detail in Ref.~\cite{jes:2013}. 

The muon momentum resolution uncertainties are split into parts specific  to the inner-detector (ID) and muon spectrometer (MS).
The \Wboson{}+jets uncertainties represent the uncertainties in the normalization of the \Wboson{}+heavy-flavor production except for \Wboson{}+jets charge asymmetry and refer to the overall data-driven normalization of the \Wboson{}+jets background, with the numbers 4 (5) referring to $=4$ ($\geq5$) jet multiplicity bins. 

\begin{table*} [htbp]
\footnotesize
\begin{center}

\noindent\makebox[\textwidth]{
\begin{tabular}{l c c c c c c c}
\hline
$\frac{1}{\sigma} \frac{\dsigma}{\dptt}$ Uncertainties [\%] $/$ Bins [GeV] &     0--50 &     50--100 &     100--150 &     150--200 &     200--250 &     250--350 &     350--800 \\ 
\hline
\hline
$b$-quark jets (JES)  & $^{  -1.81 }_{  +1.52}$ & $^{  -0.41 }_{  +0.55}$ & $^{  +1.12 }_{  -0.97}$ & $^{  +0.89 }_{  -1.10}$ & $^{  +0.46 }_{  -0.52}$ & $^{  +0.07 }_{  -0.23}$ & $^{  -0.01 }_{  -0.11}$ \\ [0.5mm] 
Close-by jets (JES) & $^{  -1.56 }_{  +1.45}$ & $^{  -0.45 }_{  +0.61}$ & $^{  +0.40 }_{  -0.77}$ & $^{  +1.39 }_{  -1.01}$ & $^{  +1.13 }_{  -1.02}$ & $^{  +1.47 }_{  -1.39}$ & $^{  +1.80 }_{  -1.65}$ \\ [0.5mm] 
Effective detector NP set 1 (JES) & $^{  -1.51 }_{  +1.06}$ & $^{  -0.47 }_{  +0.54}$ & $^{  +0.54 }_{  -0.40}$ & $^{  +1.16 }_{  -1.20}$ & $^{  +1.20 }_{  -1.04}$ & $^{  +1.17 }_{  -0.97}$ & $^{  +1.49 }_{  -0.51}$ \\ [0.5mm] 
Effective detector NP set 2 (JES) & $^{  -0.21 }_{  +0.09}$ & $^{  +0.00 }_{  +0.05}$ & $^{  -0.06 }_{  -0.08}$ & $^{  +0.24 }_{  -0.06}$ & $^{  +0.23 }_{  +0.05}$ & $^{  +0.12 }_{  -0.18}$ & $^{  +0.28 }_{  -0.26}$ \\ [0.5mm] 
Effective mixed NP set 1 (JES) & $^{  -0.03 }_{  +0.04}$ & $^{  -0.02 }_{  +0.11}$ & $^{  -0.07 }_{  -0.12}$ & $^{  +0.11 }_{  -0.09}$ & $^{  +0.07 }_{  -0.10}$ & $^{  +0.23 }_{  +0.04}$ & $^{  +0.79 }_{  +0.34}$ \\ [0.5mm] 
Effective mixed NP set 2 (JES) & $^{  -0.03 }_{  +0.03}$ & $^{  -0.03 }_{  +0.12}$ & $^{  +0.04 }_{  -0.25}$ & $^{  -0.01 }_{  +0.07}$ & $^{  +0.04 }_{  +0.00}$ & $^{  +0.08 }_{  +0.18}$ & $^{  +0.24 }_{  +0.85}$ \\ [0.5mm] 
Effective model NP set 1 (JES) & $^{  +0.28 }_{  +0.05}$ & $^{  +0.02 }_{  +0.04}$ & $^{  -0.28 }_{  -0.22}$ & $^{  +0.30 }_{  -0.24}$ & $^{  +0.19 }_{  +0.48}$ & $^{  -0.64 }_{  +0.85}$ & $^{  -1.40 }_{  +1.51}$ \\ [0.5mm] 
Effective model NP set 2 (JES) & $^{  -0.53 }_{  +0.59}$ & $^{  -0.10 }_{  +0.17}$ & $^{  -0.11 }_{  -0.22}$ & $^{  +0.62 }_{  -0.38}$ & $^{  +0.63 }_{  -0.51}$ & $^{  +0.76 }_{  -0.48}$ & $^{  +1.34 }_{  -0.48}$ \\ [0.5mm] 
Effective model NP set 3 (JES) & $^{  +0.66 }_{  -0.80}$ & $^{  +0.23 }_{  -0.14}$ & $^{  -0.34 }_{  +0.05}$ & $^{  -0.44 }_{  +0.69}$ & $^{  -0.46 }_{  +0.84}$ & $^{  -0.42 }_{  +0.77}$ & $^{  -0.28 }_{  +0.88}$ \\ [0.5mm] 
Effective model NP set 4 (JES) & $^{  -0.14 }_{  +0.16}$ & $^{  +0.03 }_{  +0.02}$ & $^{  -0.06 }_{  -0.12}$ & $^{  +0.08 }_{  +0.03}$ & $^{  +0.21 }_{  -0.06}$ & $^{  +0.10 }_{  -0.05}$ & $^{  +0.23 }_{  +0.10}$ \\ [0.5mm] 
Effective statistical NP set 1 (JES) & $^{  +0.52 }_{  -0.65}$ & $^{  +0.26 }_{  -0.07}$ & $^{  -0.39 }_{  +0.02}$ & $^{  -0.18 }_{  +0.39}$ & $^{  -0.20 }_{  +0.61}$ & $^{  -0.82 }_{  +0.92}$ & $^{  -1.16 }_{  +1.53}$ \\ [0.5mm] 
Effective statistical NP set 2 (JES) & $^{  -0.20 }_{  +0.21}$ & $^{  +0.04 }_{  +0.06}$ & $^{  -0.05 }_{  -0.14}$ & $^{  +0.15 }_{  -0.07}$ & $^{  +0.10 }_{  -0.12}$ & $^{  +0.19 }_{  -0.14}$ & $^{  +0.49 }_{  +0.04}$ \\ [0.5mm] 
Effective statistical NP set 3 (JES) & $^{  -0.44 }_{  +0.43}$ & $^{  -0.14 }_{  +0.18}$ & $^{  +0.16 }_{  -0.30}$ & $^{  +0.33 }_{  -0.20}$ & $^{  +0.39 }_{  -0.28}$ & $^{  +0.32 }_{  -0.30}$ & $^{  +0.46 }_{  -0.02}$ \\ [0.5mm] 
$\eta$-intercalibration (JES) & $^{  -0.71 }_{  +0.63}$ & $^{  -0.14 }_{  +0.26}$ & $^{  +0.24 }_{  -0.35}$ & $^{  +0.64 }_{  -0.69}$ & $^{  +0.39 }_{  -0.23}$ & $^{  -0.06 }_{  +0.00}$ & $^{  -0.29 }_{  +0.19}$ \\ [0.5mm] 
$\eta$-intercalibration statistics (JES) & $^{  -0.15 }_{  -0.16}$ & $^{  -0.06 }_{  -0.01}$ & $^{  -0.02 }_{  +0.02}$ & $^{  +0.24 }_{  +0.08}$ & $^{  +0.34 }_{  +0.03}$ & $^{  +0.02 }_{  +0.20}$ & $^{  +0.13 }_{  +0.62}$ \\ [0.5mm] 
Flavor composition (JES) & $^{  +0.02 }_{  +0.13}$ & $^{  -0.00 }_{  +0.23}$ & $^{  -0.13 }_{  -0.23}$ & $^{  +0.04 }_{  -0.15}$ & $^{  +0.25 }_{  -0.19}$ & $^{  +0.27 }_{  -0.26}$ & $^{  +0.43 }_{  +0.01}$ \\ [0.5mm] 
Flavor response (JES) & $^{  +0.41 }_{  -0.81}$ & $^{  +0.17 }_{  -0.11}$ & $^{  -0.24 }_{  +0.39}$ & $^{  -0.26 }_{  +0.21}$ & $^{  -0.15 }_{  +0.44}$ & $^{  -0.47 }_{  +0.46}$ & $^{  -0.51 }_{  +0.64}$ \\ [0.5mm] 
Pile-up offset $\mu$ (JES) & $^{  -0.02 }_{  +0.35}$ & $^{  -0.11 }_{  +0.02}$ & $^{  -0.10 }_{  -0.22}$ & $^{  +0.23 }_{  -0.19}$ & $^{  +0.50 }_{  +0.19}$ & $^{  +0.28 }_{  +0.07}$ & $^{  +0.05 }_{  +0.60}$ \\ [0.5mm] 
Pile-up offset $N_{\rm PV}$ (JES) & $^{  +0.13 }_{  -0.28}$ & $^{  +0.13 }_{  +0.08}$ & $^{  -0.20 }_{  -0.06}$ & $^{  -0.06 }_{  +0.05}$ & $^{  -0.09 }_{  +0.24}$ & $^{  -0.05 }_{  +0.31}$ & $^{  +0.25 }_{  +0.62}$ \\ [0.5mm] 
Relative non-closure MC (JES) & $^{  -0.08 }_{  +0.10}$ & $^{  -0.02 }_{  +0.13}$ & $^{  -0.13 }_{  -0.24}$ & $^{  +0.23 }_{  +0.01}$ & $^{  +0.30 }_{  +0.01}$ & $^{  +0.18 }_{  -0.05}$ & $^{  +0.25 }_{  +0.26}$ \\ [0.5mm] 
Single particle high-\pt{} (JES) & $^{  +0.02 }_{  +0.02}$ & $^{  -0.01 }_{  -0.01}$ & $^{  +0.01 }_{  -0.00}$ & $^{  +0.01 }_{  +0.01}$ & $^{  +0.01 }_{  +0.01}$ & $^{  -0.02 }_{  -0.02}$ & $^{  -0.02 }_{  -0.01}$ \\ [0.5mm] 
JES uncertainty in background & $^{  -0.05 }_{  +0.14}$ & $^{  +0.39 }_{  -0.04}$ & $^{  +0.03 }_{  -0.13}$ & $^{  -0.58 }_{  -0.07}$ & $^{  -0.98 }_{  +0.29}$ & $^{  -0.21 }_{  +0.27}$ & $^{  +0.74 }_{  +1.23}$ \\ [0.5mm] 
Jet energy resolution & $^{  +0.44 }_{  -0.44}$ & $^{  +0.12 }_{  -0.12}$ & $^{  -0.51 }_{  +0.51}$ & $^{  +0.01 }_{  -0.01}$ & $^{  +0.30 }_{  -0.31}$ & $^{  -0.07 }_{  +0.07}$ & $^{  +0.53 }_{  -0.54}$ \\ [0.5mm] 
Jet reconstruction efficiency & $^{  +0.10 }_{  -0.10}$ & $^{  +0.01 }_{  -0.01}$ & $^{  -0.05 }_{  +0.05}$ & $^{  -0.03 }_{  +0.03}$ & $^{  -0.03 }_{  +0.03}$ & $^{  -0.07 }_{  +0.07}$ & $^{  -0.14 }_{  +0.15}$ \\ [0.5mm] 
$b$-quark tagging efficiency & $^{  -1.36 }_{  +1.09}$ & $^{  -0.77 }_{  +0.59}$ & $^{  +0.35 }_{  -0.35}$ & $^{  +1.33 }_{  -1.08}$ & $^{  +2.06 }_{  -1.50}$ & $^{  +2.59 }_{  -1.61}$ & $^{  +3.00 }_{  -1.62}$ \\ [0.5mm] 
$c$-quark tagging efficiency & $^{  -0.03 }_{  +0.01}$ & $^{  +0.01 }_{  -0.01}$ & $^{  +0.07 }_{  -0.07}$ & $^{  -0.06 }_{  +0.06}$ & $^{  -0.10 }_{  +0.10}$ & $^{  -0.12 }_{  +0.12}$ & $^{  -0.23 }_{  +0.23}$ \\ [0.5mm] 
Light-jet tagging efficiency & $^{  +0.30 }_{  -0.29}$ & $^{  +0.03 }_{  -0.03}$ & $^{  -0.22 }_{  +0.22}$ & $^{  -0.03 }_{  +0.02}$ & $^{  -0.01 }_{  +0.00}$ & $^{  +0.07 }_{  -0.07}$ & $^{  +0.21 }_{  -0.18}$ \\ [0.5mm] 
$e$ energy resolution & $^{  -0.06 }_{  +0.07}$ & $^{  +0.04 }_{  -0.03}$ & $^{  +0.01 }_{  +0.02}$ & $^{  -0.05 }_{  -0.03}$ & $^{  -0.12 }_{  -0.03}$ & $^{  +0.13 }_{  +0.02}$ & $^{  +0.42 }_{  -0.11}$ \\ [0.5mm] 
$e$ energy scale & $^{  +0.22 }_{  -0.16}$ & $^{  +0.04 }_{  +0.01}$ & $^{  -0.15 }_{  +0.18}$ & $^{  +0.01 }_{  -0.07}$ & $^{  -0.03 }_{  -0.18}$ & $^{  -0.19 }_{  -0.12}$ & $^{  -0.34 }_{  +0.02}$ \\ [0.5mm] 
$\mu$ ID momentum resolution & $^{  +0.38 }_{  +0.38}$ & $^{  +0.04 }_{  +0.10}$ & $^{  -0.32 }_{  -0.40}$ & $^{  -0.02 }_{  +0.00}$ & $^{  +0.08 }_{  +0.08}$ & $^{  +0.07 }_{  +0.09}$ & $^{  +0.20 }_{  +0.23}$ \\ [0.5mm] 
$\mu$ MS momentum resolution & $^{  +0.39 }_{  +0.27}$ & $^{  +0.18 }_{  +0.02}$ & $^{  -0.43 }_{  -0.12}$ & $^{  -0.13 }_{  -0.12}$ & $^{  -0.03 }_{  -0.09}$ & $^{  +0.08 }_{  -0.04}$ & $^{  +0.32 }_{  +0.27}$ \\ [0.5mm] 
$\mu$ momentum scale & $^{  +0.27 }_{  -0.26}$ & $^{  -0.00 }_{  +0.00}$ & $^{  -0.29 }_{  +0.29}$ & $^{  +0.04 }_{  -0.04}$ & $^{  +0.27 }_{  -0.27}$ & $^{  +0.28 }_{  -0.27}$ & $^{  +0.36 }_{  -0.36}$ \\ [0.5mm] 
$\ell$ ID efficiency & $^{  +0.58 }_{  -0.56}$ & $^{  +0.08 }_{  -0.08}$ & $^{  -0.95 }_{  +0.92}$ & $^{  +0.45 }_{  -0.43}$ & $^{  +0.66 }_{  -0.63}$ & $^{  +0.84 }_{  -0.80}$ & $^{  +0.49 }_{  -0.47}$ \\ [0.5mm] 
$\ell$ reconstruction efficiency & $^{  +0.21 }_{  -0.21}$ & $^{  +0.03 }_{  -0.03}$ & $^{  -0.36 }_{  +0.35}$ & $^{  +0.17 }_{  -0.17}$ & $^{  +0.26 }_{  -0.25}$ & $^{  +0.33 }_{  -0.33}$ & $^{  +0.21 }_{  -0.21}$ \\ [0.5mm] 
$\ell$ trigger efficiency & $^{  -0.20 }_{  +0.19}$ & $^{  -0.01 }_{  +0.01}$ & $^{  +0.44 }_{  -0.43}$ & $^{  -0.28 }_{  +0.28}$ & $^{  -0.42 }_{  +0.41}$ & $^{  -0.52 }_{  +0.52}$ & $^{  -0.35 }_{  +0.35}$ \\ [0.5mm] 
\Etmiss{} unassociated cells & $^{  -0.14 }_{  +0.42}$ & $^{  -0.07 }_{  +0.05}$ & $^{  +0.24 }_{  -0.37}$ & $^{  +0.08 }_{  -0.14}$ & $^{  -0.25 }_{  +0.30}$ & $^{  -0.44 }_{  +0.29}$ & $^{  +0.07 }_{  +0.31}$ \\ [0.5mm] 
\Etmiss{} pile-up & $^{  -0.10 }_{  +0.56}$ & $^{  -0.13 }_{  +0.09}$ & $^{  +0.10 }_{  -0.59}$ & $^{  -0.02 }_{  -0.11}$ & $^{  +0.19 }_{  +0.39}$ & $^{  +0.56 }_{  +0.55}$ & $^{  +0.76 }_{  +0.71}$ \\ [0.5mm] 
MC generator & $^{  -1.52 }_{  +1.89}$ & $^{  +0.54 }_{  -0.67}$ & $^{  +0.16 }_{  -0.20}$ & $^{  +1.52 }_{  -1.89}$ & $^{  -0.11 }_{  +0.13}$ & $^{  -2.79 }_{  +3.47}$ & $^{  -8.57 }_{ +10.66}$ \\ [0.5mm] 
Fragmentation & $^{  -0.61 }_{  +0.58}$ & $^{  +0.71 }_{  -0.68}$ & $^{  -0.72 }_{  +0.70}$ & $^{  +0.88 }_{  -0.84}$ & $^{  -0.96 }_{  +0.92}$ & $^{  -0.70 }_{  +0.68}$ & $^{  +1.94 }_{  -1.87}$ \\ [0.5mm] 
IFSR & $^{  +2.23 }_{  -2.15}$ & $^{  +0.90 }_{  -0.87}$ & $^{  -0.08 }_{  +0.08}$ & $^{  -3.22 }_{  +3.11}$ & $^{  -3.22 }_{  +3.11}$ & $^{  -1.56 }_{  +1.51}$ & $^{  -0.09 }_{  +0.09}$ \\ [0.5mm] 
PDF & $^{  +0.14 }_{  -0.14}$ & $^{  +0.14 }_{  -0.14}$ & $^{  +0.04 }_{  -0.04}$ & $^{  -0.16 }_{  +0.16}$ & $^{  -0.45 }_{  +0.47}$ & $^{  -0.81 }_{  +0.84}$ & $^{  -0.79 }_{  +0.82}$ \\ [0.5mm] 
MC statistics & $^{  +1.01 }_{  -1.01}$ & $^{  +0.40 }_{  -0.40}$ & $^{  +0.67 }_{  -0.67}$ & $^{  +0.90 }_{  -0.90}$ & $^{  +1.08 }_{  -1.08}$ & $^{  +1.44 }_{  -1.44}$ & $^{  +2.60 }_{  -2.60}$ \\ [0.5mm] 
\Wboson+jets bb4 & $^{  +0.32 }_{  -0.35}$ & $^{  +0.05 }_{  -0.06}$ & $^{  -0.14 }_{  +0.15}$ & $^{  -0.12 }_{  +0.14}$ & $^{  -0.12 }_{  +0.19}$ & $^{  -0.32 }_{  +0.29}$ & $^{  -0.29 }_{  +0.38}$ \\ [0.5mm] 
\Wboson+jets bb5 & $^{  +0.32 }_{  +0.03}$ & $^{  +0.05 }_{  +0.01}$ & $^{  -0.14 }_{  -0.01}$ & $^{  -0.12 }_{  -0.02}$ & $^{  -0.12 }_{  -0.06}$ & $^{  -0.32 }_{  +0.04}$ & $^{  -0.29 }_{  -0.07}$ \\ [0.5mm] 
\Wboson+jets bbcc & $^{  -0.10 }_{  +0.09}$ & $^{  -0.06 }_{  +0.06}$ & $^{  -0.06 }_{  +0.06}$ & $^{  +0.24 }_{  -0.23}$ & $^{  +0.27 }_{  -0.26}$ & $^{  +0.12 }_{  -0.12}$ & $^{  +0.22 }_{  -0.23}$ \\ [0.5mm] 
\Wboson+jets bbccc & $^{  +1.08 }_{  -1.12}$ & $^{  +0.20 }_{  -0.22}$ & $^{  -0.48 }_{  +0.42}$ & $^{  -0.45 }_{  +0.56}$ & $^{  -0.50 }_{  +0.64}$ & $^{  -0.84 }_{  +0.99}$ & $^{  -1.00 }_{  +1.12}$ \\ [0.5mm] 
\Wboson+jets c4 & $^{  +0.48 }_{  -0.38}$ & $^{  +0.11 }_{  -0.11}$ & $^{  -0.09 }_{  +0.12}$ & $^{  -0.38 }_{  +0.29}$ & $^{  -0.44 }_{  +0.33}$ & $^{  -0.53 }_{  +0.36}$ & $^{  -0.54 }_{  +0.34}$ \\ [0.5mm] 
\Wboson+Jets c5 & $^{  +0.48 }_{  -0.16}$ & $^{  +0.11 }_{  -0.00}$ & $^{  -0.09 }_{  -0.03}$ & $^{  -0.38 }_{  +0.13}$ & $^{  -0.44 }_{  +0.16}$ & $^{  -0.53 }_{  +0.24}$ & $^{  -0.54 }_{  +0.24}$ \\ [0.5mm] 
\Wboson+jets charge asymmetry & $^{  +0.96 }_{  -1.11}$ & $^{  +0.17 }_{  -0.19}$ & $^{  -0.41 }_{  +0.51}$ & $^{  -0.44 }_{  +0.48}$ & $^{  -0.47 }_{  +0.49}$ & $^{  -0.64 }_{  +0.69}$ & $^{  -0.51 }_{  +0.57}$ \\ [0.5mm] 
Multijet normalization & $^{  +1.35 }_{  -1.36}$ & $^{  +0.01 }_{  -0.01}$ & $^{  -0.75 }_{  +0.76}$ & $^{  -0.39 }_{  +0.40}$ & $^{  -0.11 }_{  +0.11}$ & $^{  +0.66 }_{  -0.67}$ & $^{  +0.76 }_{  -0.77}$ \\ [0.5mm] 
Multijet shape & $^{  +0.11 }_{  -0.59}$ & $^{  +0.04 }_{  +0.07}$ & $^{  -0.46 }_{  +0.25}$ & $^{  +0.45 }_{  +0.12}$ & $^{  +0.46 }_{  +0.03}$ & $^{  +0.15 }_{  -0.14}$ & $^{  +0.11 }_{  -0.32}$ \\ [0.5mm] 
$Z+$jets background normalization & $^{  +0.59 }_{  -0.60}$ & $^{  +0.15 }_{  -0.15}$ & $^{  -0.42 }_{  +0.42}$ & $^{  -0.21 }_{  +0.21}$ & $^{  -0.16 }_{  +0.16}$ & $^{  -0.06 }_{  +0.06}$ & $^{  -0.13 }_{  +0.12}$ \\ [0.5mm] 
Dilepton background normalization & $^{  +0.22 }_{  -0.21}$ & $^{  +0.10 }_{  -0.09}$ & $^{  -0.05 }_{  +0.04}$ & $^{  -0.16 }_{  +0.15}$ & $^{  -0.32 }_{  +0.30}$ & $^{  -0.43 }_{  +0.41}$ & $^{  -0.50 }_{  +0.47}$ \\ [0.5mm] 
\hline
\end{tabular}}

\caption{The individual systematic uncertainties calculated as a~percentage of the normalized differential cross-section in each bin.}
\label{tab:CombSyst_1_shape}
\end{center}
\end{table*}
\begin{table*} [htbp]
\footnotesize
\begin{center}

\noindent\makebox[\textwidth]{
\begin{tabular}{l c c c c c}
\hline
$\frac{1}{\sigma} \frac{\dsigma}{\dmttbar} $ Uncertainties [\%] $/$ Bins [GeV] &     250--450 &     450--550 &     550--700 &     700--950 &     950--2700 \\ 
\hline
\hline
$b$-quark jets (JES)  & $^{  -0.32 }_{  +0.53}$ & $^{  +0.25 }_{  -0.33}$ & $^{  +0.44 }_{  -0.78}$ & $^{  +0.40 }_{  -0.80}$ & $^{  +0.35 }_{  -0.93}$ \\ [0.5mm] 
Close-by jets (JES) & $^{  -0.44 }_{  +0.45}$ & $^{  +0.12 }_{  -0.17}$ & $^{  +0.68 }_{  -0.55}$ & $^{  +1.19 }_{  -1.29}$ & $^{  +1.46 }_{  -2.03}$ \\ [0.5mm] 
Effective detector NP set 1 (JES) & $^{  -0.75 }_{  +0.47}$ & $^{  +0.44 }_{  -0.31}$ & $^{  +1.00 }_{  -0.40}$ & $^{  +1.49 }_{  -1.27}$ & $^{  +1.87 }_{  -1.61}$ \\ [0.5mm] 
Effective detector NP set 2 (JES) & $^{  -0.11 }_{  -0.05}$ & $^{  +0.08 }_{  +0.02}$ & $^{  +0.11 }_{  +0.05}$ & $^{  +0.24 }_{  +0.18}$ & $^{  +0.34 }_{  +0.19}$ \\ [0.5mm] 
Effective mixed NP set 1 (JES) & $^{  -0.03 }_{  -0.04}$ & $^{  -0.05 }_{  -0.04}$ & $^{  +0.06 }_{  +0.29}$ & $^{  +0.30 }_{  -0.13}$ & $^{  +0.23 }_{  -0.25}$ \\ [0.5mm] 
Effective mixed NP set 2 (JES) & $^{  -0.03 }_{  -0.10}$ & $^{  +0.03 }_{  +0.04}$ & $^{  +0.01 }_{  +0.21}$ & $^{  +0.07 }_{  +0.08}$ & $^{  +0.14 }_{  +0.13}$ \\ [0.5mm] 
Effective model NP set 1 (JES) & $^{  -0.22 }_{  +0.04}$ & $^{  +0.28 }_{  -0.19}$ & $^{  +0.48 }_{  +0.18}$ & $^{  -0.43 }_{  +0.11}$ & $^{  -1.00 }_{  +0.13}$ \\ [0.5mm] 
Effective model NP set 2 (JES) & $^{  -0.37 }_{  +0.13}$ & $^{  +0.24 }_{  -0.07}$ & $^{  +0.54 }_{  -0.13}$ & $^{  +0.53 }_{  -0.37}$ & $^{  +0.59 }_{  -0.49}$ \\ [0.5mm] 
Effective model NP set 3 (JES) & $^{  +0.15 }_{  -0.30}$ & $^{  -0.03 }_{  +0.15}$ & $^{  -0.18 }_{  +0.44}$ & $^{  -0.57 }_{  +0.67}$ & $^{  -0.48 }_{  +0.66}$ \\ [0.5mm] 
Effective model NP set 4 (JES) & $^{  -0.10 }_{  +0.04}$ & $^{  +0.04 }_{  -0.12}$ & $^{  +0.21 }_{  +0.13}$ & $^{  +0.15 }_{  -0.09}$ & $^{  +0.02 }_{  -0.22}$ \\ [0.5mm] 
Effective statistical NP set 1 (JES) & $^{  +0.24 }_{  -0.27}$ & $^{  -0.20 }_{  +0.05}$ & $^{  -0.27 }_{  +0.62}$ & $^{  -0.31 }_{  +0.44}$ & $^{  -0.42 }_{  +0.63}$ \\ [0.5mm] 
Effective statistical NP set 2 (JES) & $^{  -0.09 }_{  -0.02}$ & $^{  +0.03 }_{  +0.03}$ & $^{  +0.27 }_{  -0.00}$ & $^{  -0.04 }_{  -0.02}$ & $^{  -0.12 }_{  +0.20}$ \\ [0.5mm] 
Effective statistical NP set 3 (JES) & $^{  -0.11 }_{  +0.10}$ & $^{  -0.01 }_{  -0.09}$ & $^{  +0.21 }_{  -0.01}$ & $^{  +0.35 }_{  -0.37}$ & $^{  +0.32 }_{  -0.52}$ \\ [0.5mm] 
$\eta$-intercalibration (JES) & $^{  -0.84 }_{  +0.79}$ & $^{  +0.37 }_{  -0.42}$ & $^{  +1.27 }_{  -0.95}$ & $^{  +1.85 }_{  -1.87}$ & $^{  +2.11 }_{  -2.67}$ \\ [0.5mm] 
$\eta$-intercalibration statistics (JES) & $^{  -0.17 }_{  -0.01}$ & $^{  +0.15 }_{  -0.02}$ & $^{  +0.28 }_{  +0.10}$ & $^{  +0.07 }_{  -0.02}$ & $^{  +0.10 }_{  -0.15}$ \\ [0.5mm] 
Flavor composition (JES) & $^{  -0.03 }_{  +0.07}$ & $^{  -0.11 }_{  -0.08}$ & $^{  +0.13 }_{  +0.14}$ & $^{  +0.37 }_{  -0.46}$ & $^{  +0.40 }_{  -0.78}$ \\ [0.5mm] 
Flavor response (JES) & $^{  +0.05 }_{  -0.09}$ & $^{  -0.13 }_{  -0.08}$ & $^{  -0.01 }_{  +0.33}$ & $^{  +0.15 }_{  +0.26}$ & $^{  +0.27 }_{  +0.28}$ \\ [0.5mm] 
Pile-up offset $\mu$ (JES) & $^{  -0.08 }_{  +0.03}$ & $^{  +0.00 }_{  +0.07}$ & $^{  +0.33 }_{  -0.01}$ & $^{  -0.08 }_{  -0.42}$ & $^{  -0.35 }_{  -0.95}$ \\ [0.5mm] 
Pile-up offset $N_{\rm PV}$ (JES) & $^{  +0.08 }_{  -0.03}$ & $^{  -0.07 }_{  -0.08}$ & $^{  -0.02 }_{  +0.00}$ & $^{  -0.23 }_{  +0.48}$ & $^{  -0.44 }_{  +0.55}$ \\ [0.5mm] 
Relative non-closure MC (JES) & $^{  -0.18 }_{  +0.03}$ & $^{  +0.04 }_{  +0.04}$ & $^{  +0.41 }_{  -0.02}$ & $^{  +0.28 }_{  -0.29}$ & $^{  +0.15 }_{  -0.46}$ \\ [0.5mm] 
Single particle high-\pt{} (JES) & $^{  -0.01 }_{  -0.01}$ & $^{  +0.01 }_{  +0.01}$ & $^{  +0.02 }_{  +0.02}$ & $^{  -0.05 }_{  -0.04}$ & $^{  -0.09 }_{  -0.06}$ \\ [0.5mm] 
JES uncertainty in background & $^{  +0.16 }_{  -0.08}$ & $^{  -0.28 }_{  +0.13}$ & $^{  -0.12 }_{  +0.28}$ & $^{  +0.19 }_{  -0.46}$ & $^{  +0.44 }_{  -0.94}$ \\ [0.5mm] 
Jet energy resolution & $^{  -0.57 }_{  +0.58}$ & $^{  +0.90 }_{  -0.91}$ & $^{  +0.16 }_{  -0.17}$ & $^{  +0.24 }_{  -0.24}$ & $^{  -0.03 }_{  +0.03}$ \\ [0.5mm] 
Jet reconstruction efficiency & $^{  -0.03 }_{  +0.03}$ & $^{  +0.09 }_{  -0.09}$ & $^{  -0.10 }_{  +0.10}$ & $^{  +0.05 }_{  -0.05}$ & $^{  +0.23 }_{  -0.24}$ \\ [0.5mm] 
$b$-quark tagging efficiency & $^{  -0.99 }_{  +0.77}$ & $^{  +0.42 }_{  -0.39}$ & $^{  +1.62 }_{  -1.27}$ & $^{  +1.99 }_{  -1.33}$ & $^{  +2.21 }_{  -1.23}$ \\ [0.5mm] 
$c$-quark tagging efficiency & $^{  +0.07 }_{  -0.07}$ & $^{  -0.04 }_{  +0.04}$ & $^{  -0.18 }_{  +0.17}$ & $^{  +0.06 }_{  -0.05}$ & $^{  +0.12 }_{  -0.10}$ \\ [0.5mm] 
Light-jet tagging efficiency & $^{  +0.06 }_{  -0.06}$ & $^{  -0.12 }_{  +0.12}$ & $^{  +0.03 }_{  -0.04}$ & $^{  -0.03 }_{  +0.03}$ & $^{  -0.14 }_{  +0.13}$ \\ [0.5mm] 
$e$ energy resolution & $^{  -0.01 }_{  +0.01}$ & $^{  +0.02 }_{  -0.02}$ & $^{  +0.13 }_{  +0.04}$ & $^{  -0.23 }_{  -0.07}$ & $^{  -0.39 }_{  -0.11}$ \\ [0.5mm] 
$e$ energy scale & $^{  +0.07 }_{  -0.10}$ & $^{  -0.11 }_{  +0.15}$ & $^{  -0.03 }_{  +0.24}$ & $^{  +0.01 }_{  -0.37}$ & $^{  -0.08 }_{  -0.30}$ \\ [0.5mm] 
$\mu$ ID momentum resolution & $^{  +0.01 }_{  +0.11}$ & $^{  -0.05 }_{  -0.11}$ & $^{  +0.07 }_{  -0.20}$ & $^{  -0.01 }_{  +0.06}$ & $^{  -0.01 }_{  +0.10}$ \\ [0.5mm] 
$\mu$ MS momentum resolution & $^{  +0.11 }_{  +0.07}$ & $^{  +0.01 }_{  +0.03}$ & $^{  -0.32 }_{  -0.25}$ & $^{  -0.15 }_{  -0.09}$ & $^{  -0.10 }_{  -0.06}$ \\ [0.5mm] 
$\mu$ momentum scale & $^{  -0.03 }_{  +0.03}$ & $^{  -0.05 }_{  +0.05}$ & $^{  +0.12 }_{  -0.12}$ & $^{  +0.18 }_{  -0.18}$ & $^{  +0.20 }_{  -0.20}$ \\ [0.5mm] 
$\ell$ ID efficiency & $^{  +0.40 }_{  -0.38}$ & $^{  -0.69 }_{  +0.66}$ & $^{  -0.73 }_{  +0.70}$ & $^{  +1.42 }_{  -1.36}$ & $^{  +1.61 }_{  -1.54}$ \\ [0.5mm] 
$\ell$ reconstruction efficiency & $^{  +0.15 }_{  -0.15}$ & $^{  -0.26 }_{  +0.26}$ & $^{  -0.24 }_{  +0.23}$ & $^{  +0.45 }_{  -0.44}$ & $^{  +0.46 }_{  -0.46}$ \\ [0.5mm] 
$\ell$ trigger efficiency & $^{  -0.15 }_{  +0.15}$ & $^{  +0.32 }_{  -0.32}$ & $^{  +0.27 }_{  -0.27}$ & $^{  -0.74 }_{  +0.73}$ & $^{  -0.81 }_{  +0.80}$ \\ [0.5mm] 
\Etmiss{} unassociated cells & $^{  -0.03 }_{  -0.04}$ & $^{  +0.02 }_{  +0.14}$ & $^{  +0.02 }_{  -0.02}$ & $^{  +0.02 }_{  -0.21}$ & $^{  +0.50 }_{  -0.37}$ \\ [0.5mm] 
\Etmiss{} pile-up & $^{  -0.10 }_{  -0.05}$ & $^{  +0.04 }_{  -0.02}$ & $^{  +0.15 }_{  +0.19}$ & $^{  +0.19 }_{  +0.11}$ & $^{  +0.64 }_{  -0.29}$ \\ [0.5mm] 
MC generator & $^{  -2.19 }_{  +2.73}$ & $^{  +1.88 }_{  -2.34}$ & $^{  +2.57 }_{  -3.20}$ & $^{  +2.97 }_{  -3.71}$ & $^{  +2.51 }_{  -3.13}$ \\ [0.5mm] 
Fragmentation & $^{  +0.20 }_{  -0.20}$ & $^{  -0.21 }_{  +0.20}$ & $^{  +0.54 }_{  -0.52}$ & $^{  -1.75 }_{  +1.69}$ & $^{  -2.21 }_{  +2.14}$ \\ [0.5mm] 
IFSR & $^{  +0.55 }_{  -0.54}$ & $^{  -0.18 }_{  +0.18}$ & $^{  -0.92 }_{  +0.89}$ & $^{  -1.50 }_{  +1.45}$ & $^{  -0.40 }_{  +0.38}$ \\ [0.5mm] 
PDF & $^{  -0.07 }_{  +0.07}$ & $^{  -0.07 }_{  +0.08}$ & $^{  -0.02 }_{  +0.02}$ & $^{  +0.54 }_{  -0.56}$ & $^{  +2.20 }_{  -2.26}$ \\ [0.5mm] 
MC statistics & $^{  +0.41 }_{  -0.41}$ & $^{  +0.41 }_{  -0.41}$ & $^{  +0.63 }_{  -0.63}$ & $^{  +1.04 }_{  -1.04}$ & $^{  +1.60 }_{  -1.60}$ \\ [0.5mm] 
\Wboson+jets bb4 & $^{  -0.08 }_{  +0.01}$ & $^{  -0.06 }_{  +0.02}$ & $^{  +0.14 }_{  -0.04}$ & $^{  +0.48 }_{  -0.04}$ & $^{  +0.80 }_{  -0.15}$ \\ [0.5mm] 
\Wboson+jets bb5 & $^{  -0.08 }_{  +0.07}$ & $^{  -0.06 }_{  +0.03}$ & $^{  +0.14 }_{  -0.10}$ & $^{  +0.48 }_{  -0.41}$ & $^{  +0.80 }_{  -0.61}$ \\ [0.5mm] 
\Wboson+jets bbcc & $^{  -0.17 }_{  +0.16}$ & $^{  -0.09 }_{  +0.09}$ & $^{  +0.45 }_{  -0.41}$ & $^{  +0.60 }_{  -0.58}$ & $^{  +0.96 }_{  -0.91}$ \\ [0.5mm] 
\Wboson+jets bbccc & $^{  +0.03 }_{  +0.01}$ & $^{  -0.13 }_{  +0.06}$ & $^{  -0.16 }_{  +0.07}$ & $^{  +0.57 }_{  -0.39}$ & $^{  +1.03 }_{  -0.82}$ \\ [0.5mm] 
\Wboson+jets c4 & $^{  -0.05 }_{  -0.03}$ & $^{  +0.13 }_{  -0.03}$ & $^{  +0.01 }_{  +0.07}$ & $^{  -0.21 }_{  +0.17}$ & $^{  -0.17 }_{  +0.17}$ \\ [0.5mm] 
\Wboson+Jets c5 & $^{  -0.05 }_{  +0.09}$ & $^{  +0.13 }_{  -0.12}$ & $^{  +0.01 }_{  -0.09}$ & $^{  -0.21 }_{  +0.07}$ & $^{  -0.17 }_{  +0.03}$ \\ [0.5mm] 
\Wboson+jets charge asymmetry & $^{  -0.05 }_{  +0.06}$ & $^{  -0.02 }_{  +0.05}$ & $^{  +0.04 }_{  -0.00}$ & $^{  +0.34 }_{  -0.52}$ & $^{  +0.63 }_{  -0.90}$ \\ [0.5mm] 
Multijet normalization & $^{  +0.26 }_{  -0.27}$ & $^{  -0.67 }_{  +0.67}$ & $^{  -0.76 }_{  +0.75}$ & $^{  +2.11 }_{  -2.02}$ & $^{  +4.17 }_{  -3.96}$ \\ [0.5mm] 
Multijet shape & $^{  +0.04 }_{  +0.07}$ & $^{  -0.18 }_{  +0.18}$ & $^{  -0.25 }_{  +0.31}$ & $^{  +0.82 }_{  -1.52}$ & $^{  +1.45 }_{  -3.65}$ \\ [0.5mm] 
$Z+$jets background normalization & $^{  +0.14 }_{  -0.14}$ & $^{  -0.25 }_{  +0.25}$ & $^{  -0.17 }_{  +0.17}$ & $^{  +0.32 }_{  -0.32}$ & $^{  +0.49 }_{  -0.48}$ \\ [0.5mm] 
Dilepton background normalization & $^{  +0.14 }_{  -0.13}$ & $^{  -0.06 }_{  +0.06}$ & $^{  -0.22 }_{  +0.21}$ & $^{  -0.30 }_{  +0.29}$ & $^{  -0.34 }_{  +0.33}$ \\ [0.5mm] 
\hline
\end{tabular}}

\caption{The individual systematic uncertainties calculated as a~percentage of the normalized differential cross-section in each bin.}
\label{tab:CombSyst_2_shape}
\end{center}
\end{table*}

\begin{table*} [htbp]
\footnotesize
\begin{center}

\noindent\makebox[\textwidth]{
\begin{tabular}{l c c c c}
\hline
$\frac{1}{\sigma} \frac{\dsigma}{\dptttbar}$ Uncertainties [\%] $/$ Bins [GeV] &     0--40 &     40--170 &     170--340 &     340--1000 \\ 
\hline
\hline
$b$-quark jets (JES)  & $^{  -0.02 }_{  +0.03}$ & $^{  -0.01 }_{  +0.03}$ & $^{  +0.38 }_{  -0.53}$ & $^{  +0.30 }_{  -0.76}$ \\ [0.5mm] 
Close-by jets (JES) & $^{  -0.73 }_{  +0.82}$ & $^{  +0.88 }_{  -1.04}$ & $^{  +1.40 }_{  -1.19}$ & $^{  +1.40 }_{  -1.28}$ \\ [0.5mm] 
Effective detector NP set 1 (JES) & $^{  -0.53 }_{  +0.53}$ & $^{  +0.51 }_{  -0.49}$ & $^{  +2.00 }_{  -2.12}$ & $^{  +2.76 }_{  -3.18}$ \\ [0.5mm] 
Effective detector NP set 2 (JES) & $^{  +0.03 }_{  -0.05}$ & $^{  -0.01 }_{  +0.07}$ & $^{  -0.31 }_{  +0.00}$ & $^{  -0.70 }_{  -0.09}$ \\ [0.5mm] 
Effective mixed NP set 1 (JES) & $^{  -0.10 }_{  +0.03}$ & $^{  +0.13 }_{  +0.02}$ & $^{  +0.09 }_{  -0.47}$ & $^{  +0.18 }_{  -1.07}$ \\ [0.5mm] 
Effective mixed NP set 2 (JES) & $^{  -0.01 }_{  +0.04}$ & $^{  +0.06 }_{  -0.04}$ & $^{  -0.33 }_{  -0.05}$ & $^{  -0.76 }_{  -0.25}$ \\ [0.5mm] 
Effective model NP set 1 (JES) & $^{  -0.59 }_{  +0.48}$ & $^{  +0.75 }_{  -0.61}$ & $^{  +0.89 }_{  -0.72}$ & $^{  +0.39 }_{  -0.56}$ \\ [0.5mm] 
Effective model NP set 2 (JES) & $^{  +0.01 }_{  +0.21}$ & $^{  -0.03 }_{  -0.18}$ & $^{  +0.12 }_{  -0.97}$ & $^{  +0.16 }_{  -1.55}$ \\ [0.5mm] 
Effective model NP set 3 (JES) & $^{  +0.13 }_{  -0.09}$ & $^{  -0.11 }_{  +0.13}$ & $^{  -0.56 }_{  +0.05}$ & $^{  -0.70 }_{  +0.01}$ \\ [0.5mm] 
Effective model NP set 4 (JES) & $^{  -0.06 }_{  +0.09}$ & $^{  +0.09 }_{  -0.08}$ & $^{  +0.06 }_{  -0.40}$ & $^{  -0.06 }_{  -0.64}$ \\ [0.5mm] 
Effective statistical NP set 1 (JES) & $^{  -0.16 }_{  +0.25}$ & $^{  +0.22 }_{  -0.33}$ & $^{  +0.18 }_{  -0.24}$ & $^{  -0.09 }_{  -0.05}$ \\ [0.5mm] 
Effective statistical NP set 2 (JES) & $^{  +0.09 }_{  -0.01}$ & $^{  -0.11 }_{  +0.03}$ & $^{  -0.18 }_{  -0.09}$ & $^{  -0.28 }_{  -0.21}$ \\ [0.5mm] 
Effective statistical NP set 3 (JES) & $^{  -0.12 }_{  +0.05}$ & $^{  +0.15 }_{  -0.01}$ & $^{  +0.16 }_{  -0.49}$ & $^{  +0.20 }_{  -0.90}$ \\ [0.5mm] 
$\eta$-intercalibration (JES) & $^{  -0.90 }_{  +1.03}$ & $^{  +0.96 }_{  -1.14}$ & $^{  +2.71 }_{  -2.80}$ & $^{  +3.31 }_{  -3.64}$ \\ [0.5mm] 
$\eta$-intercalibration statistics (JES) & $^{  -0.16 }_{  +0.17}$ & $^{  +0.21 }_{  -0.15}$ & $^{  +0.23 }_{  -0.71}$ & $^{  -0.11 }_{  -1.17}$ \\ [0.5mm] 
Flavor composition (JES) & $^{  -0.04 }_{  +0.09}$ & $^{  +0.02 }_{  -0.09}$ & $^{  +0.26 }_{  -0.35}$ & $^{  +0.33 }_{  -0.49}$ \\ [0.5mm] 
Flavor response (JES) & $^{  -0.36 }_{  +0.42}$ & $^{  +0.45 }_{  -0.48}$ & $^{  +0.58 }_{  -0.97}$ & $^{  +0.54 }_{  -1.24}$ \\ [0.5mm] 
Pile-up offset $\mu$ (JES) & $^{  -0.19 }_{  +0.10}$ & $^{  +0.20 }_{  -0.10}$ & $^{  +0.68 }_{  -0.36}$ & $^{  +0.61 }_{  -0.34}$ \\ [0.5mm] 
Pile-up offset $N_{\rm PV}$ (JES) & $^{  -0.07 }_{  +0.14}$ & $^{  +0.09 }_{  -0.15}$ & $^{  +0.07 }_{  -0.41}$ & $^{  -0.25 }_{  -0.82}$ \\ [0.5mm] 
Relative non-closure MC (JES) & $^{  -0.04 }_{  +0.14}$ & $^{  +0.09 }_{  -0.15}$ & $^{  -0.23 }_{  -0.46}$ & $^{  -0.52 }_{  -0.74}$ \\ [0.5mm] 
Single particle high-\pt{} (JES) & $^{  +0.03 }_{  +0.03}$ & $^{  -0.04 }_{  -0.03}$ & $^{  -0.08 }_{  -0.08}$ & $^{  -0.10 }_{  -0.12}$ \\ [0.5mm] 
JES uncertainty in background & $^{  -1.20 }_{  +1.13}$ & $^{  +1.37 }_{  -1.35}$ & $^{  +2.78 }_{  -2.35}$ & $^{  +3.89 }_{  -2.60}$ \\ [0.5mm] 
Jet energy resolution & $^{  +3.42 }_{  -3.49}$ & $^{  -4.07 }_{  +4.16}$ & $^{  -7.06 }_{  +7.22}$ & $^{  -8.01 }_{  +8.19}$ \\ [0.5mm] 
Jet reconstruction efficiency & $^{  -0.04 }_{  +0.04}$ & $^{  +0.07 }_{  -0.07}$ & $^{  -0.12 }_{  +0.12}$ & $^{  -0.29 }_{  +0.29}$ \\ [0.5mm] 
$b$-quark tagging efficiency & $^{  -0.01 }_{  -0.10}$ & $^{  -0.04 }_{  +0.13}$ & $^{  +0.36 }_{  +0.17}$ & $^{  +0.98 }_{  -0.13}$ \\ [0.5mm] 
$c$-quark tagging efficiency & $^{  -0.07 }_{  +0.06}$ & $^{  +0.07 }_{  -0.07}$ & $^{  +0.23 }_{  -0.19}$ & $^{  +0.31 }_{  -0.24}$ \\ [0.5mm] 
Light-jet tagging efficiency & $^{  -0.01 }_{  +0.03}$ & $^{  +0.01 }_{  -0.03}$ & $^{  +0.06 }_{  -0.08}$ & $^{  +0.12 }_{  -0.15}$ \\ [0.5mm] 
$e$ energy resolution & $^{  +0.00 }_{  -0.05}$ & $^{  -0.01 }_{  +0.09}$ & $^{  -0.02 }_{  -0.14}$ & $^{  +0.16 }_{  -0.28}$ \\ [0.5mm] 
$e$ energy scale & $^{  -0.15 }_{  +0.19}$ & $^{  +0.28 }_{  -0.22}$ & $^{  -0.40 }_{  -0.37}$ & $^{  -0.74 }_{  -0.70}$ \\ [0.5mm] 
$\mu$ ID momentum resolution & $^{  -0.14 }_{  -0.01}$ & $^{  +0.20 }_{  +0.01}$ & $^{  +0.04 }_{  -0.03}$ & $^{  -0.15 }_{  -0.10}$ \\ [0.5mm] 
$\mu$ MS momentum resolution & $^{  -0.01 }_{  -0.08}$ & $^{  +0.00 }_{  +0.10}$ & $^{  +0.07 }_{  +0.15}$ & $^{  +0.17 }_{  +0.06}$ \\ [0.5mm] 
$\mu$ momentum scale & $^{  -0.26 }_{  +0.26}$ & $^{  +0.33 }_{  -0.33}$ & $^{  +0.35 }_{  -0.35}$ & $^{  +0.37 }_{  -0.37}$ \\ [0.5mm] 
$\ell$ ID efficiency & $^{  -0.75 }_{  +0.73}$ & $^{  +1.03 }_{  -1.00}$ & $^{  +0.49 }_{  -0.47}$ & $^{  +0.48 }_{  -0.46}$ \\ [0.5mm] 
$\ell$ reconstruction efficiency & $^{  -0.29 }_{  +0.28}$ & $^{  +0.39 }_{  -0.39}$ & $^{  +0.20 }_{  -0.20}$ & $^{  +0.21 }_{  -0.21}$ \\ [0.5mm] 
$\ell$ trigger efficiency & $^{  +0.35 }_{  -0.34}$ & $^{  -0.48 }_{  +0.48}$ & $^{  -0.17 }_{  +0.17}$ & $^{  -0.12 }_{  +0.12}$ \\ [0.5mm] 
\Etmiss{} unassociated cells & $^{  -1.56 }_{  +1.65}$ & $^{  +2.01 }_{  -2.14}$ & $^{  +2.10 }_{  -2.07}$ & $^{  +1.79 }_{  -1.81}$ \\ [0.5mm] 
\Etmiss{} pile-up & $^{  -1.16 }_{  +1.05}$ & $^{  +1.48 }_{  -1.34}$ & $^{  +1.60 }_{  -1.43}$ & $^{  +1.46 }_{  -1.59}$ \\ [0.5mm] 
MC generator & $^{  -3.46 }_{  +4.24}$ & $^{  +4.15 }_{  -5.09}$ & $^{  +7.96 }_{  -9.76}$ & $^{  -1.21 }_{  +1.49}$ \\ [0.5mm] 
Fragmentation & $^{  +0.63 }_{  -0.62}$ & $^{  -0.12 }_{  +0.12}$ & $^{  -6.90 }_{  +6.77}$ & $^{  -2.65 }_{  +2.60}$ \\ [0.5mm] 
IFSR & $^{  -1.29 }_{  +1.19}$ & $^{  +1.04 }_{  -0.96}$ & $^{  +6.22 }_{  -5.76}$ & $^{ +10.25 }_{  -9.49}$ \\ [0.5mm] 
PDF & $^{  -0.07 }_{  +0.08}$ & $^{  +0.06 }_{  -0.06}$ & $^{  +0.24 }_{  -0.25}$ & $^{  +1.30 }_{  -1.35}$ \\ [0.5mm] 
MC statistics & $^{  +0.57 }_{  -0.57}$ & $^{  +0.75 }_{  -0.75}$ & $^{  +1.66 }_{  -1.66}$ & $^{  +2.77 }_{  -2.77}$ \\ [0.5mm] 
\Wboson+jets bb4 & $^{  -0.21 }_{  -0.21}$ & $^{  +0.18 }_{  +0.23}$ & $^{  +0.99 }_{  +0.52}$ & $^{  +1.60 }_{  +0.66}$ \\ [0.5mm] 
\Wboson+jets bb5 & $^{  -0.21 }_{  +0.40}$ & $^{  +0.18 }_{  -0.40}$ & $^{  +0.99 }_{  -1.46}$ & $^{  +1.60 }_{  -2.19}$ \\ [0.5mm] 
\Wboson+jets bbcc & $^{  -0.37 }_{  +0.35}$ & $^{  +0.42 }_{  -0.39}$ & $^{  +0.93 }_{  -0.87}$ & $^{  +1.28 }_{  -1.20}$ \\ [0.5mm] 
\Wboson+jets bbccc & $^{  +0.15 }_{  -0.27}$ & $^{  -0.24 }_{  +0.40}$ & $^{  +0.14 }_{  -0.06}$ & $^{  +0.61 }_{  -0.56}$ \\ [0.5mm] 
\Wboson+jets c4 & $^{  +0.18 }_{  -0.41}$ & $^{  -0.29 }_{  +0.50}$ & $^{  +0.19 }_{  +0.77}$ & $^{  +0.54 }_{  +0.89}$ \\ [0.5mm] 
\Wboson+Jets c5 & $^{  +0.18 }_{  +0.23}$ & $^{  -0.29 }_{  -0.19}$ & $^{  +0.19 }_{  -1.06}$ & $^{  +0.54 }_{  -1.60}$ \\ [0.5mm] 
\Wboson+jets charge asymmetry & $^{  +0.02 }_{  +0.01}$ & $^{  -0.12 }_{  +0.10}$ & $^{  +0.70 }_{  -0.92}$ & $^{  +1.40 }_{  -1.81}$ \\ [0.5mm] 
Multijet normalization & $^{  -0.76 }_{  +0.76}$ & $^{  +1.03 }_{  -1.05}$ & $^{  +0.51 }_{  -0.51}$ & $^{  +0.36 }_{  -0.35}$ \\ [0.5mm] 
Multijet shape & $^{  -0.03 }_{  +0.08}$ & $^{  +0.02 }_{  -0.08}$ & $^{  +0.24 }_{  -0.27}$ & $^{  +0.51 }_{  -0.43}$ \\ [0.5mm] 
$Z+$jets background normalization & $^{  -0.22 }_{  +0.23}$ & $^{  +0.26 }_{  -0.27}$ & $^{  +0.47 }_{  -0.47}$ & $^{  +0.67 }_{  -0.68}$ \\ [0.5mm] 
Dilepton background normalization & $^{  -0.22 }_{  +0.21}$ & $^{  +0.26 }_{  -0.25}$ & $^{  +0.49 }_{  -0.46}$ & $^{  +0.64 }_{  -0.60}$ \\ [0.5mm] 
\hline
\end{tabular}}

\caption{The individual systematic uncertainties calculated as a~percentage of the normalized differential cross-section in each bin.}
\label{tab:CombSyst_3_shape}
\end{center}
\end{table*}
\begin{table*} [htbp]
\footnotesize
\begin{center}

\noindent\makebox[\textwidth]{
\begin{tabular}{l c c c}
\hline
$\frac{1}{\sigma} \frac{\dsigma}{\dabsyttbar}$ Uncertainties [\%] &     0.0--0.5 &     0.5--1.0 &     1.0--2.5 \\ 
\hline
\hline
$b$-quark jets (JES)  & $^{  +0.01 }_{  -0.04}$ & $^{  -0.02 }_{  +0.02}$ & $^{  +0.02 }_{  +0.04}$ \\ [0.5mm] 
Close-by jets (JES) & $^{  -0.24 }_{  +0.13}$ & $^{  -0.02 }_{  +0.01}$ & $^{  +0.44 }_{  -0.24}$ \\ [0.5mm] 
Effective detector NP set 1 (JES) & $^{  -0.13 }_{  +0.09}$ & $^{  +0.01 }_{  +0.01}$ & $^{  +0.20 }_{  -0.17}$ \\ [0.5mm] 
Effective detector NP set 2 (JES) & $^{  +0.00 }_{  +0.00}$ & $^{  -0.01 }_{  +0.01}$ & $^{  +0.00 }_{  -0.01}$ \\ [0.5mm] 
Effective mixed NP set 1 (JES) & $^{  -0.01 }_{  +0.01}$ & $^{  -0.03 }_{  -0.01}$ & $^{  +0.06 }_{  -0.02}$ \\ [0.5mm] 
Effective mixed NP set 2 (JES) & $^{  -0.03 }_{  +0.02}$ & $^{  -0.02 }_{  +0.01}$ & $^{  +0.08 }_{  -0.05}$ \\ [0.5mm] 
Effective model NP set 1 (JES) & $^{  -0.07 }_{  +0.04}$ & $^{  -0.01 }_{  +0.00}$ & $^{  +0.13 }_{  -0.07}$ \\ [0.5mm] 
Effective model NP set 2 (JES) & $^{  +0.01 }_{  -0.01}$ & $^{  +0.00 }_{  -0.00}$ & $^{  -0.02 }_{  +0.01}$ \\ [0.5mm] 
Effective model NP set 3 (JES) & $^{  -0.02 }_{  -0.01}$ & $^{  +0.02 }_{  +0.01}$ & $^{  +0.01 }_{  -0.00}$ \\ [0.5mm] 
Effective model NP set 4 (JES) & $^{  +0.01 }_{  +0.02}$ & $^{  -0.03 }_{  -0.02}$ & $^{  +0.01 }_{  -0.02}$ \\ [0.5mm] 
Effective statistical NP set 1 (JES) & $^{  -0.05 }_{  +0.01}$ & $^{  -0.01 }_{  +0.02}$ & $^{  +0.09 }_{  -0.05}$ \\ [0.5mm] 
Effective statistical NP set 2 (JES) & $^{  -0.02 }_{  +0.02}$ & $^{  -0.01 }_{  -0.01}$ & $^{  +0.04 }_{  -0.01}$ \\ [0.5mm] 
Effective statistical NP set 3 (JES) & $^{  -0.05 }_{  +0.01}$ & $^{  -0.01 }_{  -0.03}$ & $^{  +0.10 }_{  +0.02}$ \\ [0.5mm] 
$\eta$-intercalibration (JES) & $^{  -0.52 }_{  +0.41}$ & $^{  +0.02 }_{  -0.03}$ & $^{  +0.87 }_{  -0.67}$ \\ [0.5mm] 
$\eta$-intercalibration statistics (JES) & $^{  -0.07 }_{  -0.02}$ & $^{  -0.01 }_{  +0.01}$ & $^{  +0.13 }_{  +0.02}$ \\ [0.5mm] 
Flavor composition (JES) & $^{  -0.04 }_{  +0.02}$ & $^{  +0.00 }_{  -0.02}$ & $^{  +0.06 }_{  -0.01}$ \\ [0.5mm] 
Flavor response (JES) & $^{  -0.07 }_{  +0.07}$ & $^{  +0.01 }_{  -0.03}$ & $^{  +0.11 }_{  -0.09}$ \\ [0.5mm] 
Pile-up offset $\mu$ (JES) & $^{  -0.08 }_{  +0.05}$ & $^{  +0.00 }_{  -0.04}$ & $^{  +0.14 }_{  -0.04}$ \\ [0.5mm] 
Pile-up offset $N_{\rm PV}$ (JES) & $^{  -0.06 }_{  -0.03}$ & $^{  +0.01 }_{  +0.03}$ & $^{  +0.09 }_{  +0.01}$ \\ [0.5mm] 
Relative non-closure MC (JES) & $^{  -0.04 }_{  +0.06}$ & $^{  -0.03 }_{  +0.02}$ & $^{  +0.11 }_{  -0.13}$ \\ [0.5mm] 
Single particle high-\pt{} (JES) & $^{  -0.00 }_{  -0.00}$ & $^{  +0.00 }_{  +0.00}$ & $^{  +0.00 }_{  +0.00}$ \\ [0.5mm] 
JES uncertainty in background & $^{  -0.16 }_{  +0.21}$ & $^{  +0.04 }_{  +0.07}$ & $^{  +0.23 }_{  -0.45}$ \\ [0.5mm] 
Jet energy resolution & $^{  +0.12 }_{  -0.12}$ & $^{  +0.14 }_{  -0.14}$ & $^{  -0.38 }_{  +0.38}$ \\ [0.5mm] 
Jet reconstruction efficiency & $^{  +0.01 }_{  -0.01}$ & $^{  +0.00 }_{  -0.00}$ & $^{  -0.02 }_{  +0.02}$ \\ [0.5mm] 
$b$-quark tagging efficiency & $^{  +0.01 }_{  -0.00}$ & $^{  -0.00 }_{  +0.00}$ & $^{  -0.01 }_{  -0.00}$ \\ [0.5mm] 
$c$-quark tagging efficiency & $^{  -0.03 }_{  +0.03}$ & $^{  +0.01 }_{  -0.01}$ & $^{  +0.03 }_{  -0.03}$ \\ [0.5mm] 
Light-jet tagging efficiency & $^{  -0.01 }_{  +0.00}$ & $^{  -0.01 }_{  +0.01}$ & $^{  +0.03 }_{  -0.02}$ \\ [0.5mm] 
$e$ energy resolution & $^{  +0.03 }_{  -0.00}$ & $^{  +0.02 }_{  -0.00}$ & $^{  -0.07 }_{  +0.01}$ \\ [0.5mm] 
$e$ energy scale & $^{  -0.06 }_{  +0.08}$ & $^{  -0.02 }_{  +0.01}$ & $^{  +0.14 }_{  -0.15}$ \\ [0.5mm] 
$\mu$ ID momentum resolution & $^{  +0.01 }_{  +0.03}$ & $^{  -0.01 }_{  -0.02}$ & $^{  +0.01 }_{  -0.01}$ \\ [0.5mm] 
$\mu$ MS momentum resolution & $^{  -0.04 }_{  -0.06}$ & $^{  +0.02 }_{  +0.03}$ & $^{  +0.05 }_{  +0.07}$ \\ [0.5mm] 
$\mu$ momentum scale & $^{  +0.03 }_{  -0.03}$ & $^{  -0.04 }_{  +0.04}$ & $^{  +0.00 }_{  -0.00}$ \\ [0.5mm] 
$\ell$ ID efficiency & $^{  -0.35 }_{  +0.34}$ & $^{  -0.10 }_{  +0.09}$ & $^{  +0.73 }_{  -0.70}$ \\ [0.5mm] 
$\ell$ reconstruction efficiency & $^{  -0.12 }_{  +0.12}$ & $^{  -0.02 }_{  +0.02}$ & $^{  +0.23 }_{  -0.23}$ \\ [0.5mm] 
$\ell$ trigger efficiency & $^{  +0.21 }_{  -0.20}$ & $^{  +0.02 }_{  -0.02}$ & $^{  -0.38 }_{  +0.38}$ \\ [0.5mm] 
\Etmiss{} unassociated cells  & $^{  +0.13 }_{  -0.08}$ & $^{  -0.06 }_{  +0.04}$ & $^{  -0.15 }_{  +0.09}$ \\ [0.5mm] 
\Etmiss{} pile-up & $^{  +0.07 }_{  +0.01}$ & $^{  -0.01 }_{  -0.01}$ & $^{  -0.11 }_{  -0.01}$ \\ [0.5mm] 
MC generator & $^{  -2.00 }_{  +2.48}$ & $^{  -1.19 }_{  +1.48}$ & $^{  +4.97 }_{  -6.19}$ \\ [0.5mm] 
Fragmentation & $^{  -1.89 }_{  +1.81}$ & $^{  -0.83 }_{  +0.80}$ & $^{  +4.31 }_{  -4.15}$ \\ [0.5mm] 
IFSR & $^{  -0.11 }_{  +0.11}$ & $^{  +0.08 }_{  -0.07}$ & $^{  +0.09 }_{  -0.09}$ \\ [0.5mm] 
PDF & $^{  -1.06 }_{  +1.08}$ & $^{  -0.07 }_{  +0.08}$ & $^{  +1.92 }_{  -1.97}$ \\ [0.5mm] 
MC statistics & $^{  +0.19 }_{  -0.19}$ & $^{  +0.03 }_{  -0.03}$ & $^{  +0.29 }_{  -0.29}$ \\ [0.5mm] 
\Wboson+jets bb4 & $^{  -0.08 }_{  +0.04}$ & $^{  +0.01 }_{  -0.00}$ & $^{  +0.11 }_{  -0.07}$ \\ [0.5mm] 
\Wboson+jets bb5 & $^{  -0.08 }_{  +0.03}$ & $^{  +0.01 }_{  -0.01}$ & $^{  +0.11 }_{  -0.04}$ \\ [0.5mm] 
\Wboson+jets bbcc & $^{  -0.25 }_{  +0.24}$ & $^{  +0.04 }_{  -0.03}$ & $^{  +0.39 }_{  -0.37}$ \\ [0.5mm] 
\Wboson+jets bbccc & $^{  -0.05 }_{  +0.01}$ & $^{  -0.02 }_{  +0.01}$ & $^{  +0.11 }_{  -0.03}$ \\ [0.5mm] 
\Wboson+jets c4 & $^{  +0.11 }_{  -0.06}$ & $^{  +0.00 }_{  +0.00}$ & $^{  -0.19 }_{  +0.10}$ \\ [0.5mm] 
\Wboson+Jets c5 & $^{  +0.11 }_{  -0.06}$ & $^{  +0.00 }_{  -0.00}$ & $^{  -0.19 }_{  +0.11}$ \\ [0.5mm] 
\Wboson+jets charge asymmetry & $^{  -0.05 }_{  +0.07}$ & $^{  +0.01 }_{  -0.01}$ & $^{  +0.08 }_{  -0.11}$ \\ [0.5mm] 
Multijet normalization & $^{  -0.35 }_{  +0.35}$ & $^{  -0.13 }_{  +0.13}$ & $^{  +0.77 }_{  -0.77}$ \\ [0.5mm] 
Multijet shape & $^{  -0.09 }_{  +0.09}$ & $^{  -0.05 }_{  +0.05}$ & $^{  +0.22 }_{  -0.22}$ \\ [0.5mm] 
$Z+$jets background normalization & $^{  -0.14 }_{  +0.14}$ & $^{  -0.05 }_{  +0.05}$ & $^{  +0.30 }_{  -0.30}$ \\ [0.5mm] 
Dilepton background normalization & $^{  -0.01 }_{  +0.01}$ & $^{  -0.00 }_{  +0.00}$ & $^{  +0.02 }_{  -0.02}$ \\ [0.5mm] 
\hline
\end{tabular}}

\caption{The individual systematic uncertainties calculated as a~percentage of the normalized differential cross-section in each bin.}
\label{tab:CombSyst_4_shape}
\end{center}
\end{table*}
\clearpage

\section{Statistical Correlations Among Variables}
\label{Sec:Appendix:CorrVars}

\begin{figure*}[!ht]
\centering
{ \includegraphics[width=0.45\textwidth]{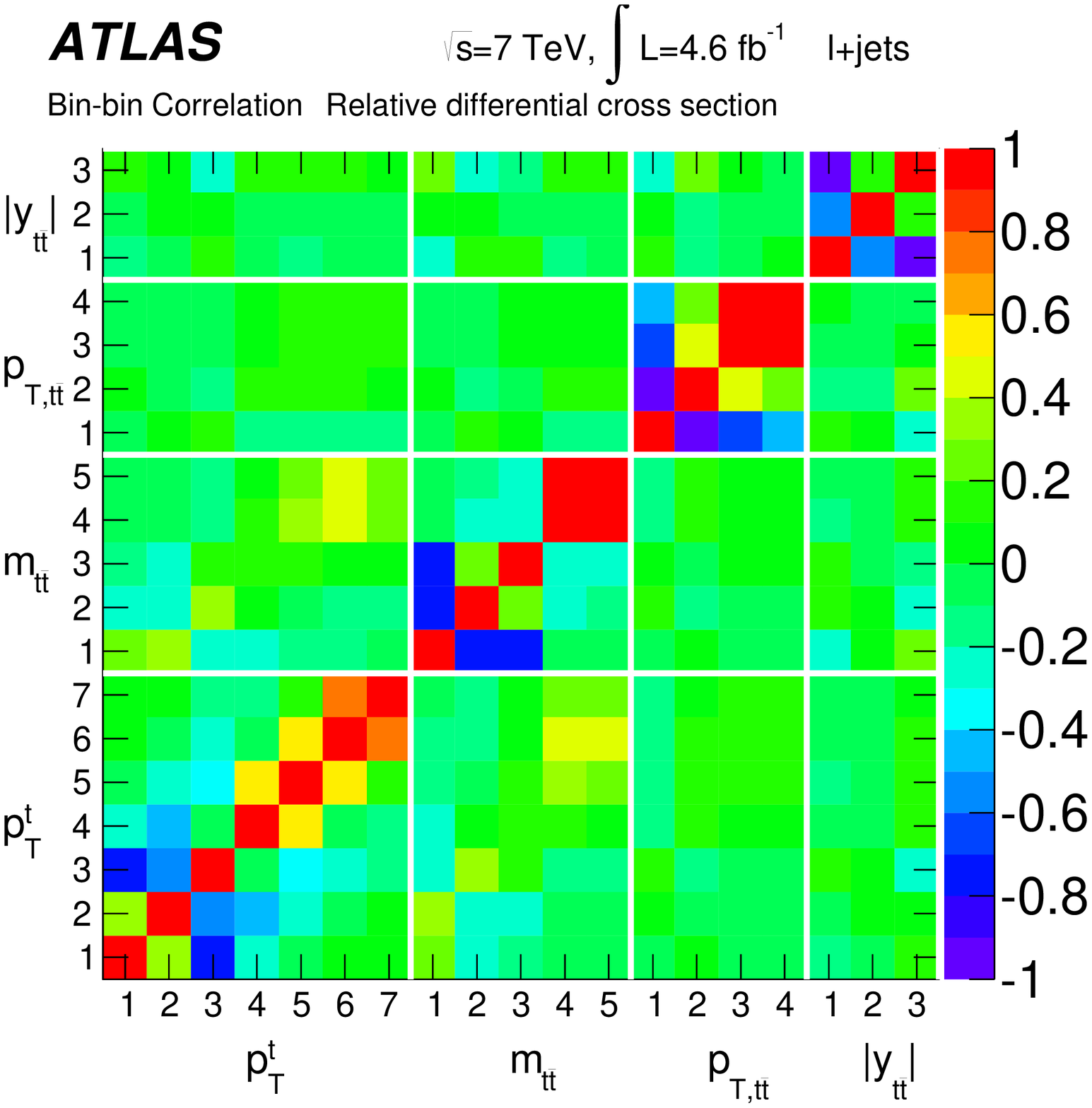}} 
\caption{(Color online) Graphical representation of the statistical correlation matrix between the normalized differential cross-sections. All variables are included to show the correlations between different bins of different variables. From left to right and bottom to top the rows and columns are labeled by bin number for each variable and the variables are ordered: $\ptt$, $\mttbar$, $\ptttbar$, and $\absyttbar$.}
\label{fig:Corr:StatVars}
\end{figure*}

Statistical correlations among the variables are evaluated by unfolding 
statistically coupled (co-varied) replicas of individual spectra in data 
using the ``bootstrap'' method~\cite{bootstrap_bohm2010}. 
The result is obtained by unfolding the separate \ejets{} and \mujets{} spectra, combining with the same procedure used for the nominal result, and normalizing each replica to obtain the normalized differential cross-section. The results are tabulated in Table~\ref{tab:Corr:StatVars} and presented graphically in~Fig.~\ref{fig:Corr:StatVars}.

\begin{table*}[htbp]
\begin{center}
{ \noindent\makebox[\textwidth]{ \begin{tabular}{l|ccccccc|ccccc|cccc|ccc}
 3 & $\phantom{-}$ 0.14 & $\phantom{-}$ 0.02 & $-$0.22 & $\phantom{-}$0.14 & $\phantom{-}$0.16 & $\phantom{-}$0.13 & $\phantom{-}$0.03 & $\phantom{-}$0.23 & $-$0.25 & $-$0.19 & $\phantom{-}$0.18 & $\phantom{-}$0.12 & $-$0.24 & $\phantom{-}$0.28 & $\phantom{-}$0.02 & $-$0.01 & $-$0.90 & $\phantom{-}$0.11 & $\phantom{-}$1.00\\
 2 & $-$0.01 & $\phantom{-}$0.02 & $\phantom{-}$0.04 & $-$0.05 & $-$0.07 & $-$0.05 & $-$0.02 & $\phantom{-}$0.02 & $\phantom{-}$0.07 & $-$0.06 & $-$0.06 & $-$0.04 & $\phantom{-}$0.10 & $-$0.11 & $-$0.02 & $-$0.01 & $-$0.53 & $\phantom{-}$1.00 & $\phantom{-}$0.11\\
 1 & $-$0.12 & $-$0.03 & $\phantom{-}$0.17 & $-$0.10 & $-$0.11 & $-$0.09 & $-$0.02 & $-$0.21 & $\phantom{-}$0.19 & $\phantom{-}$0.19 & $-$0.12 & $-$0.09 & $\phantom{-}$0.16 & $-$0.19 & $-$0.01 & $\phantom{-}$0.01 & $\phantom{-}$1.00 & $-$0.53 & $-$0.90\\
\hline
 4 & $-$0.03 & $-$0.06 & $-$0.02 & $\phantom{-}$0.04 & $\phantom{-}$0.10 & $\phantom{-}$0.14 & $\phantom{-}$0.15 & $-$0.03 & $-$0.01 & $\phantom{-}$0.04 & $\phantom{-}$0.03 & $\phantom{-}$0.02 & $-$0.45 & $\phantom{-}$0.23 & $\phantom{-}$0.92 & $\phantom{-}$1.00 & $\phantom{-}$0.01 & $-$0.01 & $-$0.01\\
 3 & $-$0.03 & $-$0.08 & $-$0.02 & $\phantom{-}$0.08 & $\phantom{-}$0.12 & $\phantom{-}$0.14 & $\phantom{-}$0.14 & $-$0.03 & $-$0.02 & $\phantom{-}$0.04 & $\phantom{-}$0.04 & $\phantom{-}$0.02 & $-$0.65 & $\phantom{-}$0.44 & $\phantom{-}$1.00 & $\phantom{-}$0.92 & $-$0.01 & $-$0.02 & $\phantom{-}$0.02\\
 2 & $\phantom{-}$0.05 & $-$0.06 & $-$0.14 & $\phantom{-}$0.17 & $\phantom{-}$0.18 & $\phantom{-}$0.14 & $\phantom{-}$0.08 & $\phantom{-}$0.09 & $-$0.18 & $-$0.05 & $\phantom{-}$0.19 & $\phantom{-}$0.13 & $-$0.97 & $\phantom{-}$1.00 & $\phantom{-}$0.44 & $\phantom{-}$0.23 & $-$0.19 & $-$0.11 & $\phantom{-}$0.28\\
 1 & $-$0.04 & $\phantom{-}$0.07 & $\phantom{-}$0.12 & $-$0.17 & $-$0.19 & $-$0.16 & $-$0.11 & $-$0.07 & $\phantom{-}$0.16 & $\phantom{-}$0.03 & $-$0.17 & $-$0.12 & $\phantom{-}$1.00 & $-$0.97 & $-$0.65 & $-$0.45 & $\phantom{-}$0.16 & $\phantom{-}$0.10 & $-$0.24\\
\hline 5 & $-$0.01 & $-$0.06 & $-$0.10 & $\phantom{-}$0.03 & $\phantom{-}$0.21 & $\phantom{-}$0.42 & $\phantom{-}$0.27 & $-$0.09 & $-$0.19 & $-$0.24 & $\phantom{-}$0.92 & $\phantom{-}$1.00 & $-$0.12 & $\phantom{-}$0.13 & $\phantom{-}$0.02 & $\phantom{-}$0.02 & $-$0.09 & $-$0.04 & $\phantom{-}$0.12\\
 4 & $-$0.02 & $-$0.07 & $-$0.15 & $\phantom{-}$0.10 & $\phantom{-}$0.33 & $\phantom{-}$0.49 & $\phantom{-}$0.29 & $-$0.06 & $-$0.28 & $-$0.23 & $\phantom{-}$1.00 & $\phantom{-}$0.92 & $-$0.17 & $\phantom{-}$0.19 & $\phantom{-}$0.04 & $\phantom{-}$0.03 & $-$0.12 & $-$0.06 & $\phantom{-}$0.18\\
 3 & $-$0.14 & $-$0.24 & $\phantom{-}$0.12 & $\phantom{-}$0.18 & $\phantom{-}$0.16 & $\phantom{-}$0.03 & $\phantom{-}$0.09 & $-$0.73 & $\phantom{-}$0.23 & $\phantom{-}$1.00 & $-$0.23 & $-$0.24 & $\phantom{-}$0.03 & $-$0.05 & $\phantom{-}$0.04 & $\phantom{-}$0.04 & $\phantom{-}$0.19 & $-$0.06 & $-$0.19\\
 2 & $-$0.25 & $-$0.26 & $\phantom{-}$0.32 & $\phantom{-}$0.09 & $-$0.05 & $-$0.14 & $-$0.14 & $-$0.72 & $\phantom{-}$1.00 & $\phantom{-}$0.23 & $-$0.28 & $-$0.19 & $\phantom{-}$0.16 & $-$0.18 & $-$0.02 & $-$0.01 & $\phantom{-}$0.19 & $\phantom{-}$0.07 & $-$0.25\\
 1 & $\phantom{-}$0.27 & $\phantom{-}$0.36 & $-$0.24 & $-$0.22 & $-$0.19 & $-$0.12 & $-$0.08 & $\phantom{-}$1.00 & $-$0.72 & $-$0.73 & $-$0.06 & $-$0.09 & $-$0.07 & $\phantom{-}$0.09 & $-$0.03 & $-$0.03 & $-$0.21 & $\phantom{-}$0.02 & $\phantom{-}$0.23\\
\hline
 7 & $\phantom{-}$0.05 & $\phantom{-}$0.02 & $-$0.20 & $-$0.17 & $\phantom{-}$0.11 & $\phantom{-}$0.74 & $\phantom{-}$1.00 & $-$0.08 & $-$0.14 & $\phantom{-}$0.09 & $\phantom{-}$0.29 & $\phantom{-}$0.27 & $-$0.11 & $\phantom{-}$0.08 & $\phantom{-}$0.14 & $\phantom{-}$0.15 & $-$0.02 & $-$0.02 & $\phantom{-}$0.03\\
 6 & $\phantom{-}$0.04 & $-$0.09 & $-$0.28 & $-$0.03 & $\phantom{-}$0.50 & $\phantom{-}$1.00 & $\phantom{-}$0.74 & $-$0.12 & $-$0.14 & $\phantom{-}$0.03 & $\phantom{-}$0.49 & $\phantom{-}$0.42 & $-$0.16 & $\phantom{-}$0.14 & $\phantom{-}$0.14 & $\phantom{-}$0.14 & $-$0.09 & $-$0.05 & $\phantom{-}$0.13\\
 5 & $-$0.01 & $-$0.25 & $-$0.36 & $\phantom{-}$0.57 & $\phantom{-}$1.00 & $\phantom{-}$0.50 & $\phantom{-}$0.11 & $-$0.19 & $-$0.05 & $\phantom{-}$0.16 & $\phantom{-}$0.33 & $\phantom{-}$0.21 & $-$0.19 & $\phantom{-}$0.18 & $\phantom{-}$0.12 & $\phantom{-}$0.10 & $-$0.11 & $-$0.07 & $\phantom{-}$0.16\\
 4 & $-$0.26 & $-$0.41 & $-$0.07 & $\phantom{-}$1.00 & $\phantom{-}$0.57 & $-$0.03 & $-$0.17 & $-$0.22 & $\phantom{-}$0.09 & $\phantom{-}$0.18 & $\phantom{-}$0.10 & $\phantom{-}$0.03 & $-$0.17 & $\phantom{-}$0.17 & $\phantom{-}$0.08 & $\phantom{-}$0.04 & $-$0.10 & $-$0.05 & $\phantom{-}$0.14\\
 3 & $-$0.73 & $-$0.60 & $\phantom{-}$1.00 & $-$0.07 & $-$0.36 & $-$0.28 & $-$0.20 & $-$0.24 & $\phantom{-}$0.32 & $\phantom{-}$0.12 & $-$0.15 & $-$0.10 & $\phantom{-}$0.12 & $-$0.14 & $-$0.02 & $-$0.02 & $\phantom{-}$0.17 & $\phantom{-}$0.04 & $-$0.22\\
 2 & $\phantom{-}$0.32 & $\phantom{-}$1.00 & $-$0.60 & $-$0.41 & $-$0.25 & $-$0.09 & $\phantom{-}$0.02 & $\phantom{-}$0.36 & $-$0.26 & $-$0.24 & $-$0.07 & $-$0.06 & $\phantom{-}$0.07 & $-$0.06 & $-$0.08 & $-$0.06 & $-$0.03 & $\phantom{-}$0.02 & $\phantom{-}$0.02\\
 1 & $\phantom{-}$1.00 & $\phantom{-}$0.32 & $-$0.73 & $-$0.26 & $-$0.01 & $\phantom{-}$0.04 & $\phantom{-}$0.05 & $\phantom{-}$0.27 & $-$0.25 & $-$0.14 & $-$0.02 & $-$0.01 & $-$0.04 & $\phantom{-}$0.05 & $-$0.03 & $-$0.03 & $-$0.12 & $-$0.01 & $\phantom{-}$0.14\\
\hline  & $\phantom{-}$1 & 2 & 3 & 4 & 5 & 6 & 7 & $\phantom{-}$1 & 2 & 3 & 4 & 5 & $\phantom{-}$1 & 2 & 3 & 4 & $\phantom{-}$1 & 2 & 3\\
 \end{tabular}
}}
\caption{Statistical correlation matrix between the normalized differential cross-sections. All variables are included to show the correlations between different bins of different variables. From left to right and bottom to top the rows and columns are labeled by bin number for each variable and the variables are ordered: $\ptt$, $\mttbar$, $\ptttbar$, and $\absyttbar$.}
\label{tab:Corr:StatVars}
\end{center}
\end{table*}
 
\clearpage
\label{app:References}
\bibliographystyle{atlasBibStyleWithTitle}
\bibliography{main.bib}

\clearpage

\onecolumngrid
\clearpage 
\begin{flushleft}
{\Large The ATLAS Collaboration}

\bigskip

G.~Aad$^{\rm 84}$,
T.~Abajyan$^{\rm 21}$,
B.~Abbott$^{\rm 112}$,
J.~Abdallah$^{\rm 152}$,
S.~Abdel~Khalek$^{\rm 116}$,
O.~Abdinov$^{\rm 11}$,
R.~Aben$^{\rm 106}$,
B.~Abi$^{\rm 113}$,
M.~Abolins$^{\rm 89}$,
O.S.~AbouZeid$^{\rm 159}$
% ...
\bigskip
\\
% ...
$^{11}$ Institute of Physics, Azerbaijan Academy of Sciences, Baku, Azerbaijan\\
% ...
$^{21}$ Physikalisches Institut, University of Bonn, Bonn, Germany\\
% ...
$^{84}$ CPPM, Aix-Marseille Universit{\'e} and CNRS/IN2P3, Marseille, France\\
% ...
$^{89}$ Department of Physics and Astronomy, Michigan State University, East Lansing MI, United States of America\\
% ...
$^{106}$ Nikhef National Institute for Subatomic Physics and University of Amsterdam, Amsterdam, Netherlands\\
% ...
$^{112}$ Homer L. Dodge Department of Physics and Astronomy, University of Oklahoma, Norman OK, United States of America\\
$^{113}$ Department of Physics, Oklahoma State University, Stillwater OK, United States of America\\
% ...
$^{116}$ LAL, Universit{\'e} Paris-Sud and CNRS/IN2P3, Orsay, France\\
% ...
$^{152}$ Institute of Physics, Academia Sinica, Taipei, Taiwan\\
% ...
$^{159}$ Department of Physics, University of Toronto, Toronto ON, Canada\\
% ...
\end{flushleft}


\end{document}